\documentclass[aps,twocolumn,superscriptaddress,floatfix]{revtex4-2}
\usepackage{graphicx}
\usepackage{latexsym}
\usepackage{amsmath}
\usepackage{amsfonts}
\usepackage{amssymb}
\usepackage{multirow}
\usepackage{bm}
\usepackage{txfonts}
\usepackage{color}

\def\bk{{\textbf{k}}}
\def\bM{{{\textbf{M}}}}
\def\br{{\textbf{r}}}
\def\bq{{\textbf{q}}}
\def\bQ{{\textbf{Q}}}
\def\bl{{\textbf{l}}}
\def\bG{{\textbf{G}}}
\def\ba{{{\bm a}}}

\usepackage{appendix}
\usepackage[colorlinks = true,
            linkcolor = blue,
            urlcolor  = blue,
            citecolor = blue,
            anchorcolor = blue]{hyperref}
\usepackage{ulem}
\usepackage{epstopdf}
\usepackage{multirow}

\begin{document}

\title{Exotic charge density waves and superconductivity on the Kagome Lattice}
	
\author{Rui-Qing Fu }
\affiliation{CAS Key Laboratory of Theoretical Physics, Institute of Theoretical Physics, Chinese Academy of Sciences, Beijing 100190, China}
\affiliation{School of Physical Sciences, University of Chinese Academy of Sciences, Beijing 100049, China}

\author{Jun Zhan}
\affiliation{ Institute of Physics, Chinese Academy of Sciences, Beijing 100190, China}
\affiliation{School of Physical Sciences, University of Chinese Academy of Sciences, Beijing 100049, China}

\author{Matteo D\"{u}rrnagel}
\affiliation{Institut f\"{u}r Theoretische Physik und Astrophysik, Universit\"{a}t W\"{u}rzburg, Am Hubland Campus S\"{u}d, W\"{u}rzburg 97074, Germany}
\affiliation{Institute for Theoretical Physics, ETH Z\"{u}rich, 8093 Z\"{u}rich, Switzerland}

\author{Hendrik Hohmann}
\affiliation{Institut f\"{u}r Theoretische Physik und Astrophysik, Universit\"{a}t W\"{u}rzburg, Am Hubland Campus S\"{u}d, W\"{u}rzburg 97074, Germany}

\author{Ronny Thomale}
\affiliation{Institut f\"{u}r Theoretische Physik und Astrophysik, Universit\"{a}t W\"{u}rzburg, Am Hubland Campus S\"{u}d, W\"{u}rzburg 97074, Germany}

\author{Jiangping Hu}
\affiliation{ Institute of Physics, Chinese Academy of Sciences, Beijing 100190, China}
\affiliation{New Cornerstone Science Laboratory, Beijing 100190, China}

\author{Ziqiang Wang }
\thanks{wangzi@bc.edu}
\affiliation{Department of Physics, Boston College, Chestnut Hill, Massachusetts 02467, USA}

\author{Sen Zhou}
\thanks{zhousen@itp.ac.cn}
\affiliation{CAS Key Laboratory of Theoretical Physics, Institute of Theoretical Physics, Chinese Academy of Sciences, Beijing 100190, China}
\affiliation{School of Physical Sciences, University of Chinese Academy of Sciences, Beijing 100049, China}
\affiliation{CAS Center for Excellence in Topological Quantum Computation, University of Chinese Academy of Sciences, Beijing 100049, China}

\author{Xianxin Wu}
\thanks{xxwu@itp.ac.cn}
\affiliation{CAS Key Laboratory of Theoretical Physics, Institute of Theoretical Physics, Chinese Academy of Sciences, Beijing 100190, China}

%Recent experiments have identified fascinating electronic orders in kagome materials, including intriguing superconductivity, charge density wave (CDW) and nematicity. 	
\begin{abstract}
Loop current order has been long-pursued in various electronic systems, including cuprates and honeycomb lattice materials, but its realization remains elusive in both experiment and theory. 
Intriguingly, recent experimental evidence for AV$_3$Sb$_5$ (A = K,Rb,Cs) and related kagome metals hints at the formation of orbital currents in the charge density wave ordered regime, providing a mechanism for spontaneous time-reversal symmetry breaking in the absence of local moments. However, concrete theoretical model realizations of the loop current order in the kagome lattice has been very challenging and remain an outstanding unresolved problem.
In this work, we comprehensively explore the competitive charge instabilities of the spinless kagome lattice with inter-site Coulomb interactions at the pure-sublattice van Hove filling. 
From the analysis of the charge susceptibility, we find that, at the nesting vectors, while the onsite charge order is dramatically suppressed, the bond charge orders are substantially enhanced owing to the sublattice texture on the hexagonal Fermi surface. 
Furthermore, we demonstrate that nearest-neighbor and next nearest-neighbor
bonds are characterized by significant intrinsic real and imaginary bond fluctuations, respectively.
The 2$\times$2 loop current order is thus favored by the next nearest-neighbor Coulomb repulsion. 
Interestingly, increasing interactions further leads to a nematic state with intra-cell sublattice density modulation that breaks the $C_6$ rotational symmetry.
We further explore superconducting orders descending from onsite and bond charge fluctuations, and discuss our model's implications on the experimental status quo.
\end{abstract}
\maketitle
	
\section{Introduction}
Exploring novel quantum states has been a central theme in contemporary condensed matter physics. One long-sought-after state is the exotic staggered flux phase, proposed from different perspectives in the same year: for strongly correlated systems on a square lattice~\cite{PhysRevB.37.3774} and for topological phases of matter on a  honeycomb lattice~\cite{PhysRevLett.61.2015}. In hole doped cuprates, the staggered flux phase is associated with a staggered electric current order on the square lattice~\cite{Wangkotliarwang-PRB90,Zhang-PRL90,HsuMarstonAfflect-PRB91}, which is closely related to the intra-unit-cell copper-oxygen loop current order \cite{PhysRevB.55.14554} and the d-density wave order~\cite{PhysRevB.63.094503}. 
This order is a charge bond order with imaginary order parameters and features looped current patterns along the bonds in lattices.
These proposed exotic phases of matter spontaneously break the time-reversal symmetry (TRS) due to orbital circulating currents without spin-related magnetism. When the lattice transition symmetry is additionally broken~\cite{Wangkotliarwang-PRB90,Zhang-PRL90,HsuMarstonAfflect-PRB91,PhysRevB.63.094503}, they describe the highly unconventional TRS breaking loop-current charge density wave (CDW) states. The loop current order was suggested as a hidden order in the pseudogap phase in underdoped cuprates~\cite{PhysRevLett.83.3538,PhysRevB.63.094503,ZhouWang-PRB04} and loop current fluctuations have been shown to provide d-wave pairing instability~\cite{PhysRevB.55.14554,Wangkotliarwang-PRB90}.
In Haldane model for spinless fermions on the honeycomb lattice~\cite{PhysRevLett.61.2015}, this flux phase breaks the time-reversal symmetry and leads to quantum anomalous Hall (QAH) insulator, realizing quantum Hall effect without Landau levels. It has become a prototype example for exploring topological phases of matter~\cite{Qi2011,Kane2010}. 

Despite its physical significance, concrete theoretical model realizations of loop current order in the ground state have been very challenging. In the limit of strong electron-electron correlation, achieving loop current order beyond the mean-field level remains controversial, partly due to partially conflicting many-body numerical evidence from the analysis of finite size systems~\cite{PhysRevLett.102.017005,PhysRevLett.99.027005,PhysRevB.77.094511,PhysRevLett.112.117001,PhysRevB.90.224507,PhysRevB.70.113105,PhysRevB.71.075103}. While the induced orbital magnetism has uniquely testable experimental signatures, the experimental status of loop currents in the strongly correlated cuprates, in particular whether they live up to energy scales relevant to High-Tc superconductivity, is still heavily debated~\cite{CRPHYS_2021__22_S5_7_0,PhysRevB.96.214504}. For the honeycomb lattice at Dirac filling, the next nearest-neighbor repulsion was suggested to promote a loop current order, realizing an interaction-driven topological Haldane model~\cite{PhysRevLett.100.156401}. However, unbiased numerical calculations, such as exact diagonalization, density-matrix renormalization-group and functional renormalization group studies~\cite{PhysRevB.88.245123,PhysRevB.89.035103,PhysRevB.92.085147,PhysRevB.92.085146,PhysRevB.92.155137}, reveal that the true ground state is a charge order state with trivial topology. 
Additionally, while quadratic band touching point was proposed to generate a QAH state from weak-coupling limit ~\cite{PhysRevLett.103.046811}, its realization in lattice model requires intricate interaction setting~\cite{PhysRevB.82.075125,PhysRevLett.117.096402}. In both square and honeycomb lattice systems, the loop current bond order always faces strong competition from onsite orders either in charge or spin channel, making it less dominant. Therefore, devising a solid microscopic foundation for such a state beyond mean-field theory and biased variational methods has been particularly challenging.

The recent discovery of kagome metals AV$_3$Sb$_5$ (A = K,Rb,Cs)~\cite{AV3Sb5_Ortiz_first_paper,AV3Sb5_nature_review,jiangping_hu_review} and FeGe~\cite{PhysRevLett.129.166401,TengXK2022,TengXK2023}, with the Fermi level close to van Hove singularities (VHSs)~\cite{YHu2022,MKang2022,TengXK2022,TengXK2023}, offers new opportunities and fresh new ideas to explore this important issue on the geometrically frustrated kagme lattice. In non-magnetic AV$_3$Sb$_5$, a CDW order with translational symmetry breaking (TSB) occurs~\cite{YJiang2021,PhysRevX.11.031026} and exhibits signatures of TRS from various experimental measurements~\cite{Mielke2022,YuL2021,yang2020giant,xu2022three,2021arXiv211011306W,GuoCY2022,2022arXiv220410116X,xing2024optical}. 
In the kagome magnet FeGe, where magnetic moments align ferromagnetically within each layer and antiferromagnetically between layers, the emergence of CDW with TSB  induces an enhancement in the magnetic moment and anomalous Hall effect~\cite{TengXK2022}. This evidence implies the potential presence of loop current order in kagome metals with a signal strength and data diversity far beyond existing evidence in cuprates~\cite{FENG20211384,PhysRevLett.127.217601,PhysRevB.104.045122,PhysRevB.104.035142,SZhou2022,Chistensen2022}. 
The Fermi surface nesting vector associated with these VHSs is consistent with the in-plane wavevector of CDW, implying their crucial role in promoting CDW. 

The VHSs in the kagome lattice exhibit a unique sublattice texture, leading to matrix element reduction effects of scattering channels between van Hove (VH) points, dubbed as sublattice interference (SI)~\cite{PhysRevB.86.121105, wu_PRL_21}. Whenever the accumulation of electronic density of states at the VH points represents a relevant contribution to the formation of Fermi surface instabilities, it is to be expected that SI could have a crucial impact on the nature of electronic order. Despite intensive theoretical studies, the question of whether electronic interactions within the kagome lattice, intertwined with SI, can give rise to loop current states at VH filling remains an open issue. Studies using the functional renormalization group approach applied to the t-U-V model have not identified the presence of such orders~\cite{PhysRevB.86.121105,SYu2012,WWang2013,PhysRevLett.110.126405,PhysRevB.109.075127,2024arXiv240211916P}. Mean-field analyses, however, suggest that longer-range interactions, in particular the next nearest-neighbor interaction, could have a crucial impact on loop current order~\cite{PhysRevB.107.045127}. Additionally, the topological loop current order can be promoted by bond order fluctuations~\cite{Tazai2022,Tazai2023}. Few physical insights into this order within the kagome lattice have been elucidated.

In this article, we seek to perform model building that aims at providing a microscopic foundation for loop current order. Our work is guided by the core idea that SI in kagome metals could set the stage for loop current order reachable through a full scale many-body analysis beyond {mean field approximation} and finite size studies. As we delve into the intrinsic charge orders of the kagome lattice, we further reduce complexity by examining a single orbital spinless fermion model on the kagome lattice, similar to the Haldane model on the honeycomb lattice, in the presence of inter-site Coulomb interactions. To understand the nature of the loop current order, we employ the random phase approximation (RPA) approach, which allow us to treat both onsite and bond order on equal footing.
{To go beyond mean-field approximation that neglects charge fluctuations,} we employ the random phase approximation (RPA) approach, which allows us to treat both onsite and bond order on equal footing. Our analysis of charge susceptibilities reveals that while the onsite charge order will be suppressed, the bond charge order gets enhanced owing to SI from the sublattice texture associated with the $p$-type VHS. Moreover, facilitated by the unique lattice's geometry, we observe that the nearest-neighbor (NN) and next NN (NNN) bonds exhibit pronounced intrinsic real and imaginary bond charge fluctuations, respectively. The emergence of a $2\times2$ loop current order CDW as the ground state is favored when the NNN repulsion is sufficiently strong, while dominant NN repulsion favors a $2\times2$ trihexagonal (inverse star of David) charge bond order. Strong inter-site replusion can stabilize a novel nematic state characterized by charge density modulations coupled with bond charge modulations  within the unit cell. We further explore the nature of triplet superconductivity away from VH filling descending from such exotic charge orders within our model, where $p$- and $f$-wave pairings emerge due to bond charge fluctuations. Finally, we discuss possible experimental implications and contemplate on future quantitative theoretical studies that build upon the conceptual narrative outlined in our work. 

\begin{figure}
\centering
\includegraphics[width=1.0\linewidth]{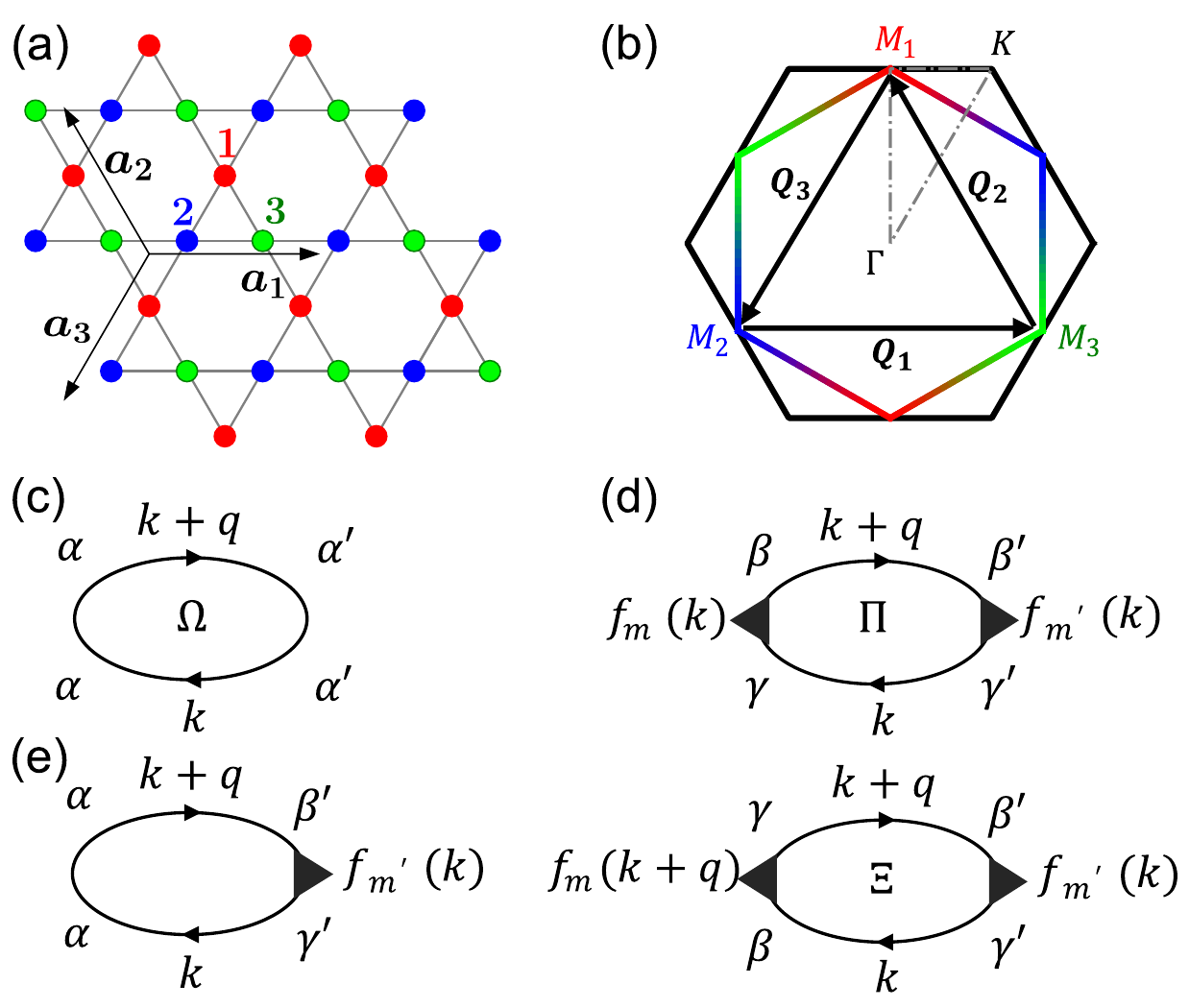}
\caption{Kagome lattice and charge susceptibility bubbles. (a) Kagome lattice and (b) the sublattice-resolved Fermi surface at the $p$-type van Hove filling, with three sublattice indicated by red (1), blue (2) and green (3) circles.
The two basis lattice vectors are $\ba_1=(1,0)$, $\ba_2=(-\frac{1}{2},\frac{\sqrt{3}}{2})$, with the third vector $\ba_3=-\ba_1-\ba_2$. 
Feynman diagrams for susceptibilities in the onsite (c), bond (d), and mixed (e) channels. 
$\Omega$ is the onsite susceptibility and $\Pi$ and $\Xi$ are two bond susceptibilities, with $f_m(\bk)$ describing the corresponding form factors.}
\label{Fig1}
\end{figure}

\section{tight-binding model and susceptibilities of onsite and bond charge orders}
The kagome lattice consists of corner-sharing triangles with three sublattices, as shown in Fig.~\ref{Fig1}(a). 
In analogous to the Haldane model, we consider the spinless kagome model and there are three motivations. First, as we are avoiding complicated magnetic orders via a frozen spin scenario, we can thoroughly study the effect of sublattice texture on charge fluctuations. Second, given the absence of magnetic phases for most kagome metals of our interest, the amount of competing density wave orders removed through this simplification is highly limited, and hence allows to draw rather accurate implications for the spinful model. Third, the model is directly relevant to experiments, like antiferromagnet FeGe.
The kinetic energy is described by the tight-binding Hamiltonian, 
\begin{eqnarray}
\mathcal{H}_0=-t\sum_{\langle \br \br'\rangle,\alpha\neq\beta} c^\dag_{\alpha}(\br) c_{\beta}(\br') -\mu \sum_{\br, \alpha} n_{\alpha}(\br),
\end{eqnarray}
where $c^\dag_{\alpha}(\br)$ and $c_{\alpha}(\br)$ are the creation and annihilation operators of an electron at the lattice site $\br$, $\alpha=1,2,3$ is the sublattice index, $\langle \br\br'\rangle$ denotes the NN sites, and $n_{\alpha}(\br)=c^\dag_{\alpha}(\br) c_{\alpha}(\br)$ is the electron density operator. 
$t$ and $\mu$ are the NN hopping parameter and chemical potential, respectively. 
Defining $t$ as the unit of energy, we set $t=1$ from now on. 
The tight-binding band structures feature two VHSs, one Dirac cone, and a flat band. 
In particular, the Fermi surfaces at the two van Hove (VH) fillings are characterized by distinct sublatice textures \cite{PhysRevB.86.121105, wu_PRL_21}. 
In this work, we focus on the upper VH case at the pristine filling, i.e. the $p$-type VHS \cite{PhysRevB.86.121105, wu_PRL_21}, and the sublattice-resolved Fermi surface is displayed in Fig.~\ref{Fig1}(b).
Clearly, the wavefunction at each saddle point (i.e. $\bM$ point) is attributed to a single sublattice, while at the midpoint between two saddle points, the wavefunction exhibits a mixture of two sublattices. 
The Fermi surface nesting vectors $\bQ_{1,2,3}$ always connect distinct sublattice characters of states around the three saddle points, leading to substantial bond fluctuations, as will be demonstrated in the subsequent analysis.

To examine the intrinsic fluctuations, we consider the relevant charge orders on the kagome lattice. 
The first one is charge modulation, i.e., the onsite charge order, and it is described by the operator $n_{\alpha}(\br)$ in real space. 
In momentum space, this operator reads $n_{\alpha}(\bq)=\frac{1}{\sqrt{N}} \sum_{\br} e^{-i\bq\cdot\br} n_{\alpha}(\br) =\frac{1}{\sqrt{N}} \sum_{\bk} c^\dag_{\alpha} (\bk+\bq) c_{\alpha}(\bk)$. 
We additionally consider bond charge modulation, i.e. charge bond order, on NN and NNN bonds.
Due to the unique geometry of the kagome lattice, within each unit cell there are two NN (NNN) bonds along the direction parallel (perpendicular) to each basis vector $\ba_\alpha$, and they all connect two distinct sublattices $\beta$ and $\gamma$, with the Levi-civta symbol satisfying $\epsilon_{\alpha\beta\gamma}=1$, i.e., $(\alpha, \beta, \gamma)=$ (1, 2, 3), (2, 3, 1), and (3, 1, 2).
This allows us to define the symmetric ($+$) and antisymmetric ($-$) bond operators \cite{SZhou2022, PhysRevB.107.045127}
\begin{align}
B_{\alpha,+,\eta} (\br)&=\frac{1}{2} \left[ c_\beta^\dagger(\br) c_\gamma(\br+\bl_{\alpha,\eta}) + c_\beta^\dagger(\br) c_\gamma(\br-\bl_{\alpha,\eta}) \right], \\
B_{\alpha,-,\eta} (\br)&=\frac{i}{2} \left[ c_\beta^\dagger(\br) c_\gamma(\br+\bl_{\alpha,\eta}) - c_\beta^\dagger(\br) c_\gamma(\br-\bl_{\alpha,\eta}) \right], \label{antibond}
\end{align}
where $\eta$ denotes NN and NNN bonds, and the corresponding displacement vectors connecting the two sites of $\beta$ and $\gamma$ sublattices are $\bl_{\alpha,\text{NN}}=\frac{1}{2} \ba_\alpha$ and $\bl_{\alpha,\text{NNN}}=\frac{1}{2} (\ba_\beta- \ba_\gamma)$, respectively. 
Performing Fourier transformation, the bond order operators in the momentum space can be written as,
\begin{equation}
B_{\alpha,\pm,\eta}(\bq)=\frac{1}{\sqrt{N}} \sum_{\bk} f_{\alpha,\pm,\eta} (\bk) c^\dagger_{\beta}(\bk+\bq) c_{\gamma}(\bk), \label{Bk}
\end{equation}
with $f_{\alpha,+,\eta}(\bk) =\cos(\mathbf{k\cdot l}_{\alpha,\eta})$ and $f_{\alpha,-,\eta}(\bk) = \sin(\mathbf{k \cdot l}_{\alpha,\eta})$ being the form factors of symmetric and antisymmetric bonds, respectively. 
We note that the antisymmetric bond defined in this work differs from that in Refs. \cite{SZhou2022} and \cite{PhysRevB.107.045127} by a factor of $i$, which leads to a real form factor in Eq. (\ref{Bk}).
Clearly, including NN and NNN bonds, there are in total 12 independent bond orders indexed by $(\alpha,\pm,\eta)$ within each unit cell.
To simplify their indices, we introduce one-dimensional indices $m,n= \{1,2,\cdots,12\}$ $=\{({1,+,\text{NN}})$, $({1,-,\text{NN}})$, $1\rightarrow2,3$, $\text{NN}\rightarrow\text{NNN}\big\}$.
Note that the bond orders are complex in general, we thus introduce their conjugate partners as well, $B^\dag_{\alpha,\pm,\eta}(\br) \equiv [B_{\alpha,\pm,\eta}(\br)]^\dag$ in real space and, consequently, 
$[B_{\alpha, \pm, \eta} (\bq)]^\dagger =B_{\alpha, \pm, \eta}^\dagger (-\bq)$ in momentum space.

To investigate the intrinsic fluctuations of different charge orders, we calculate the corresponding susceptibilities defined as,
\begin{equation}
\chiup_{pq} (\bq,i\omega_n) =  \int_0^\beta d \tau e^{i\omega_n \tau} \langle T_\tau \mathcal{O}_p (\bq,\tau) [\mathcal{O}_q (\bq,0)]^\dagger \rangle.
\end{equation}
Here the operator $\mathcal{O}_p$ runs over the 27 charge orders mentioned above, consisting of 24 bond orders in the order of $\{B_1, B^\dagger_1, B_2, B^\dagger_2, \cdots B_{12}, B^\dagger_{12}\}$ followed by the 3 onsite charge orders \{$n_1$, $n_2$, $n_3$\}. The bare static susceptibility is given by $\chiup^0_{pq} (\bq)\equiv\chiup_{pq} (\bq,0)$.
For the convenience of discussion, we use different notations to distinguish the susceptibilities of onsite and bond charge orders in the following, and we further note that the latter can be categorized into two types.
Explicitly, $3\times3$ susceptibility matrix for onsite charge orders $\Omega^0_{\alpha \beta}$ = $\chiup^0_{24+\alpha, 24+\beta}$, and $24\times24$ susceptibilities for bond charge orders $\Pi^0_{mn}$ = $\chiup^0_{2m-1,2n-1}$ = $\chiup^0_{2m,2n}$ and $\Xi^0_{mn}$ = $\chiup^0_{2m-1,2n}$ = $\chiup^0_{2m,2n-1}$ with $m,n=1,\cdots,12$ {corresponding to the 12 independent bond order indices introduced in the above paragraph}.  {$\Omega$ operators represent onsite charge fluctuations, while $\Pi$ and $\Xi$ operators represent bond fluctuations characterized by the correlation between $B/B^\dag$ and $B^\dag/B$, and between $B/B^\dag$ and $B/B^\dag$, respectively.}
The corresponding Feynman diagrams for $\Omega$ are just the normal bubbles while those for $\Pi$ and $\Xi$ carry two additional vertices of form factors, as depicted in Fig.~\ref{Fig1}(c) and Fig.~\ref{Fig1}(d).
The analytical expressions of $\Omega$, $\Pi$, and $\Xi$ are given by
\begin{align}
\Omega^0_{\alpha\beta}(\bq)= -\frac{T}{N} \sum_{\bk,l} & G^0_{\beta \alpha} (\bk+\bq, i\omega_l) G^0_{\alpha \beta} (\bk,i\omega_l), \label{Omega}\\
\Pi^0_{mn}(\bq)= -\frac{T}{N} \sum_{\bk,l} & f_m(\bk) f_n(\bk) \nonumber \\
 \times& G^0_{\beta_n \beta_m} (\bk+\bq, i\omega_l)  G^0_{\gamma_m \gamma_n}(\bk, i\omega_l), \label{Pi} \\
\Xi^0_{mn} (\bq)= -\frac{T}{N} \sum_{\bk,l}&  f_m(\bk+\bq) f_n(\bk) \nonumber \\ 
 \times & G^0_{\beta_n \gamma_m} (\bk+\bq, i\omega_l) G^0_{\beta_m \gamma_n} (\bk,i\omega_l), \label{Xi}
\end{align}
where the noninteracting Green's function $G^0_{\beta \gamma} (\bk,i\omega_l)=\sum_\nu a_{\beta\nu} (\bk) a^*_{\gamma\nu} (\bk) /(i\omega_l-\epsilon_{\nu\bk})$ with $\epsilon_{\nu\bk}$ being the $\nu$-th eigen energy of $\mathcal{H}_0$ and $a_{\beta\nu}(\bk)$ the corresponding eigen state. The summation over the fermionic Matsubara frequency $\omega_l=(2l+1)\pi k_BT$ at temperature $T$ yields the lindhard function and sublattice-associated matrix elements with the detailed formulas given in the supplementary material (SM). 
These sublattice characters embedded in the noninteracting Green's functions play a predominant role in determining the behavior of suscpetibilities.
The diagram with only one vertex displayed in Fig.~\ref{Fig1}(e) represents the susceptibility in the mixed channel that couples the onsite and bond charge orders.

Before presenting the numerical data, we analyse the contributions to the bare susceptibilities from the VH points.
At the $p$-type VH filling, the hexagonal Fermi surface encompasses the three inequivalent VH points labeled by $\bM_{1,2,3}$ at the zone boundary and features perfect nesting with three  wave vectors $\bQ_\alpha = {1\over 2}\bG_\alpha$, where $\bG_\alpha$ denotes the reciprocal wave vector of the kagome lattice. 
These VH points with diverging density of states (DOS) are expected to contribute dominantly to the susceptibilities, especially at the two pertinent vectors, $\bq=0$ and $\bq= \bQ_\alpha  \equiv \bM_\alpha$.
Furthermore, as mentioned before and shown in Fig.~\ref{Fig1}(b), the Bloch states at $\bM_\alpha$ points are exclusively localized on the $\alpha$th sublattice.
Consequently, the Green's functions at $\bM_\alpha$ are non vanishing only for $G^0_{\alpha\alpha}(\bM_\alpha)$.
It is thus straightforward to show that these VH points contribute only to the diagonal elements of the onsite charge susceptibilities at $\bq=0$, $\Omega_{\alpha \alpha}(0)$, while their contributions to off-diagonal elements of $\Omega(0)$ and all bond susceptibilities, $\Pi(0)$ and $\Xi(0)$, vanish. This results a dominant onsite charge fluctuation at $\bq=0$.
The situation is, however, completely the opposite for wave vector $\bq=\bM_\alpha$.
The two VH points $\bM_\beta$ and $\bM_\gamma$ connected by $\bq=\bQ_\alpha$ feature pure $\beta$th and $\gamma$th sublattice, respectively. This feature leads to the vanishing contribution of VH points in the onsite charge fluctuation $\Omega^0_{\alpha'\alpha''}(\bM_\alpha)$. But these two VH points can thus dominantly contribute to the susceptibilities $\Pi(\bM_\alpha)$ of the bonds that connecting $\beta$ and $\gamma$ sublattices.
We note that they have no contributions to $\Xi(\bM_\alpha)$ since at least one of the two Green's function in Eq. (\ref{Xi}) involves mixed sublattices. This indicates predominant bond fluctuations at $\bq=\bM_\alpha$ rather onsite charge fluctuations.
Furthermore, since $\bM_{\beta/\gamma} \cdot \bl_{\alpha,\text{NN}} ={\pi \over 2}$ and $\bM_{\beta/\gamma} \cdot \bl_{\alpha,\text{NNN}} =\pm {\pi \over 2}$, the contribution from these two VH points to the susceptibilities of symmetric bonds with form factors $\cos(\mathbf{k\cdot l}_{\alpha,\eta})$ also vanishes.
Therefore, at the wave vector $\bq=\bM_\alpha$, the two connected VH points at $\bM_\beta$ and $\bM_\gamma$, with $\epsilon_{\alpha\beta\gamma}=1$, contribute only to the elements of $\Pi(\bM_\alpha)$ associated with the antisymmetric bonds that connects $\beta$ and $\gamma$ sublattices.
Explicitly, taking $\bq=\bM_1$ as an example, the VH points contribute only to susceptibility $\Pi_{22}(\bM_1)$ for bond $B_2 =B_{1,-,\text{NN}}$, $\Pi_{88}(\bM_1)$ for bond $B_8=B_{1,-,\text{NNN}}$, and $\Pi_{28} (\bM_1)$ that couples $B_2$ and $B_8$.
Clearly, the unique sublattice texture at the $p$-type VH filling plays a pivotal role in suppressing the onsite charge fluctuations at wavevector $\bq=\bM_\alpha$ but significantly promoting  the bond charge fluctuations in the antisymmetric channel. 
This behavior in the kagome lattice markedly differs from what is observed in the triangular and honeycomb lattices, where onsite charge fluctuations are dominant \cite{PhysRevB.103.235150, Gneist2022}. 
Additionally, it is readily shown that the contribution from the VH points to the susceptibilities in the mixed channels depicted in Fig.~\ref{Fig1}(e) vanishes as well at both $\bq=0$ and $\bq=\bM_\alpha$.

\begin{figure}
\centering
\includegraphics[width=1.0\linewidth]{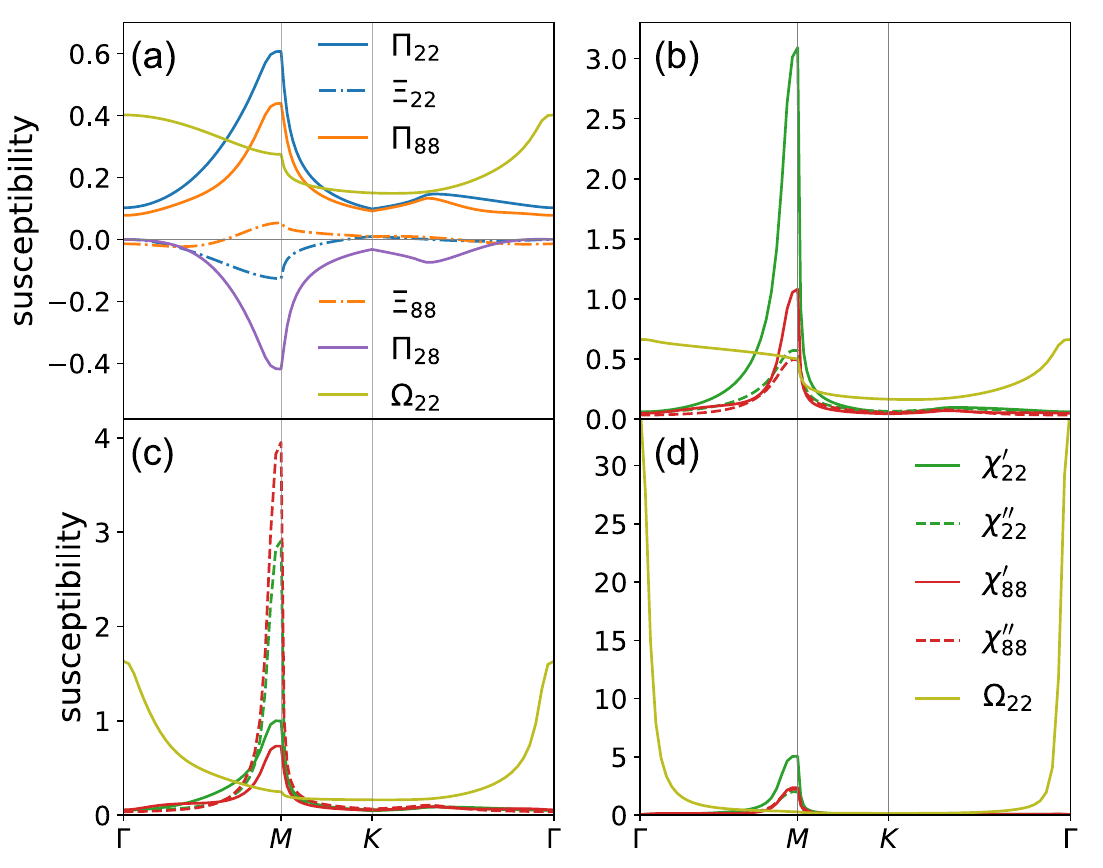}
\caption{Bare and RPA susceptibilities along the high-symmetry path. (a) Representative components of the bare susceptibility for onsite and bond charge orders. 
Representative components of the RPA susceptibilities for onsite and bond charge orders at various inter-site Coulomb interactions: (b) $V_{\text{NN}}=0.6$, $V_{\text{NNN}}=0.0$, (c) $V_{\text{NN}}=0.0$, $V_{\text{NNN}}=0.95$, and (d) $V_{\text{NN}}=0.5$, $V_{\text{NNN}}=0.75$. 
We adopt $k_B T = 0.005$ in a-c, while a higher temperature $k_B T = 0.01$ in d is used to avoid divergence. {Explicit expressions of susceptibilities in the legend are listed  in abbreviated form: $\Pi_{22}=\left\langle B_{1,-,NN} B^\dag_{1,-,NN} \right\rangle$, $\Xi_{22}=\left\langle B_{1,-,NN} B_{1,-,NN} \right\rangle$, $\Pi_{88}=\left\langle B_{1,-,NNN} B^\dag_{1,-,NNN} \right\rangle$, $\Xi_{88}=\left\langle B_{1,-,NNN} B_{1,-,NNN} \right\rangle$, $\Pi_{28}=\left\langle B_{1,-,NN} B^\dag_{1,-,NNN} \right\rangle$, $\Omega_{22}=\left\langle n_2 n_2 \right\rangle$.}
}
\label{Fig2}
\end{figure}

The calculated bare susceptibilities are presented in detail in the SM, with the representative elements displayed in Fig.~\ref{Fig2}(a) along the high-symmetry path $\Gamma$-$\bM_1$-$\textbf{K}$-$\Gamma$ depicted in Fig.~\ref{Fig1}(b).
A temperature of $k_BT = 0.005$ is applied in the calculation, under which the dominant fluctuations are reflected by the peaks at $\bq=0$ and $\bq=\bM_\alpha$.
Indeed, as suggested by the above analysis of the contribution from the VH points, the diagonal elements of onsite charge susceptibilities $\Omega_{\alpha\alpha}$ dominate at $\bq =0$, while the leading susceptibilities at $\bq=\bM_1$ are $\Pi_{22}$, $\Pi_{88}$, and $\Pi_{28}$ of the antisymmetric bonds connecting the two sites of 2nd and 3rd sublattices.
The large $\Pi^0_{28}(\bM_1)$ shown in Fig.~\ref{Fig2}(a) indicates the strong coupling between the NN and NNN antisymmetric bond orders. 
These results promote the leading fluctuations in  the antisymmetric bond channel, instead of symmetric bond or onsite charge channels.
However, the nature of the antisymmetric bond order is yet to be explored.

To reveal the nature of the antisymmetric bond order which exhibits the leading fluctuation at the nesting wavevector $\bM_\alpha$, we further separate the bond orders into their real and imaginary parts, 
\begin{align}
B'_m(\bq) & = \frac{1}{2} [B_m + B^\dagger_m], \text{  }
B''_m(\bq)  = \frac{1}{2i} [B_m - B^\dagger_m], \text{  }
 m\in \text{odd}, \\ 
B'_m(\bq) & = \frac{1}{2i} [B_m - B^\dagger_m], \text{  }
B''_m(\bq)  = \frac{1}{2} [B_m + B^\dagger_m], \text{  }
 m\in\text{even}. 
\end{align}
Here the definition in the antisymmetric bond channel is different because of the additional factor $i$ used in Eq. (\ref{antibond}).
$B'_m$ and $B''_m$ represent the hopping and current modulation on the bonds, respectively. 
It is straightforward to show that the static susceptibilities of the real and imaginary bond charge orders can be respectively rewritten as,
\begin{align}
\chiup^{\prime}_{mm} (\bq)&= [\Pi_{mm}(\bm{q}) - (-1)^m \Xi_{mm}(\bm{q})]/2 ,\nonumber\\
\chiup^{\prime\prime}_{mm}(\bm{q})&= [\Pi_{mm}(\bm{q}) + (-1)^m \Xi_{mm}(\bm{q})]/2. \label{chiRI}
\end{align}
Clearly, the relative strength of real and imaginary bond fluctuations is dictated by the sign of $\Xi(\bM_\alpha)$. According to previous line of reasoning, the VH points contribute nothing to  $\Xi$, rendering these fluctuations degenerate when considering only states at VH points.
As a result, one has to go \textit{beyond} the VH points and consider the contributions from other portions of the hexagonal FS to determine the sign of $\Xi(\bM_\alpha)$.

We consider the diagonal elements of $\Xi$ at $\bq=\bM_\alpha$ for the antisymmetric bonds connecting $\beta$ and $\gamma$ sublattice, i.e., NN bond $B_{2\alpha}$ and NNN bond $B_{6+2\alpha}$, that are tied to the leading fluctuation.
Because of the unique geometry of the kagome lattice, the nesting and connecting vectors satisfy $\bl_{\alpha,\text{NN}}\parallel \bQ_{\alpha}$ and $\bl_{\alpha,\text{NNN}}\perp \bQ_{\alpha}$, leading to $\bQ_\alpha \cdot \bl_{\alpha,\text{NN}} =\pi$ and $\bQ_\alpha \cdot \bl_{\alpha,\text{NNN}} =0$.
Consequently, for $\bk$ on the hexgonal FS, the two form factors in $\Xi_{2\alpha,2\alpha}(\bQ_\alpha)$ for the NN antisymmetric bond read $f_{2\alpha}(\bk+\bQ_\alpha) f_{2\alpha}(\bk)= -\sin^2(\bk\cdot \bl_{\alpha,\text{NN}}) \leq 0$, whereas the two form factors in $\Xi_{6+2\alpha,6+2\alpha}(\bQ_\alpha)$ for the NNN antisymmetric bond are given by $f_{6+2\alpha}(\bk+\bQ_\alpha) f_{6+2\alpha}(\bk)= \sin^2(\bk\cdot \bl_{\alpha,\text{NNN}}) \geq 0$.
These distinctive characteristics suggest that $\Xi_{2\alpha,2\alpha} (\bQ_\alpha)$ and $\Xi_{6+2\alpha,6+2\alpha}(\bQ_\alpha)$ have the opposite signs, pointing to the different nature of bond fluctuations on the NN and NNN bonds.
Indeed, as shown in Fig.~\ref{Fig1}(a), $\Xi_{22}$ is negative while $\Xi_{88}$ is positive at $\bq=\bM_1$ due to the positive Green's function related sublattice factors.
From the Eq. (\ref{chiRI}), it is apparent that, in the antisymmetric channel, the bare real bond fluctuation on the NN bonds is more pronounced, whereas the bare imaginary bond fluctuation is stronger on the NNN bonds. These distinctive characteristics, determined by the sublattice texture and unique geometry in the kagome lattice, open up the possibility of realizing exotic electronic orders, such as loop current ground states.

\section{competing electronic states with inter-site Coulomb interactions}

After studying the intrinsic charge fluctuations at the VH filling, we turn to explore the effect of electronic interactions on them. In the spinless scenario, the onsite Coloumb repulsion is absent by the Pauli exclusion and we consider the NN and NNN inter-site Coloumb repulsions,
\begin{align}
\mathcal{H}_{\text{int}}  =&\sum_\eta V_\eta \sum_{\alpha, \br} \left[ n_\beta (\br) n_\gamma (\br+\bl_\eta)+n_\beta (\br) n_\gamma (\br-\bl_\eta) \right] \label{HI} \\ 
 =& \frac{1}{N} \sum_{ \eta, \alpha} \sum_{\bk \bk' \bq}2 V_\eta(\bq) c^\dagger_\beta (\bk) c_\beta (\bk+\bq) c^\dagger_\gamma (\bk'+\bq) c_\gamma (\bk'), \nonumber
\end{align}
with $V_\eta(\bq)= V_\eta \cos(\bq \cdot \bl_{\alpha,\eta})$. 
The interactions can be decoupled in terms of onsite charge orders
\begin{equation}
\mathcal{H}_{\text{int}} = \sum_{\eta,\alpha,\bq}2 V_\eta (\bq) [n_\beta (\bq)]^\dagger n_\gamma(\bq),
\end{equation}
or offsite bond orders
\begin{equation}
\mathcal{H}_{\text{int}} = -\sum_{\eta,\alpha,\bq} \sum_{s=\pm} 2 V_\eta [B_{\alpha,s,\eta} (\bq)]^\dagger B_{\alpha,s,\eta}(\bq).
\label{intBO}
\end{equation}

Once these interactions are introduced, both onsite and bond charge susceptibilities get renormalized. The first-order ladder of bond susceptibilities can be decomposed into the product of bond susceptibilities in different channels, as the interaction carrying internal momentum of fermion propogators can be decoupled owing to Eq.\ref{intBO}, derived from $V_\eta(\bk-\bk')=V_\eta \sum_{s=\pm} f_{\alpha,s,\eta} (\bk) f_{\alpha,s,\eta} (\bk')$. 
The first-order bubble of bond susceptibilities will introduce a susceptibility in the mixed channel with only one vertex (as shown in Fig.~\ref{Fig1}(e)), which is the coupling between bond and onsite charge order. Then, the first-order bubble of this mixed susceptibility will involve the onsite susceptibility. This hierarchy structure can be treated within the susceptibility matrix $\chiup$, which involves both onsite and bond charge orders. We employ the random phase approximation (RPA) summation of all bubble and ladder diagrams (details in SM), that yields the renormalized susceptibility matrix,
\begin{eqnarray}
\chiup_{\text{RPA}}(\bq)=[1+\chiup^{0}(\bq)\mathcal{U}_c(\bm{\bq})]^{-1}\chiup^{0}(\bq).
\end{eqnarray}
The interaction matrix $\mathcal{U}_c(\bq)$ is given by,
 \begin{eqnarray}
\mathcal{U}_c(\bq)&=& \left(\begin{array}{ccc}  V^c_{\text{NN}}& 0 &0    \\ 
 0 & V^c_{\text{NNN}} & 0  \\  
 0 & 0 & U^c(\bq) \\ \end{array}\right),\nonumber\\
{U}^c(\bq)&=& \left(\begin{array}{ccc}  0 & V_{12} &V_{13}     \\ 
V_{12}  & 0 & V_{23}   \\  
V_{13}  & V_{23}  & 0\\ \end{array}\right), \nonumber\\
V^c_{\eta}&=&-\text{diag}\{2V_{\eta},2V_{\eta},...\}, \nonumber\\
V_{\beta\gamma}&=&2V_{\text{NN}}f_{\alpha,+,\text{NN}}(\bq)+2V_{\text{NNN}}f_{\alpha,+,\text{NNN}}(\bq),
\end{eqnarray}
with the indices $\alpha,\beta,\gamma$ being in $( \alpha,\beta,\gamma)$. As the temperature decreases, an eigenvalue of the RPA susceptibility $\chiup_{\text{RPA}}$ at a specific momentum $\bq$ turns negative, signaling an instability at this $\bq$ vector. The associated eigenvector contains the structure of the charge instability, i.e. the CDW pattern. 

\begin{figure}[tb]
    \centering
    \includegraphics[width=1.0\linewidth]{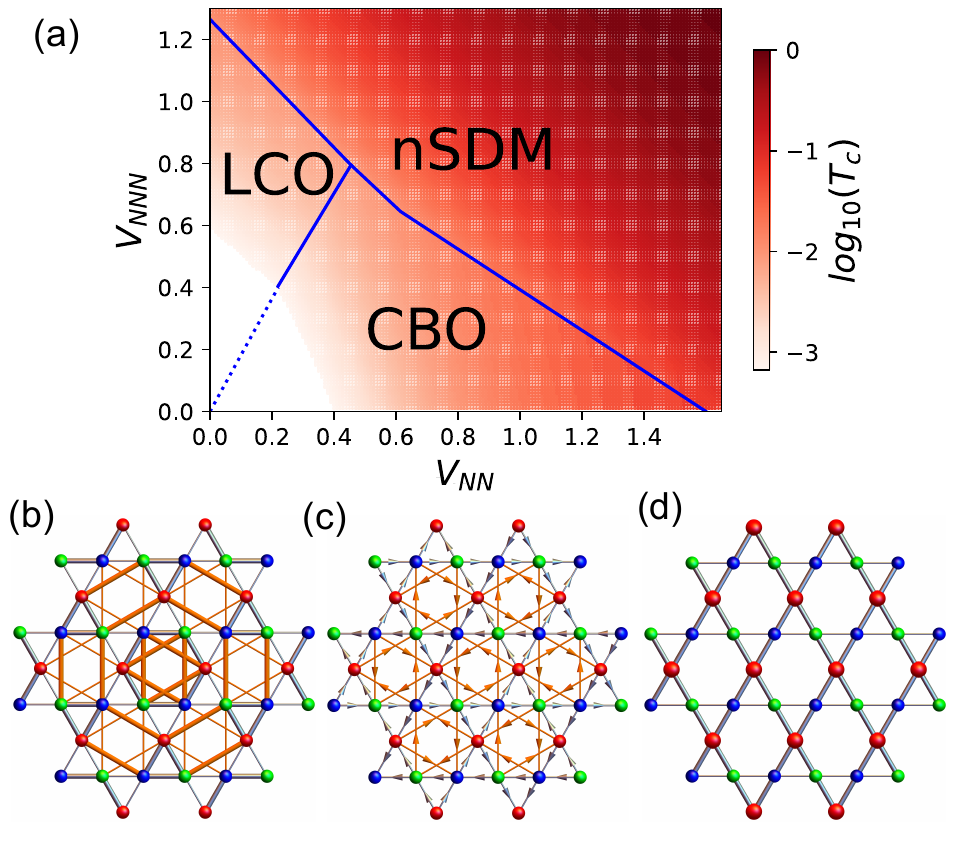}
    \caption{Phase diagram and real-space patterns of relevant charge orders. (a) Phase diagram of the spinless kagome lattice with inter-site Coulomb interactions at the p-type VH filling.  Real-space patterns of three orders: (b) trihexagonal pattern of CBO. (c) bond pattern of LCO, (d) nematic SDM. The thick (thin) bond represents a strong (weak) hopping, the arrow represents the direction of current and the size of sphere denotes the charge density. The transition temperature in the white region is below $10^{-3}$. }
    \label{Fig3}
\end{figure}

According to our previous analysis, the relevant fluctuations are in the onsite and anti-symmetric bond channels and we thus study the effect of inter-site Coulomb interactions on them. With a typical NN repulsion of $V_{\text{NN}}=0.6$, the susceptibilities in various channels $\chiup^{\prime,\prime\prime}/\Omega$ are displayed in Fig.~\ref{Fig2}(b). The NN bond fluctuations are significantly enhanced at the $\bM$ point but the NN real bond susceptibility is dominant, consistent with previous studies~\cite{WWang2013,PhysRevLett.110.126405}. The NN imaginary bond fluctuation (green dashed line) is the subdominant while the onsite charge fluctuation is quite weak. In contrast, with a moderate NNN repulsion $V_{\text{NN}}=0.95$, the susceptibilities of imaginary bond orders are significant at the $\textbf{M}$ point and the NNN imaginary bond susceptibility is much larger than the others, as shown in Fig.~\ref{Fig2}(c). This indicates that the NNN repulsion can promote the imaginary bond fluctuation on the NNN bond. When both NN and NNN repulsions are substantial, the onsite charge fluctuation at $\bf{q}=0$ exceeds the bond fluctuations at the $\textbf{M}$ point and becomes dominant, as shown in Fig.~\ref{Fig2}(d). Meanwhile, the enhancement of the onsite charge susceptibility at the $\textbf{M}$ point always remains weak, due to the aforementioned sublattice interference effect.

We further scrutinize the eigenvalues of $\chiup_{\text{RPA}}(\bq)$ to study the particle-hole instabilities with decreasing temperature. The obtained phase diagram is displayed in Fig.~\ref{Fig3}(a), with color representing the transition temperatures. When the NN repulsion is dominant and the NNN repulsion is weak (region I), the susceptibility of real bond order at three $\textbf{M}$ points first diverges as the temperature decreases and the system favors the charge bond order (CBO). For a dominant NNN repulsion (region II), the imaginary bond order, i.e. loop current order (LCO), is the leading instability. Due to the coupling between bond order on the NN and NNN bonds, both CBO and LCO exhibit a sizable mixture between these bonds, as shown in Fig.~\ref{Fig3} (b) and (c). These particle-hole instabilities are consistent with our weak-coupling analysis, which indicates that the real bond order generates uniform larger gaps on the Fermi surface for the NN channel and the imaginary bond order produces larger gaps on the Fermi surface for the NNN channel (details in SM).
When both NN and NNN repulsions are strong (region III), the two-fold charge order with $\bq=0$ is favored and characterized by a mixture of onsite and symmetric bond orders (shown in Fig.~\ref{Fig3} (d)). The onsite order, characterized by distinct occupations on three sublattices, is dubbed as nematic sublattice density modulation (nSDM) and exhibits an electrostatic energy gain that scales linearly with the increasing  inter-sublattice repulsion. When the Coulomb repulsion is relatively weak, this energy gain is small and charge bond orders predominate. However, as the repulsion strengthens, the energy benefit of the onsite charge order increases rapidly, making it the dominant configuration under conditions of strong repulsion (details in SM).

\begin{figure}[t]
    \centering
    \includegraphics[width=1.0\linewidth]{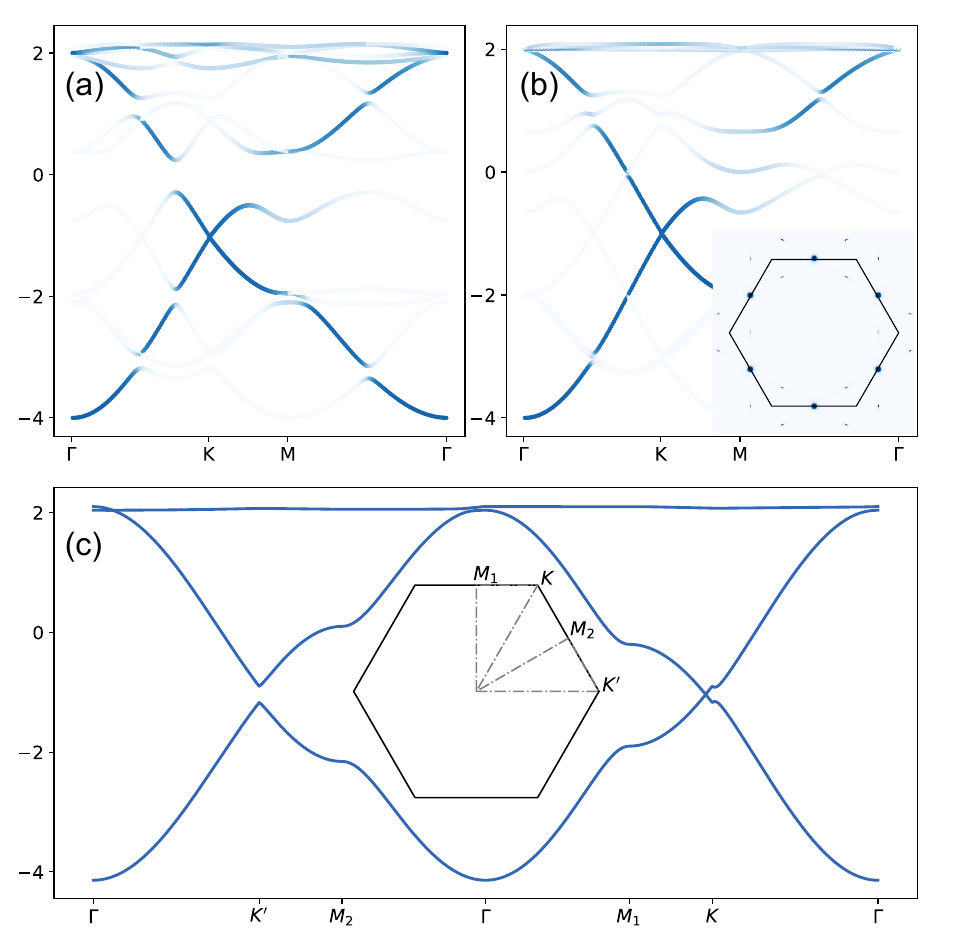} 
    \caption{Unfolded band structures for relevant charge orders: (a) $2\times2$ CBO (a), (b) $2\times2$ LCO and (c) nematic order with charge density modulations. The NN and NNN bond order parameters in the anti-symmetric channel are $\Delta_{-,\text{NN}}=-0.12$ and $\Delta_{-,\text{NNN}}=0.07$ for the CBO and $\Delta_{-,\text{NN}}=-0.09i$ and $\Delta_{-,\text{NNN}}=0.10i$ for the LCO. For the nematic CDM phase, the adopted parameters are $\Delta_0(2,-1,-1)$ in the onsite channel with $\Delta_0=0.1$ and $\Delta_{2(3),+,\text{NN}}=0.05$ in the NN symmetrical channel. The detailed Hamiltonian is provided in SM.}
    \label{unfoldedband}
\end{figure}
For both CBO and LCO, the instability occurs simultaneously at three symmetry related $\bf{M}$ points and the ground state can be determined by the analysis of Ginzburg-Landau free energy. In the CBO, the trilinear term favors the triple-$\bf{M}$ phase with $2\times2$ reconstructions and the corresponding sign of its coefficient determines the real-space pattern~\cite{PhysRevB.104.035142}: a negative sign favors the trihexagonal pattern and a positive sign favors the Star of David pattern. The real-space trihexagonal configuration involving NN and NNN bonds is displayed in Fig.~\ref{Fig3} (b), where the thick (thin) bond represents a strong (weak) hopping. With typical order parameters, the corresponding unfolded band structure is shown in Fig.~\ref{unfoldedband} (a) and the Fermi surface is fully gapped with a maximum gap occurring around the VHSs. For the LCO, the trilinear term vanishes due the time-reversal symmetry and the free energy up to quartic terms reads,
\begin{eqnarray}
F_{\text{LCO}}=a\Psi^2+b\Psi^4+c(\psi^2_1\psi^2_2+\psi^2_2\psi^2_3+\psi^2_3\psi^2_1),
\end{eqnarray}   
where $\psi_i$ is the order parameter of LCO with the vector $\text{M}_i$ and $\Psi^2=\sum_i\psi^2_i$. The quadratic coefficient is $a=a_0(T-T_c)$ with $a_0>0$. The coefficient of the coupling term determines the ground state. A large positive $c$ usually favors the single-$\bf{M}$ phase with $1\times2$ reconstructions but a negative $c$ favors the triple-$\bf{M}$ phase with $2\times2$ reconstructions.  Fig.~\ref{Fig3} (c) illustrates the real-space pattern of triple-$\bf{M}$ $2\times2$ LCO with the six-fold rotational symmetry, where the arrows denote the direction of the current pattern emerging in both NN and NNN bonds.
 Within this phase, the time-reversal symmetry is broken and the occupied band features a nontrivial Chern number and an orbital magnetism occurs. The orbital magnetic moment is correlated with the order parameter of LCO (details are in Sec. VII of SM). The unfolded band structure is displayed in Fig.~\ref{unfoldedband} (b), and the gap opening is anisotropic: the gap along $\Gamma$-K almost vanishes but reaches the maximum at VHSs. Distinct from the CBO, there is an additional state located at the Fermi level around VHSs. These lead to finite spectral weight at the Fermi energy along the $\Gamma$-K line and around $\textbf{M}$, as observed from the Fermi surface shown in the inset of Fig.~\ref{unfoldedband} (b). For the two-fold sublattice density modulation order, the free energy reads,
 \begin{eqnarray}
F_{\text{CDW}}=a'(\rho^2_1+\rho^2_2)+c'(\rho^3_+ +\rho^3_-),
\end{eqnarray} 
where $\rho_{1,2}$ are the two-fold order parameters and $\rho_{\pm}=\rho_1\pm i\rho_2$. Assuming $(\rho_1,\rho_2)=\rho(sin2\theta, cos2\theta)$, the cubic term can be written as $\rho^3cos(6\theta)$, which is minimized by $2\theta=2n\pi/3$ for $c'<0$ and  $2\theta=(2n+1)\pi/3$ for $c'>0$. The resulting order will break the six-fold rotational symmetry and thus is nematic. The order mainly involves charge density modulations within the unit cell and a representative nematic real-space configuration is shown in Fig.~\ref{Fig3} (d), where the large red spheres denote larger occupation and the red sublattice related bonds have stronger hopping amplitude. As shown in Fig.~\ref{unfoldedband} (c), this nSDM order will not introduce any band fold and gap opening around the Fermi level but introduce anisotropic energy shifts for the VHSs.

\begin{figure}[t]
    \centering
    \includegraphics[width=1.0\linewidth]{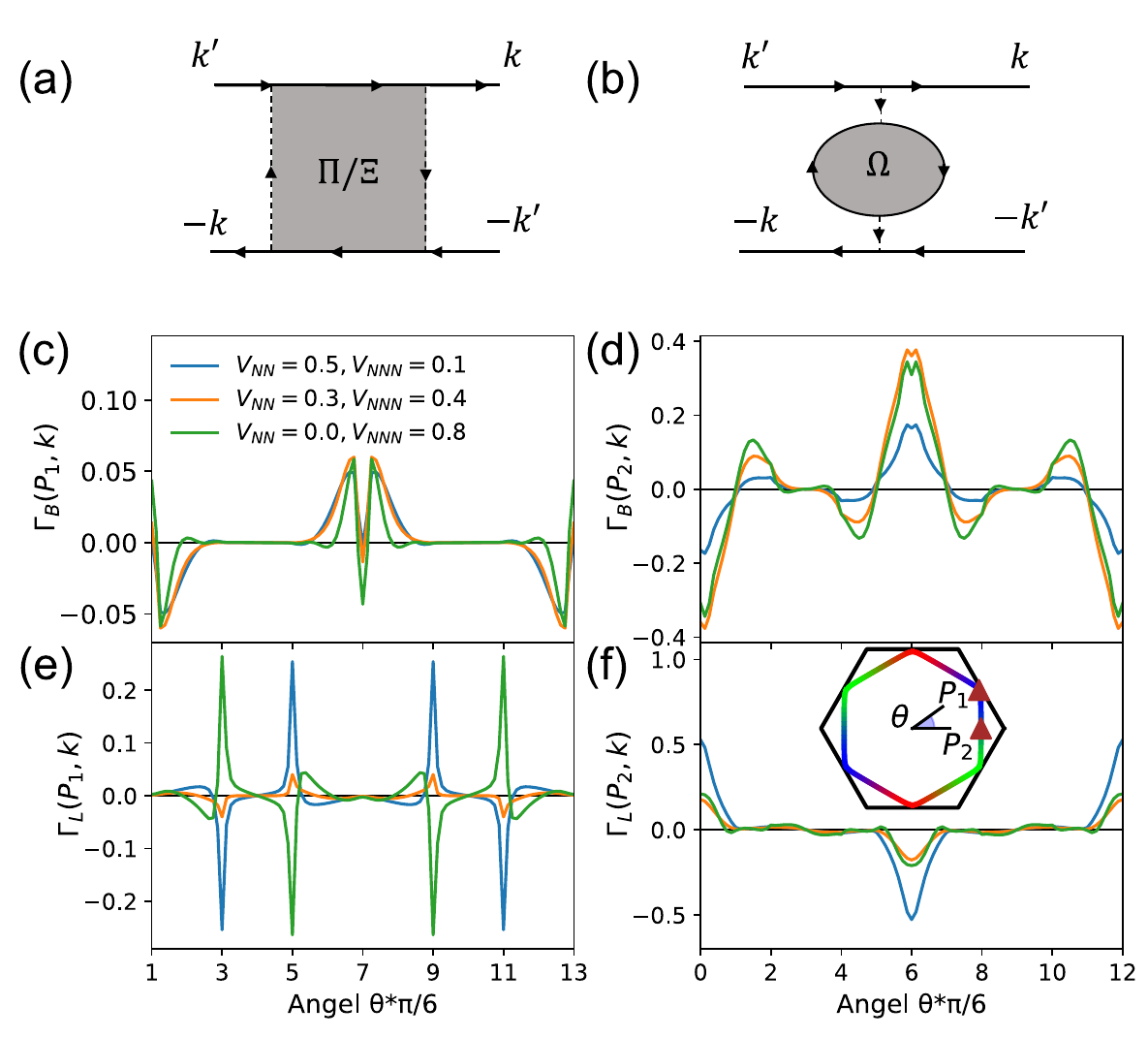}
    \caption{ 
    Effective Cooper pair scattering mediated by onsite and bond charge fluctuations. Feynman diagrams of effective pairing interaction from bond (a) and onsite (b) charge fluctuations. Effective pairing interaction on the Fermi surface from bubbles (c),(d) and ladders (e),(f). Two reference points $\text{P}_1$ and $\text{P}_2$ are marked in the insert of (f) and $\theta$ is measured in the counterclockwise direction from the horizontal axis.  The adopted chemical potential and temperature are $\mu=0.01$ and $k_B T=0.007$, respectively.    }
    \label{Fig4}
\end{figure}

\section{Superconductivity mediated by onsite and bond charge density fluctuations}

\begin{figure*}
    \centering
    \includegraphics[width=0.8\textwidth]{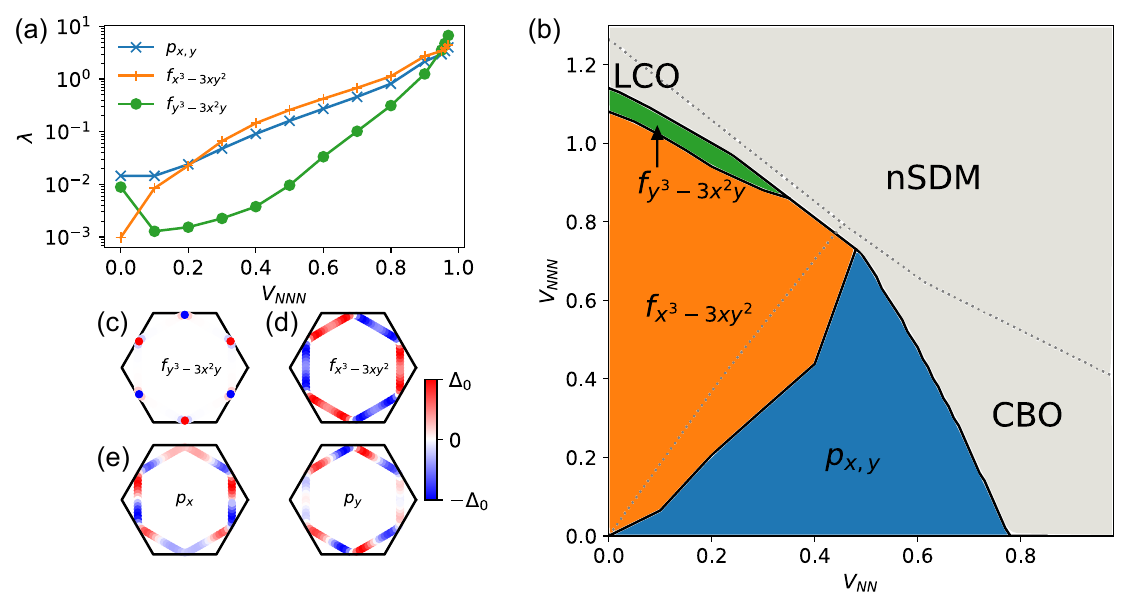}
    \caption{
  Superconducting phase diagram and leading gap functions away from the p-type VH filling. (a) Pairing eigenvalues $\lambda$ with a variation of $V_{\text{NNN}}$ and a fixed $V_{\text{NN}}=0.2$. (b) Phase diagram of superconductivity from the inter-site repulsions. In upper right region, RPA charge susceptibilities diverge at the adopted temperature $k_BT=0.007$ and thus particle-hole instabilities are leading. Representative superconducting gap functions: (c) $f_{y^3-3x^2y}$-wave (d) $f_{x^3-3xy^2}$-wave (e) $p$-wave.
    }
    \label{Fig5}
    
\end{figure*}

\noindent When the Fermi level moves away from VHSs, the Fermi surface nesting weakens, leading to the suppression of both onsite and bond charge orders. However, these charge fluctuations can promote particle-particle instabilities, i.e. superconductivity. In this section, we explore the induced superconducting pairing when these particle-hole orders become unstable. Based on the Feynman diagrams in Fig.~\ref{Fig4}(a) and (b), the onsite charge fluctuation ($\Omega$) peaking at $\bq=0$ dominantly contributes to the forward Cooper pair scattering. While, the bond charge fluctuation ($\Pi / \Xi$) peaking at $\mathbf{M}$ points contributes to the Cooper pair scattering with a large momentum transfer. The effective pairing interaction vertex $\Gamma \left(\bm{k}, \bm{k}'\right)$ can be expressed as the onsite and bond charge fluctuations in the RPA approximation (details in SM). We tune the chemical potential slightly away from VH filling and the corresponding Fermi surface is shown in the inset of Fig.~\ref{Fig4}(f), where there are two representative points $\text{P}_1$ and $\text{P}_2$.  We plot the effective interaction from RPA bubbles $\Gamma_{\text{B}}$ ($\Omega$) and ladders $\Gamma_{\text{L}}$ ($\Pi / \Xi$) with different inter-site Coulomb interactions in Fig.~\ref{Fig4} (c)-(f), respectively. When one momentum is fixed at the point $\text{P}_1$, whose eigenvector is dominantly contributed by one sublattice, the effective interaction $\Gamma_{\text{B}}(\text{P}_1,\bm{k})$ is weak and nonzero only when $\bm{k}$ is close to $\pm \text{P}_1$ due to the sublattice texture on the Fermi surface~\cite{PhysRevB.109.014517,Romer2022}. In contrast, the effective interaction $\Gamma_{\text{L}}(\text{P}_1,\bm{k})$ is substantial and exhibits sharp peaks when $\bm{k}$ is in proximity to the other two VHSs. Moreover, the effective interactions display opposite signs in two cases where $V_{\text{NN}}$ and $V_{\text{NNN}}$ are dominant. The Cooper pair scattering between different VHSs can  be exclusively mediated by the bond fluctuation $\Xi$. The NN and NNN $\Xi$ at the nesting vector features the opposite sign and thus real and imaginary bond fluctuations generate the opposite effective interactions (details in SM), featuring distinct pairing states. When one momentum is fixed at the point $\text{P}_2$,  whose eigenvector is attributed to a mixture of two sublattices, the effective interaction $\Gamma_{\text{B}}(\text{P}_2,\bm{k})$ is large and peaks at $\bm{k}=\pm \text{P}_2$. Due to its anti-symmetric nature, it turns repulsive around $\bm{k}=-\text{P}_2$ ($\theta=\pi$), as indicated from Fig.~\ref{Fig4} (d). While, the effective interaction $\Gamma_{\text{L}}(\text{P}_2,\bm{k})$ mainly mediated by the bond fluctuation $\Pi$ is also significant around $\bm{k}=\pm \text{P}_2$ and becomes attractive around $\bm{k}=-\text{P}_2$ but drops to zero when the momentum transfer is large, as shown in Fig.~\ref{Fig4} (f). Intriguingly, the total effectively interaction $\Gamma_{\text{T}}(\text{P}_2,\bm{k})=\Gamma_{\text{B}}(\text{P}_2,\bm{k})+\Gamma_{\text{L}}(\text{P}_2,\bm{k})$ for $\bm{k}$ around $-\text{P}_2$  is attractive with a dominant $V_{\text{NN}}$ but repulsive with a dominant $V_{\text{NNN}}$, which determines the pairing gap functions.

Near the transition temperature, the gap function can be obtained by solving the linearized gap equation,
\begin{equation}
  -\int_{\rm FS} \frac{\sqrt{3}d {\bf k'}}{2(2 \pi)^2|v_{\text{F} }({\bf k'})|}V^{\rm t}\left(\mathbf{k},\mathbf{k}^\prime\right) \Delta_{i}(\mathbf{k}^{\prime})=\lambda_i\Delta_{i}(\mathbf{k}),
  \label{eq:gapeigenvalues}
\end{equation}
where $v_{\text{F}}(\textbf{k})$ is the Fermi velocity at the momentum ${\mathbf k}$ on the Fermi surface (FS). $\lambda_{i}$ denotes the pairing strength for the gap function $\Delta_{i}(\textbf{k})$ from the pairing interaction vertex in the triplet (t) channel,  with $V^{\text{t}}(\bm{k},\bm{k}')=\frac{1}{2}[\Gamma_{\text{T}}(\bm{k},\bm{k}')- \Gamma_{\text{T}}(\bm{k},-\bm{k}')]$ (for details see SM). {Here, only triplet pairing is allowed due to the spinless model.} We study the dominant pairing states based on the above equation and Fig.~\ref{Fig5} (a) displays the leading pairing eigenvalues with a variation of $V_{\text{NNN}}$ and a fixed $V_{\text{NN}}=0.2$. The $p$-wave state is favored for $V_{\text{NNN}}<0.2$. For an intermediate $V_{\text{NNN}}$, the $f_{x^3-3xy^2}$-wave pairing is dominant. Increasing $V_{\text{NNN}}$ further, the eigenvalue of the $f_{y^3-3yx^2}$-wave pairing increases rapidly and becomes the leading around $V_{\text{NNN}}=1.0$. The comprehensive $V_{\text{NN}}-V_{\text{NNN}}$ phase diagram is illustrated in Fig.~\ref{Fig5} (b). Here, the two dominant $p$-wave and $f_{x^3-3xy^2}$-wave pairings lie adjacent to CBO and LCO/nSDM, implying that their emergence are facilitated by the corresponding charge fluctuations. The corresponding gap functions are displayed in the Fig.~\ref{Fig5} (c)-(e), where the $f$-wave gaps feature a sign change with a six-fold rotation and $p_{x,y}$-wave state is two-fold degenerate. The $p_{x,y}$-wave pairing tends to form a $p_x+ip_y$ state to maximize the superconducting condensation energy. The $f_{y^3-3yx^2}$-wave state existing in a narrow region is extremely close to LCO, indicating that it is dominantly promoted by loop current fluctuations. The superconducting gaps around saddle points connected by the nesting vectors have same sign, which is dictated by the attractive pairing interaction between different VHSs. The $f_{x^3-3xy^2}$-wave pairing is generated by both LCO and nSDM fluctuations and the sign-reversed gaps between two opposite edges are attributed to the repulsive nature of $\Gamma_{\text{T}}(\text{P}_2,\bm{k})$ around $\bm{k}=-\text{P}_2$. The $p$-wave pairing is driven by the CBO fluctuations and the attractive total interaction $\Gamma_{\text{T}}(\text{P}_2,\bm{k})$ around $\bm{k}=-\text{P}_2$ ensures that the gaps maintains the same sign on two opposite edges. In addition, its odd-parity nature will introduce line nodes along $\Gamma$-K line.  {Non-magnetic disorders will suppress the transition temperatures of these signed-changing pairing states, but will not change their symmetries (see SM). In our spinless model, spin-single pairing is absent, thus our results of pairing states cannot be directly applied to realistic kagome materials. However, the discussed effective Cooper scattering between VHSs can play an essential role in determining pairing symmetry within kagome materials.}

\section{Discussions and conclusions}

At the p-type VH filling, the associated sublattice texture on the Fermi surface plays pivotal role in determining the correlated states in the kagome lattice. The real-space $2\times2 $ modulated onsite charge order is significantly suppressed and the bond charge order gets promoted due to the sublattice interference.  Owing to the unique geometry of the kagome lattice, the NN and NNN bonds are characterized by strong intrinsic real and imaginary bond fluctuations, respectively. The loop current state can naturally emerge when there is a strong NNN repulsion. Our work demonstrates that the kagome lattice is an ideal platform to realize such loop current state. The obtained $2\times2$ loop current order is in the anti-symmetric channel and breaks the translational symmetry derived from the Fermi surface nesting. It is distinct from the loop current order in the symmetric channel with $\bm{q}=0$ at 1/3 or 2/3 fillings, where quadratic band touching is believed to be essential~\cite{PhysRevLett.103.046811,PhysRevB.82.075125,PhysRevLett.117.096402}. The inclusion of NNN hopping will deform the Fermi surface at VH filling, weakening Fermi surface nesting. But the associated sublattice texture remains and thus the discussed mechanism is robust. A weak NNN hopping will slightly suppress the bond order but enhance nSDM, only leading to quantitative changes in the phase diagram according to our calculations (see Sec. VIII of SM).

In the nonmagnetic kagome materials, there is an additional degree of freedom, i.e. electron's spin. In a spinful model, the onsite Coulomb repulsion becomes relevant and the onsite charge fluctuation gets enhanced as the charge density doubles. A strong NNN repulsion will further enhance the onsite CDW, rendering the LCO subleading. However, the third NN repulsion acting on the same subalttice can suppress the CDW and LCO may still be stabilized in certain parameter space. A strong onsite Coulomb interaction can enhance the spin bond order and complicates the phase diagram, which deserves future investigation.    

We discuss the potential experimental implications of the correlated states in our calculations. The obtained $2\times2$ LCO state can be relevant in two types of kagome materials. In AV$_3$Sb$_5$, the CDW exhibits an in-plane $2\times2$ reconstruction and TRS breaking. There are both p-type and m-type VHSs in the vicinity of the Fermi level and the multi-orbital nature and strong hybridization between V d orbitals and Sb p orbitals can enhance the inter-site repulsion in the kagome lattice~\cite{YHu2022,MKang2022}. These are consistent with our setting in our model calculations. Moreover, the multiple types of VHSs may be helpful to stabilize LCO in the spinful case~\cite{LiHQ2024}. The delocalized Wannier functions on the kagome lattice due to strong d-p hybridization enhances the inter-site interaction and the multi-fold VHSs could reduce the critical interaction for LCO. {Indeed, constrained RPA calculations suggest that the NN and NNN repulsion in AV$_3$Sb$_5$ and FeGe are comparable in strength~\cite{PhysRevB.111.125163}.}
The CDW observed in AV$_3$Sb$_5$ may be attributed to the LCO and driven by inter-site Coulomb interactions. Another relevant kagome material is FeGe, which exhibits both antiferromagnetic and CDW orders. Each kagome layer is ferromagnetic and ferromagnetic splitting is large, resulting  multiple spin-polarized VHSs in proximity to the Fermi level~\cite{TengXK2022,TengXK2023}. This spin-polarized band is close to the adopted spinless kagome model here. The orbital magnetism associated with LCO could account for the change of magnetic moment upon the CDW transition in FeGe~\cite{TengXK2022,TengXK2023}. The nematic SDM involving onsite and symmetric bond orders in our calculations can account for the nematicity in 
CsTi$_3$Bi$_5$ observed by the quasi-particle interference in STM measurements~\cite{2022arXiv221112264Y,HLi2023}. Especially, the observed anisotropic symmetry-breaking feature in momentum space can be attributed to the nematic bond order. {For other kagome materials like ScV$_6$Sn$_6$ and LaRu$_3$Si$_2$, the charge orders are not correlated with VHS-related Fermi surface nesting, thus are out of the scape of this work~\cite{Hu2024,mielke2024}. In the 3D frustrated pyrochlore lattice, a 3D counterpart of the kagome lattice comprising a network of corner-sharing tetrahedra, the revealed sublattice interference may also appear and contribute to promote exotic charge orders, which can be relevant for 3D materials like CeRu$_2$ and CuV$_2$S$_4$~\cite{Guguchia2023,Huang2024,PhysRevB.99.195140}.}

In summary, our study demonstrates that the loop current state can be stabilized within the spinless kagome lattice, driven by the pronounced imaginary bond fluctuations on next-nearest-neighbor (NNN) bonds. The uncovered sublattice texture plays a pivotal role in the formation of bond charge orders, with the accompanying sublattice interference being deeply connected to the emergence of exotic correlated states. Our findings shed light on the unique character of the kagome lattice and propose a new mechanism for realizing exotic orders in kagome-based materials, which can be applied \textbf{to} other lattices with multiple sublattcies.

\section{Acknowledgments}
R.F., S.Z., and X.W. are supported by the National Key R\&D Program of China (Grants No. 2023YFA1407300 and 2022YFA1403800) and the National Natural Science Foundation of China (Grants No. 12374153, 12047503, and 11974362). 
J.Z. and J.P. are supported by the Ministry of Science and Technology (Grant No. 2022YFA1403901), the National Natural Science Foundation of China (Grant No. NSFC- 11888101, and the New Cornerstone Investigator Program. Z.W. is supported by U.S. Department of Energy, Basic Energy Sciences Grant No. DE-FG02-99ER45747 and the Cottrell SEED Award No. 27856 from Research Corporation for Science Advancement. R.T., M.D. and H.H acknowledge funding by the Deutsche Forschungsgemeinschaft (DFG, German Research Foundation) through Project-ID 258499086 - SFB 1170, and through the research unit QUAST, FOR
5249, project ID 449872909, and through the W\"{u}rzburg-Dresden Cluster of Excellence on Complexity and Topology in Quantum Matter – ct.qmat Project-ID 390858490- EXC 2147. Numerical calculations in this work were performed on the HPC Cluster of ITP-CAS.	
\bibliography{references_loopcurrent0515}

% ---------------- 切换到补充材料版式 ----------------
\clearpage % 强制换页，确保SM从新页面开始
\onecolumngrid % 【关键】RevTeX专用命令：切换到单栏模式

% 3. 重置计数器 + 重新定义编号格式（加“S”前缀）
\setcounter{section}{0} % 重置章节计数器
\setcounter{figure}{0} % 重置图表计数器
\setcounter{table}{0}
\setcounter{equation}{0}

% 重新定义编号为“S1, S2...”样式
\renewcommand{\thesection}{S\arabic{section}}
\renewcommand{\thefigure}{S\arabic{figure}}
\renewcommand{\thetable}{S\arabic{table}}
\renewcommand{\theequation}{S\arabic{equation}}

\section*{Supplementary Materials}
\title{Supplemental Material for "Exotic charge density waves and superconductivity on the Kagome Lattice"}
\author{Rui-Qing Fu }
\affiliation{CAS Key Laboratory of Theoretical Physics, Institute of Theoretical Physics, Chinese Academy of Sciences, Beijing 100190, China}
\affiliation{School of Physical Sciences, University of Chinese Academy of Sciences, Beijing 100049, China}

\author{Jun Zhan}
\affiliation{ Institute of Physics, Chinese Academy of Sciences, Beijing 100190, China}
\affiliation{School of Physical Sciences, University of Chinese Academy of Sciences, Beijing 100049, China}

\author{Hendrik Hohmann}
\affiliation{Institut f\"{u}r Theoretische Physik und Astrophysik, Universit\"{a}t W\"{u}rzburg, Am Hubland Campus S\"{u}d, W\"{u}rzburg 97074, Germany}

\author{Matteo D\"{u}rrnagel}
\affiliation{Institut f\"{u}r Theoretische Physik und Astrophysik, Universit\"{a}t W\"{u}rzburg, Am Hubland Campus S\"{u}d, W\"{u}rzburg 97074, Germany}

\author{Ronny Thomale}
\affiliation{Institut f\"{u}r Theoretische Physik und Astrophysik, Universit\"{a}t W\"{u}rzburg, Am Hubland Campus S\"{u}d, W\"{u}rzburg 97074, Germany}

\author{Jiangping Hu}
\affiliation{ Institute of Physics, Chinese Academy of Sciences, Beijing 100190, China}

\author{Ziqiang Wang }
\thanks{wangzi@bc.edu}
\affiliation{Department of Physics, Boston College, Chestnut Hill, Massachusetts 02467, USA}

\author{Sen Zhou}
\thanks{zhousen@itp.ac.cn}
\affiliation{CAS Key Laboratory of Theoretical Physics, Institute of Theoretical Physics, Chinese Academy of Sciences, Beijing 100190, China}
\affiliation{School of Physical Sciences, University of Chinese Academy of Sciences, Beijing 100049, China}
\affiliation{CAS Center for Excellence in Topological Quantum Computation, University of Chinese Academy of Sciences, Beijing 100049, China}

\author{Xianxin Wu}
\thanks{xxwu@itp.ac.cn}
\affiliation{CAS Key Laboratory of Theoretical Physics, Institute of Theoretical Physics, Chinese Academy of Sciences, Beijing 100190, China}

\maketitle

%%%%%%%%%% Merge with supplemental materials %%%%%%%%%%
%%%%%%%%%% Prefix a "S" to all equations, figures, tables and reset the counter %%%%%%%%%%
\setcounter{equation}{0}
\setcounter{figure}{0}
\setcounter{table}{0}
\setcounter{page}{1}
\setcounter{section}{0}
%\makeatletter
\renewcommand{\theequation}{S\arabic{equation}}
\renewcommand{\thefigure}{S\arabic{figure}}
\renewcommand{\bibnumfmt}[1]{[S#1]}
\renewcommand{\citenumfont}[1]{S#1}

In this Supplementary Material, we discuss the details about calculations of the onsite and bond charge susceptibility, RPA formalism for analysis of charge instabilities and charge fluctuation mediated pairing interaction. 
The lattice convention in this SM is the same with the main text. 
%superconductivity and some analytical analysis about charge order physics.

\section{Onsite and bond charge operators and susceptibility }
\subsection{Onsite and bond charge operators  }

%Our full susceptibility is defined by onsite and bond charge operators. 
We explore the competing charge order in the kagome lattice and the relevant orders are in the onsite and bond channels. The onsite charge operator is
\begin{eqnarray}
n_\alpha(\bm{r})=c_{\alpha,\bm{r}}^\dagger c_{\alpha,\bm{r}},
\end{eqnarray}
with $\alpha$ being the sublattice index. After the Fourier transformation, the operation in the momentum space reads,
\begin{eqnarray}
n_\alpha(\bm{q})=\frac{1}{\sqrt{N}}\sum_{\bm{k}} c_{\alpha,\bm{k+q}}^\dagger c_{\alpha,\bm{q}}.
\end{eqnarray}
We additionally consider bond charge modulation, i.e. charge bond order, on NN and NNN bonds.
Due to the unique geometry of the kagome lattice, within each unit cell there are two NN (NNN) bonds along the direction parallel (perpendicular) to each basis vector $\ba_\alpha$, and they all connect two distinct sublattices $\beta$ and $\gamma$, with the Levi-civta symbol satisfying $\epsilon_{\alpha\beta\gamma}=1$, i.e., $(\alpha, \beta, \gamma)=$ (1, 2, 3), (2, 3, 1), and (3, 1, 2). The bond operator involving $\beta,\gamma$ sublattices and its complex conjugate are defined as,
\begin{eqnarray}
B_{\alpha,s,\eta}(\bm{r_\beta})&=&\frac{1}{2}(c_{\beta,r_\beta}^\dagger c_{\gamma,r_\beta+l_{\alpha,\eta}}+s c_{\beta,r_\beta}^\dagger c_{\gamma,r_\beta-l_{\alpha,\eta}}),\nonumber\\
B_{\alpha,s,\eta}^\dagger(\bm{r_\beta})&=&\frac{1}{2}(c_{\gamma,r_\beta+l_{\alpha,\eta}}^\dagger c_{\beta,r_\beta} +s c_{\gamma,r_\beta l_{\alpha,\eta}}^\dagger c_{\beta,r_\beta}),
\end{eqnarray}
with $s=+/-$ denoting the symmetric/antisymmetric channel and the connecting vectors being $l_{\alpha,1}=\frac{1}{2}\bm{a}_\alpha$, $l_{\alpha,2}=\frac{1}{2}(\bm{a}_\beta-\bm{a}_\gamma)$. \if This definition in the antisymmetric channel is slightly different from that in the main text.\fi
The Fourier transformation of $B_{\alpha,s,\eta}(\bm{r})$ and $B_{\alpha,s,\eta}^\dagger(\bm{r})$ are given by
\begin{eqnarray}
\tilde{B}_{\alpha,s,\eta}(\bm{q})&=&\frac{1}{\sqrt{N}}\sum_{\bm{k}} \tilde{f}_{\alpha,s,\eta}(\bm{k})  c_{\beta,\bm{k+q}}^\dagger c_{\gamma,\bm{k}},\nonumber\\
\tilde{B}_{\alpha,s,\eta}^\dagger (\bm{q})&=& \frac{1}{\sqrt{N}}\sum_{\bm{k}} \tilde{f}^*_{\alpha,s,\eta}(\bm{k+q}) c_{\gamma,\bm{k+q}}^\dagger c_{\beta,\bm{k}}.
\end{eqnarray}
%Here, index $\alpha,\beta,\gamma \in {1,2,3}$ are orbital index and they are set in order to satisfy $\epsilon_{\alpha\beta\gamma}=1$, $s=\pm $ is symmetric/anti-symmetric symbol, $\eta=1,2$ are NN and NNN bond symbol. 
Here, the complex form factors are given by $\tilde{f}_{\alpha,+,\eta}(\bm{k})=\cos(\bm{k \cdot l_{\alpha,\eta}})$ and $\tilde{f}_{\alpha,-,\eta}(\bm{k})=-i\sin(\bm{k \cdot l_{\alpha,\eta}})$. The tilde symbol on $\tilde{B}_{\alpha,s,\eta}$ and $\tilde{f}_{\alpha,s,\eta}$ represents the complex form factor obtained directly from the Fourier transform. For the convenience of susceptibility calculations in the following, we further introduce the real form factors  $f_{\alpha,+,\eta}(\bm{k})=\cos(\bm{k \cdot l_{\alpha,\eta}})$ and  $f_{\alpha,-,\eta}(\bm{k})=\sin(\bm{k \cdot l_{\alpha,\eta}})$ without the tilde symbol. The bond operators defined by these real form factors are
\begin{eqnarray}
B_{\alpha,s,\eta}(\bm{q})&=&\frac{1}{\sqrt{N}}\sum_{\bm{k}} f_{\alpha,s,\eta}(\bm{k})  c_{\beta,\bm{k+q}}^\dagger c_{\gamma,\bm{k}},\nonumber\\
B_{\alpha,s,\eta}^\dagger (\bm{q})&=& \frac{1}{\sqrt{N}}\sum_{\bm{k}} f_{\alpha,s,\eta}(\bm{k+q}) c_{\gamma,\bm{k+q}}^\dagger c_{\beta,\bm{k}}.
\end{eqnarray}
The relation between bond operators with and without tilde symbol is given by,
\begin{eqnarray}
\tilde{B}_{\alpha,+,\eta}(\bm{q}) & = &B_{\alpha,+,\eta}(\bm{q}), \nonumber \\
\tilde{B}^\dagger_{\alpha,+,\eta}(\bm{q}) & = &B^\dagger_{\alpha,+,\eta}(\bm{q}), \nonumber \\
\tilde{B}_{\alpha,-,\eta}(\bm{q}) & = &-i B_{\alpha,-,\eta}(\bm{q}), \nonumber \\
\tilde{B}^\dagger_{\alpha,-,\eta}(\bm{q}) & = &i B^\dagger_{\alpha,-,\eta}(\bm{q}).
\end{eqnarray}
Note that only $\tilde{B}$ operators are physical and the sign difference in the anti-symmetric channel should be noticed.
%We dropped the $-i$ coefficient in anti-symmetric form factor to calculate RPA susceptibility in the following content. 
From the above definitions, it can be easily shown that
\begin{eqnarray}
[n(\bm{q})]^\dagger & =&n(\bm{-q}),\nonumber\\
\ [B_{\alpha,s,\eta}(\bm{q})]^\dag &=& B_{\alpha,s,\eta}^\dagger(\bm{-q}).
\end{eqnarray}

%In this SM the lattice vector in real space are $\bm{a_1}=(1,0)$ and $\bm{a_{2,3}}=(-\frac{1}{2},\pm \frac{\sqrt{3}}{2})$. In reciprocal space, vH vectors $\bm{M_1}=(0,\frac{2\pi}{\sqrt{3}})$ and $M_{2,3}=(\pm \pi, -\frac{\pi}{\sqrt{3}})$

\subsection{Bare Susceptibilities for the relevant charge orders}
\begin{figure}
    \centering
    \includegraphics[width=1.0\textwidth]{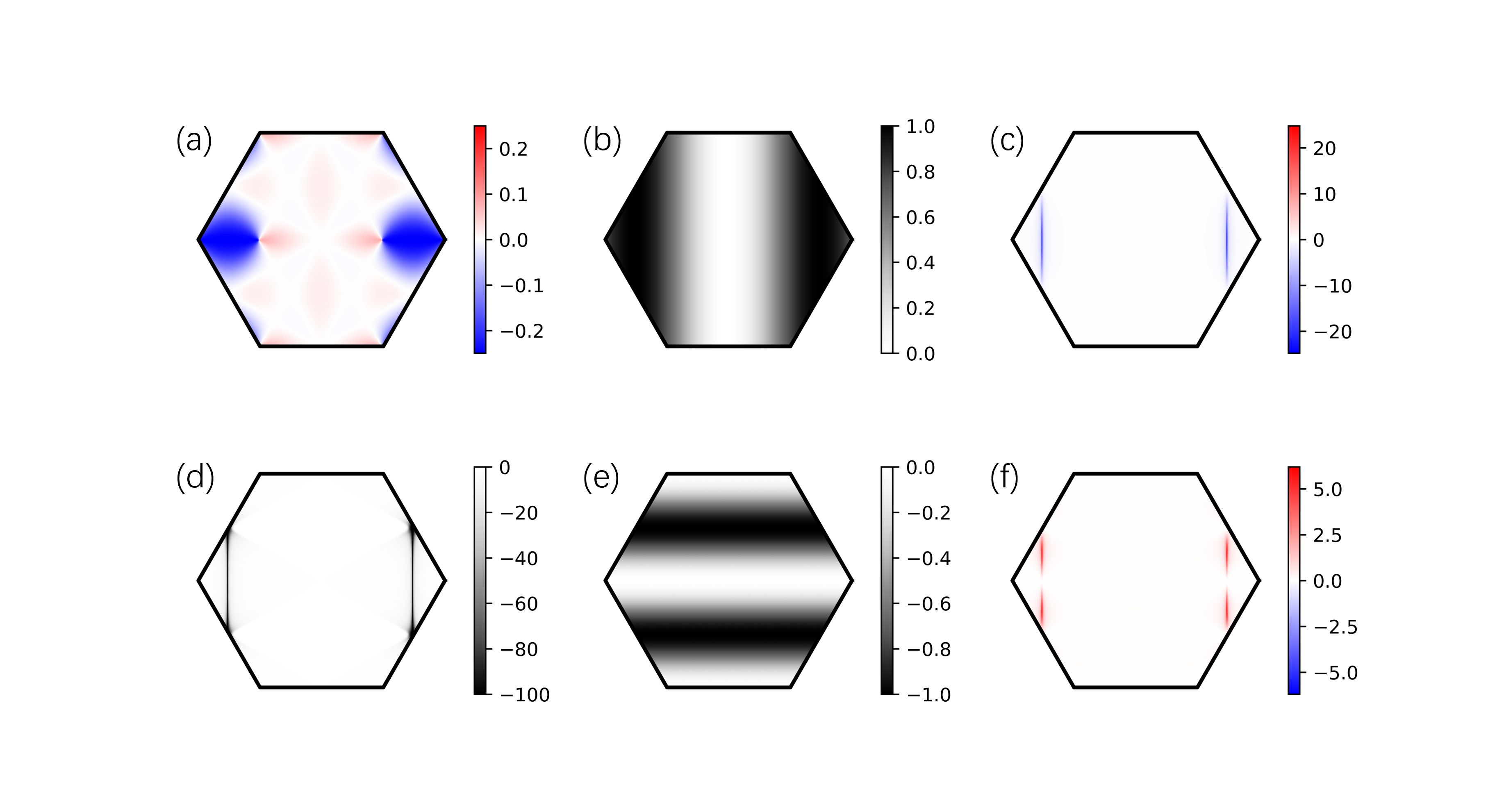}
    \caption{Detailed analysis of the sign of $\Xi$-type susceptibility on the NN and NNN bond: (a) $a_2^\gamma(k) a_2^\beta (k+\textbf{M}_1) a_2^\gamma ((k+\textbf{M}_1) a_2^\beta (k)$. (d) $\frac{n_F(E_2(k))-n_F(E_2(k+\textbf{M}_1))}{E_2(k)-E_2((k+\textbf{M}_1))}$. (b) NN form factor $\sin(\mathbf{k\cdot l_{\alpha,nn}})\sin(k+\textbf{M}_1)\cdot l_{\alpha,nn})$ (e) NNN form factor $\sin(\mathbf{k\cdot l_{\alpha,nnn}})\sin(\mathbf{(k+\textbf{M}_1)\cdot l_{\alpha,nnn}})$. (c) Contribution to $\Pi_{2;nn,nn}$ from each k-point in 1st BZ. (f) Contribution to $\Pi_{2;nnn,nnn} $}
    \label{FigS2_2}
\end{figure}

\begin{figure}
    \centering
    \includegraphics[width=1.0\textwidth]{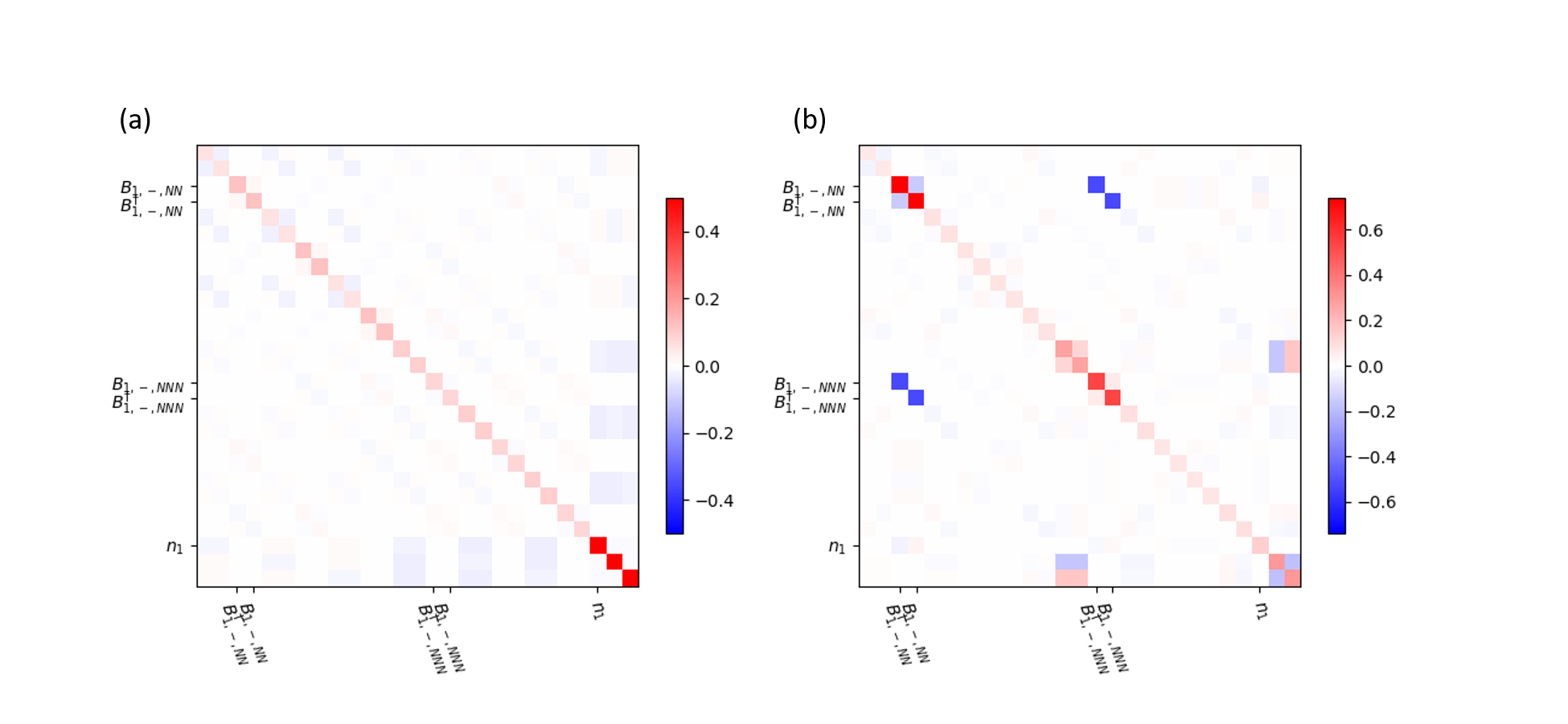}
    \caption{Bare $27\times27$ susceptibility matrix $\chiup(q)$ at $\mathbf{\Gamma}$ point(a) and  $\mathbf{M_1}$ point(b) with the operator order $\mathcal{O}_o=[(B_{1,+,\text{NN}},B_{1,+,\text{NN}}^\dagger,B_{1,-,\text{NN}},B_{1,-,\text{NN}}^\dagger),(1\rightarrow 2,3),(\text{NN}\rightarrow \text{NNN}),n_1,n_2,n_3]$.}
    \label{fig2}
\end{figure}

 %defined the above operators, we can write down all kinds of susceptibility in combination of these operators. We will arrange these susceptibilities in a $27\times27$ matrix whose row indices are defined as $[(B_{1,+,\text{NN}},B_{1,+,\text{NN}}^\dagger,B_{1,-,\text{NN}},B_{1,-,\text{NN}}^\dagger),(1\rightarrow 2,3),(\text{NN}\rightarrow \text{NNN}),n_1,n_2,n_3](\bm{q})$ and column indices are defined as Hermitian conjugate of the row. The detailed expression of all type of bare susceptibility matrix elements are listed below.

To investigate the intrinsic fluctuations of different charge orders and their coupling, we calculate the corresponding susceptibilities defined as,
\begin{equation}
\chiup_{pq} (\bq,i\omega_n) =  \int_0^\beta d \tau e^{i\omega_n \tau} \langle T_\tau \mathcal{O}_o (\bm{q},\tau) [\mathcal{O}_{o'} (\bm{q},0)]^\dagger \rangle= \int_0^\beta d \tau e^{i\omega_n \tau} \langle T_\tau \mathcal{O}_o (\bq,\tau) \mathcal{O}^\dagger_{o'} (-\bm{q},0) \rangle.
\end{equation}
Here the operator $\mathcal{O}_o$ runs over the 27 charge orders mentioned above, i.e.  $\mathcal{O}_o=[(B_{1,+,\text{NN}},B_{1,+,\text{NN}}^\dagger,B_{1,-,\text{NN}},B_{1,-,\text{NN}}^\dagger),(1\rightarrow 2,3),(\text{NN}\rightarrow \text{NNN}),n_1,n_2,n_3]$. The analytical expressions of the bare susceptibilities in all channels are given by,
\begin{align}
\label{onsiteC}
        & \left\langle n_\alpha [n_{\alpha'}]^\dagger \right\rangle(\bm{q}, i\omega_n)= -\frac{1}{N} \sum_{k\mu\nu}  a^{\alpha*}_\mu(\bm{k+q}) a^{\alpha}_{\nu}({\bm{k}}) a^{\alpha'*}_{\nu}({\bm{k}}) a^{\alpha'}_{\mu}(\bm{k+q})  \frac{n_F(E_\mu(\bm{k+q}))-n_F(E_\nu(\bm{k}))}{i\omega_n+E_\mu(\bm{k+q})-E_\nu(\bm{k})}, \\
        & \left\langle n_\alpha [B_{2m'-1}]^\dagger \right\rangle(\bm{q}, i\omega_n)= -\frac{1}{N} \sum_{k\mu\nu} a^{\alpha*}_\mu(\bm{k+q}) a^{\alpha}_{\nu}({\bm{k}}) a^{\gamma'*}_{\nu}({\bm{k}}) a^{\beta'}_{\mu}(\bm{k+q}) f_{q}(\bm{k}) \frac{n_F(E_\mu(\bm{k+q}))-n_F(E_\nu(\bm{k}))}{i\omega_n+E_\mu(\bm{k+q})-E_\nu(\bm{k})}, \\
        & \left\langle n_\alpha [B_{2m'}]^\dagger \right\rangle(\bm{q}, i\omega_n)= -\frac{1}{N} \sum_{k\mu\nu} a^{\alpha*}_\mu(\bm{k+q}) a^{\alpha}_{\nu}({\bm{k}}) a^{\beta'*}_{\nu}({\bm{k}}) a^{\gamma'}_{\mu}(\bm{k+q}) f_{q}(\bm{k+q}) \frac{n_F(E_\mu(\bm{k+q}))-n_F(E_\nu(\bm{k}))}{i\omega_n+E_\mu(\bm{k+q})-E_\nu(\bm{k})}, \\
        \label{bondC1}
        & \left\langle B_{2m-1} [B_{2m'-1}]^\dagger \right\rangle(\bm{q}, i\omega_n)= -\frac{1}{N} \sum_{k\mu\nu} f_{p}(\bm{k}) a^{\beta*}_\mu(\bm{k+q}) a^{\gamma}_{\nu}({\bm{k}}) a^{\gamma'*}_{\nu}({\bm{k}}) a^{\beta'}_{\mu}(\bm{k+q}) f_{q}(\bm{k}) \frac{n_F(E_\mu(\bm{k+q}))-n_F(E_\nu(\bm{k}))}{i\omega_n+E_\mu(\bm{k+q})-E_\nu(\bm{k})}, \\
        \label{bondC2}
        & \left\langle B_{2m-1} [B_{2m'}]^\dagger \right\rangle(\bm{q}, i\omega_n)= -\frac{1}{N} \sum_{k\mu\nu} f_{p}(\bm{k}) a^{\beta*}_\mu(\bm{k+q}) a^{\gamma}_{\nu}({\bm{k}}) a^{\beta'*}_{\nu}({\bm{k}}) a^{\gamma'}_{\mu}(\bm{k+q}) f_{q}(\bm{k+q}) \frac{n_F(E_\mu(\bm{k+q}))-n_F(E_\nu(\bm{k}))}{i\omega_n+E_\mu(\bm{k+q})-E_\nu(\bm{k})}, \\
        \label{bondC3}
        & \left\langle B_{2m} [B_{2m'-1}]^\dagger \right\rangle(\bm{q}, i\omega_n)= -\frac{1}{N} \sum_{k\mu\nu} f_{p}(\bm{k+q}) a^{\gamma*}_\mu(\bm{k+q}) a^{\beta}_{\nu}({\bm{k}}) a^{\gamma'*}_{\nu}({\bm{k}}) a^{\beta'}_{\mu}(\bm{k+q}) f_{q}(\bm{k}) \frac{n_F(E_\mu(\bm{k+q}))-n_F(E_\nu(\bm{k}))}{i\omega_n+E_\mu(\bm{k+q})-E_\nu(\bm{k})}, \\
        \label{bondC4}
        & \left\langle B_{2m} [B_{2m'}]^\dagger \right\rangle(\bm{q}, i\omega_n)= -\frac{1}{N} \sum_{k\mu\nu} f_{p}(\bm{k+q}) a^{\gamma*}_\mu(\bm{k+q}) a^{\beta}_{\nu}({\bm{k}}) a^{\beta'*}_{\nu}({\bm{k}}) a^{\gamma'}_{\mu}(\bm{k+q}) f_{q}(\bm{k+q}) \frac{n_F(E_\mu(\bm{k+q}))-n_F(E_\nu(\bm{k}))}{i\omega_n+E_\mu(\bm{k+q})-E_\nu(\bm{k})},
\end{align} 
where $\langle \mathcal{O}_{o} [\mathcal{O}_{o'}]^\dagger \rangle$ denotes the corresponding bare susceptibility.
Here $\mu/\nu$ is the band index and $a_\mu^\alpha(\bm{k})$ is the $\alpha$-th element of the $\mu$-th eigenvector with corresponding index obtained from tight-binding Hamiltonian. $B_{o/o'}$ is the bond operator in $o/o'$-th row(column) basis, which corresponds to ($\beta,\gamma,p$)/($\beta',\gamma',q$) indices on the right site of these equations. Under this convention, $B_{2m}$ is of the $B^\dagger$ type and $B_{2m-1}$ is of the $B$ type bond operator. The susceptibility of onsite charge orders belongs to the $\Omega$-type term in main text. The operators of charge bond order (CBO) and loop current order (LCO) phase are the combination of above bond operators.
%For the band crossing Fermi energy, this index is $\mu=2$. For static susceptibility, we set $i\omega_n=i0^+$. 
The real ($B'$) and imaginary ($B''$) bond operators can be expressed using defined operators, 
\begin{eqnarray}
    B'_{\alpha,+,\eta}(\bm{q})& =&\frac{1}{2}(\tilde{B}_{\alpha,+,\eta}^\dagger(\bm{q})+\tilde{B}_{\alpha,+,\eta}(\bm{q})) =\frac{1}{2}(B_{\alpha,+,\eta}^\dagger(\bm{q})+B_{\alpha,+,\eta})(\bm{q}), \\
    B''_{\alpha,+,\eta}(\bm{q}) & =&\frac{1}{2}(\tilde{B}_{\alpha,+,\eta}(\bm{q})-\tilde{B}^\dagger_{\alpha,+,\eta}(\bm{q})) =\frac{1}{2}(B_{\alpha,+,\eta}(\bm{q})-B^\dagger_{\alpha,+,\eta}(\bm{q})), \\
    B'_{\alpha,-,\eta}(\bm{q})& =&\frac{1}{2}(\tilde{B}_{\alpha,-,\eta}^\dagger(\bm{q})+\tilde{B}_{\alpha,-,\eta}(\bm{q}))  =-\frac{i}{2}(B_{\alpha,-,\eta}(\bm{q})-B^\dagger_{\alpha,-,\eta}(\bm{q})), \\
    B''_{\alpha,-,\eta}(\bm{q}) & =&\frac{1}{2}(\tilde{B}_{\alpha,-,\eta}(\bm{q})-\tilde{B}^\dagger_{\alpha,-,\eta}(\bm{q}))  =\frac{i}{2}(B_{\alpha,-,\eta}(\bm{q})+B^\dagger_{\alpha,-,\eta} (\bm{q})).
\end{eqnarray}
The sign in anti-symmetric order is different from symmetric one because we dropped the $-i$ coefficient in antisymmetric bond operators. Therefore, their static susceptibilities can be expressed using the above $\chiup_{pq}$ terms,
%n  So the suscepibility of the antisymmetric LCO and CBO can be expreessed as 
\begin{align}
    \chi'_{mm'}(\bm{q}) & =\frac{1}{4}(\langle B_{2m-1} [B_{2m'-1}]^\dagger \rangle+\langle B_{2m} [B_{2m'}]^\dagger \rangle + \langle B_{2m-1} [B_{2m'}]^\dagger \rangle + \langle B_{2m} [B_{2m'-1}]^\dagger \rangle)(\bm{q}), \\
    \chi''_{mm'}(\bm{q}) & =\frac{1}{4}(\langle B_{2m-1} [B_{2m'-1}]^\dagger \rangle+\langle B_{2m} [B_{2m'}]^\dagger \rangle -\langle B_{2m-1} [B_{2m'}]^\dagger \rangle - \langle B_{2m} [B_{2m'-1}]^\dagger \rangle)(\bm{q}),
\end{align}
when $m,m'\in \text{odd}$ represent symmetric bond operators and
\begin{align}
    \chi'_{mm'}(\bm{q}) & =\frac{1}{4}(\langle B_{2m-1} [B_{2m'-1}]^\dagger \rangle+\langle B_{2m} [B_{2m'}]^\dagger \rangle-\langle B_{2m-1} [B_{2m'}]^\dagger \rangle-\langle B_{2m} [B_{2m'-1}]^\dagger \rangle)(\bm{q}), \\
    \chi''_{mm'}(\bm{q}) & =\frac{1}{4}(\langle B_{2m-1} [B_{2m'-1}]^\dagger \rangle+\langle B_{2m} [B_{2m'}]^\dagger \rangle+\langle B_{2m-1} [B_{2m'}]^\dagger \rangle+\langle B_{2m} [B_{2m'-1}]^\dagger \rangle)(\bm{q}),
\end{align}
when $m,m'\in \text{even}$ represent anti-symmetric bond operators.
The first two terms on the right hand side of the equation belong to the $\Pi$-type terms in main text, while the latter two terms are $\Xi$-type terms. For the susceptibility of a specific bond order, $m=m'$, two $\Pi$ are equal, as are the two $\Xi$ terms. It is evident that the relative strength of susceptibilities for real and imaginary bond orders are fully determined by the sign of the $\Xi$.
%Two $\Pi$ and $\Xi$ terms are equal if $m=m'$. So the leading bond instability is determined by the sign of the $\Xi$ operator when $\Pi \gg \Xi$ and $\Pi$ operator is large enough to make the bond fluctuation dominate.

Due to the perfect Fermi surface nesting at the p-type VH filling, we examine the characteristics of bare susceptibilities at two pertinent vectors $\bm{q}=0,\textbf{M}$ based on the above expressions. For $\bm{q}=0$, the six edges of the Fermi surface in Fig.1(b) mainly contribute to the susceptibilities (as indicated by matrix elements in Eq.\ref{onsiteC}) and $\Omega^0_{\alpha\alpha}$ is dominant due to the pure sublattice nature at three VHSs with diverging density of states (DOS). In contrast, each vertex of the bond susceptibility bubble involves two different sublattices (as indicated by matrix elements in Eq.\ref{bondC1},\ref{bondC2},\ref{bondC3},\ref{bondC4}) and the contribution from VHSs vanishes due to the sublattice-resolved eigenvectors in Green functions, leading to minimal values for $\Pi^0_{mn}(0)$. For $\bm{q}=\textbf{M}$, however, the behaviors are the opposite. Taking $\bm{q}=\bm{Q}_1$ as an example, the susceptibilities are predominantly contributed by two edges connected by $\bm{Q}_1$ shown in Fig.1(b), as dictated by the Lindhard function (see Eq.\ref{bondC1},\ref{bondC2},\ref{bondC3},\ref{bondC4}).  As the $\bm{Q}_1$ always connects two VHSs with distinct sublattices,  the contribution of these VHSs in $\Omega^0_{\alpha\alpha'}$ vanishes and only other segments away from VHSs with small DOS can contribute, resulting a small $\Omega^0_{\alpha\alpha'}(\bm{Q}_1)$. On the contrary, VHSs can contribute to $\Pi^0_{mm}$, leading to a significant $\Pi^0_{mm}(\bm{Q}_1)$. As the Green functions in the $\Xi^0_{mm}$ involves mixed sublattices in both electron lines,  the magnitude of $\Xi^0_{mm}(\bm{Q}_1)$ is small. Consequently, this unique sublattice texture at the p-type VH filling suppresses the onsite charge fluctuations with $\bm{q}=\textbf{M}$ but promote pronounced bond charge fluctuations. This behavior in the kagome lattice markedly differs from that observed in the triangular and honeycomb lattices, where onsite charge fluctuations are dominant. Moreover, for $\bm{k}$ points on the two edges connected by the nesting vectors, $\bm{k} \cdot \bm{l}_{\alpha,\text{NN}}=\pm\pi/2$ and thus the NN symmetric form factor vanishes and NN anti-symmetric one reaches the maximum. The NNN symmetric form factor also vanishes at the VHSs due to $\bm{k}_{\beta,\gamma} \cdot \bm{l}_{\alpha,\text{NNN}}=\pm\pi/2$, weakening the corresponding bond fluctuations. Consequently, the anti-symmetric bond fluctuations on both NN and NNN bonds at the $\textbf{M}$ point dominate over the corresponding symmetric conterparts.

% the p-type vH filling, the nesting vectors $\bm{Q}_{1,2,3}$ connect VHSs, whose wavefunctions are attributed to different sublattices. Thus, onsite charge susceptibility at these nesting vectors will be greatly suppressed, as the product of matrix elements $a_\mu^\alpha(\bm{k})$ in Eq.\ref{onsiteC} is vanishing at VHSs. In contrast, the bond susceptibilities $\Pi(\bm{Q}_{1,2,3})$ get enhanced as the VHSs can contribute to them (as indicated by the matrix elements Eq.\ref{bondC1},\ref{bondC4}). The bond susceptibilities $\Pi(\bm{Q}_{1,2,3})$ are also small (as indicated by the matrix elements Eq.\ref{bondC2},\ref{bondC3})

%when compared to the bond susceptibility as a $\Pi(\bm{M_1})$ type susceptibility include Green's function $G^{22}(\bm{M_2})$ and $G^{33}(\bm{M_3})$, which match the p-type vHSs perfectly. So bond fluctuation is important when considering about Fermi surface nesting.
%in Fig.\ref{fig1}, the detailed contribution to $\Xi_{1,-,\eta}(\bm{M_1})$ from each k point on 2nd band is plotted and is divided into wave functions, Lindhard function and form factors.
%Interestingly, the form factors for the NN and NNN bonds on these edges have opposite signs, as shown in Fig.\ref{fig1} (c) and (d). This difference results in opposite $\Xi_{1,-,\eta}(\bm{M_1})$ values for the NN and NNN bonds.

We further analyze the $\Xi$ susceptibility in details. From the definition of bare susceptibilities we find that the only difference between NN and NNN bond order is the form factor. The susceptibility is mainly contributed by the band close to the Fermi level. To reveal the difference between susceptibilites $\Xi$ on NN and NNN bonds, we plot the product of matrix elements, Lindhard function and form factors of the susceptibility in the 2D Brillouin zone, respectively. From the product of matrix elements and the Lindhard function, it is evident that the $\Xi^0_{1,-,\eta}(\textbf{M}_1)$ susceptibility are mainly contributed by $\bm{k}$ points near the two opposite Fermi surface segments connected by the nesting vector $\bm{Q}_1$. Interestingly, the form factors for the NN and NNN bonds on these edges have the opposite sign are different so that NN and NNN, as shown in Fig.\ref{FigS2_2} (c),(d). This difference results in opposite $\Xi^0_{1,-,\eta}(\textbf{M}_1)$ values for the NN and NNN bonds, determined by the unique kagome geometry. Note for $\bm{q}=\textbf{M}_1$ and $\beta / \gamma=2/3 \text{ or } 3/2$, the product of matrix element $a_2^{\gamma*}(\bm{k}+\textbf{M}_1) a_2^\beta (\bm{k}) a_2^{\gamma*} (\bm{k}) a_2^\beta (\bm{k}+\textbf{M}_1)=0$ at two VHSs $\bm{k}=\textbf{M}_2,\textbf{M}_3$ and thus VHS makes no contribution to the $\Xi$ susceptibility, as mentioned before. Therefore, the patch model involving only VHSs can fail to capture the difference between fluctuation between real and imaginary bond orders. 
%some patch theory may fail to describe this FS nesting physics in kagome lattice.

After above analysis, we numerically illustrate the full bare susceptibility matrix elements at $\textbf{M}_1$ and $\Gamma$ point in Fig.\ref{fig2} (a) and (b), respectively. 
At the $\Gamma$ point the susceptibility matrix is dominated by its onsite component ($\Omega$ type) while bond susceptibilities are much weaker. The onsite order also exhibit some coupling with symmetric bond orders. At the $\bm{M_1}$ point, the susceptibilities of symmetric bond orders are vanishing small and those of anti-symmetric bond orders are dominant (larger than the onsite susceptibility $\Omega$). It is apparent the susceptibilities $\Xi$ on the NN and NNN bonds have the opposite signs, consistent with our above analysis. In addition, the coupling between NN and NNN bonds are also strong, indicating a strong mixing between them. 
%We can see that at $\bm{M_1}$ point the fluctuation properties are dominated by anti-symmetric NN, NNN and their coupling susceptibilities in the matrix. This negative coupling will lead to the same instability of NN and NNN bond. Other matrix elements are irrelevant. In both cases, the coupling between onsite and bond operators is relatively weak. This allows us to analyse these two type of susceptibilities separately. 
 %Because of the sign difference of form factor of NN and NNN bond shown ,  the non-diagnoal operators of NN and NNN bond must have different sign and driven the system to different phases. 
 
 \section{Classification of CDW order with $\bf{Q}=0,M$ in kagome lattice}
\begin{figure}
    \centering
    \includegraphics[scale=0.7]{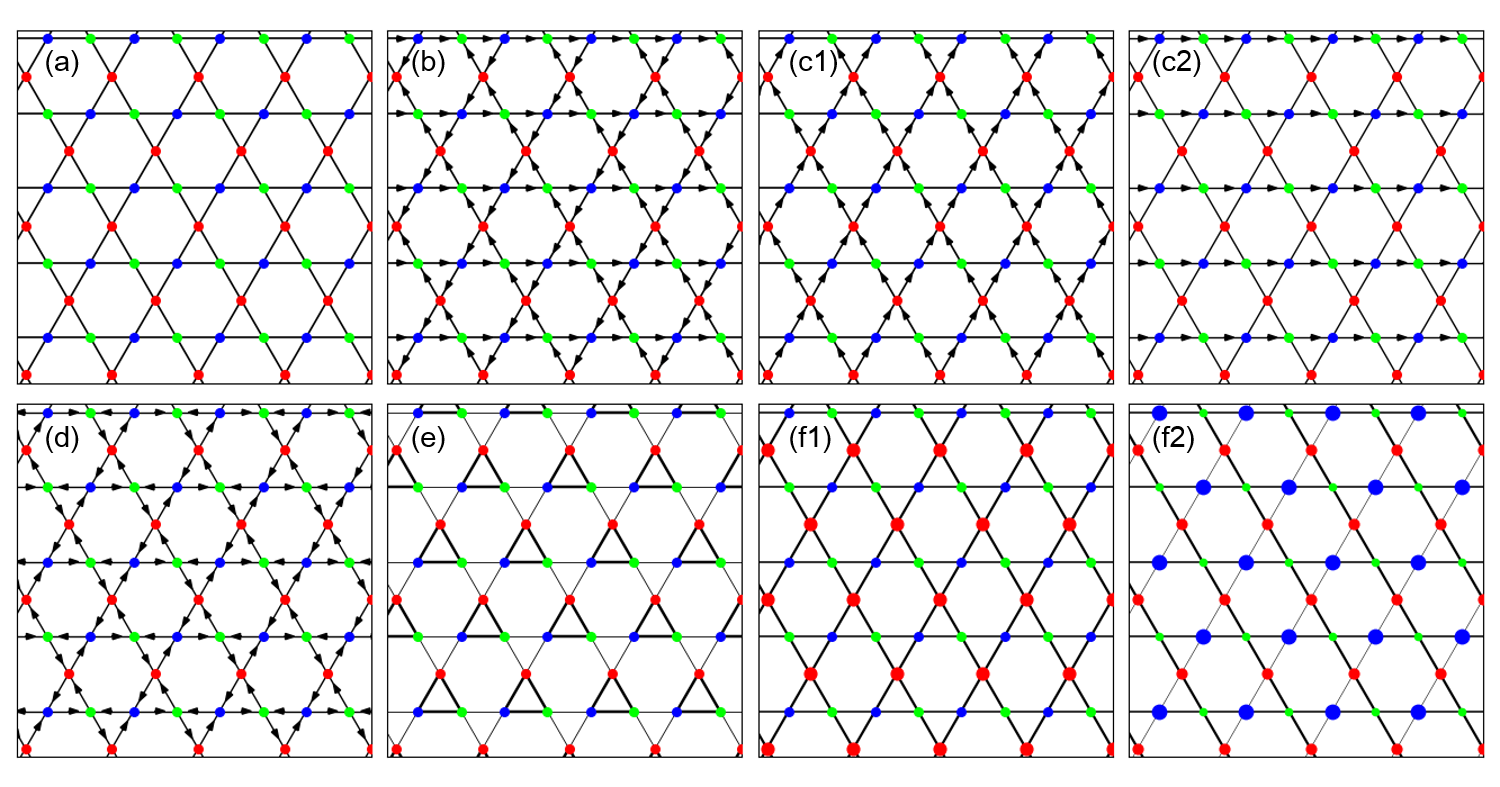}
    \caption{Possible charge orders at $\bm{q=0}$ for each irreducible representation of $C_{6v}$ group (a) $A_1$ (b) $B_1$ (c) $E_1$ (d) $A_2$ (e) $B_2$ (f) $E_2$. The radius of red, blue and green circles are onsite charge density of $1,2,3$ sublattices. The thickness of black bonds show the hopping amplitude. Arrows represent spontaneous current inter sublattices.}
    \label{gammaorder}
\end{figure}
\begin{figure}
    \centering
    \includegraphics[scale=0.5]{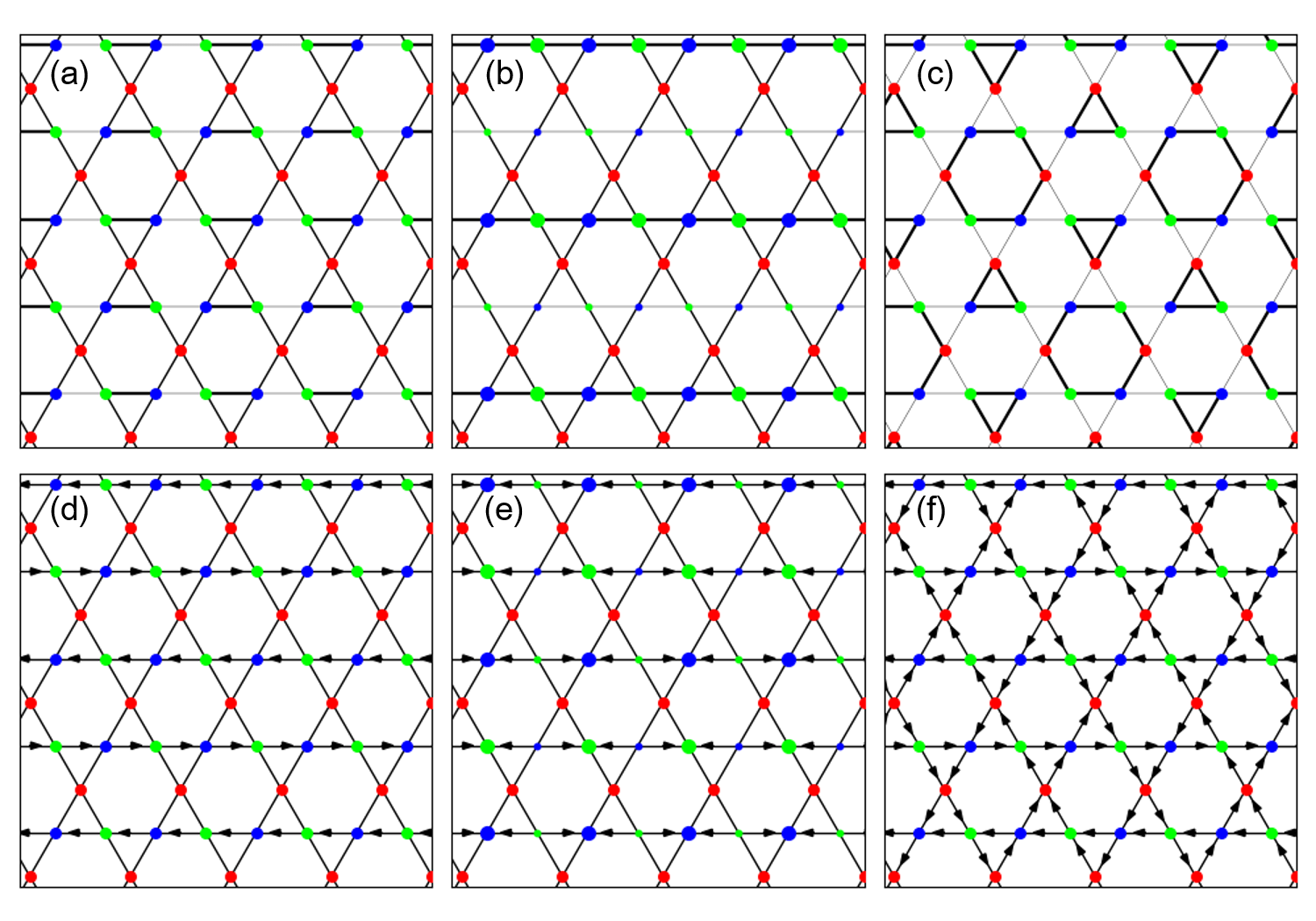}
    \caption{Possible charge orders at $\bm{q=M}$ for each irreducible representation of $C_{2v}$ group (a) $A_1$ (b) $B_1$ (d) $A_2$ (e) $B_2$. (c) and (f) are $3\bm{Q}$ orders of (a) and (d), respectively.}
    \label{Morder}
\end{figure}
\begin{table}[]
\centering
\begin{tabular}{|c|c|c|c|c|c|c|cccccccccccc}
\cline{1-7} \cline{15-19}
      & $E$ & $2C_6$ & $2C_3$ & $C_2$ & $3\sigma_v$ & $3\sigma_d$ &  &  &  &  &  &  & \multicolumn{1}{c|}{} & \multicolumn{1}{c|}{}      & \multicolumn{1}{c|}{$E$} & \multicolumn{1}{c|}{$C_2$} & \multicolumn{1}{c|}{$\sigma_v$} & \multicolumn{1}{c|}{$\sigma_d$} \\ \cline{1-7} \cline{15-19} 
$A_1$ & 1 & 1      & 1      & 1     & 1           & 1           &  &  &  &  &  &  & \multicolumn{1}{c|}{} & \multicolumn{1}{c|}{$A_1$} & \multicolumn{1}{c|}{1}   & \multicolumn{1}{c|}{1}     & \multicolumn{1}{c|}{1}          & \multicolumn{1}{c|}{1}          \\ \cline{1-7} \cline{15-19} 
$A_2$ & 1 & 1      & 1      & 1     & -1          & -1          &  &  &  &  &  &  & \multicolumn{1}{c|}{} & \multicolumn{1}{c|}{$A_2$} & \multicolumn{1}{c|}{1}   & \multicolumn{1}{c|}{1}     & \multicolumn{1}{c|}{-1}         & \multicolumn{1}{c|}{1}          \\ \cline{1-7} \cline{15-19} 
$B_1$ & 1 & -1     & 1      & -1    & 1           & -1          &  &  &  &  &  &  & \multicolumn{1}{c|}{} & \multicolumn{1}{c|}{$B_1$} & \multicolumn{1}{c|}{1}   & \multicolumn{1}{c|}{-1}    & \multicolumn{1}{c|}{1}          & \multicolumn{1}{c|}{-1}         \\ \cline{1-7} \cline{15-19} 
$B_2$ & 1 & -1     & 1      & -1    & -1          & 1           &  &  &  &  &  &  & \multicolumn{1}{c|}{} & \multicolumn{1}{c|}{$B_2$} & \multicolumn{1}{c|}{1}   & \multicolumn{1}{c|}{-1}    & \multicolumn{1}{c|}{-1}         & \multicolumn{1}{c|}{1}          \\ \cline{1-7} \cline{15-19} 
$E_1$ & 2 & 1      & -1     & -2    & 0           & 0           &  &  &  &  &  &  &                       &                            &                          &                            &                                 &                                 \\ \cline{1-7}
$E_2$ & 2 & -1     & -1     & 2     & 0           & 0           &  &  &  &  &  &  &                       &                            &                          &                            &                                 &                                 \\ \cline{1-7}
\end{tabular}
\label{table1}
\caption{Character tables of $C_{6v}$ at $\bm{Q}=0$ and $C_{2v}$ at $\bm{Q}=\bm{M}$ point groups. }
\end{table}

In this section, we discussion the classification of relevant charge order in kagome lattice within our calculations. For $\bm{Q=\Gamma/M}$ point, the little group is $C_{6v}/ C_{2v}$ and the corresponding character tables are shown in Table\ref{table1}. For $\bm{Q}=0$, there are five nontrivial irrep. except $A_1$ and the corresponding pattern of charge orders are displayed in Fig.\ref{gammaorder}. Two of them (e,f) belong to the real bond orders and three of them (b,c,d) belong to the imaginary bond orders. While, for $\bm{Q}=\textbf{M}$, there are four kinds of pattern of charge orders, shown in Fig.\ref{Morder}(a),(b),(d),(e). In the symmetric and antisymmetric channels, there are both real and imaginary bond orders. Further combining order at three $\textbf{M}$ points, we will have  $3\bm{Q}$ order and the corresponding pattern are displayed in Fig.\ref{Morder} (c) and (f).

%And the possible charge order for each irreducible representation is show in Fig.\ref{gammaorder} and Fig.\ref{Morder}, up to NN bond.

\section{RPA formalism for the onsite and bond charge suscepbilities}

\begin{figure}
    \centering
    \includegraphics[scale=0.6]{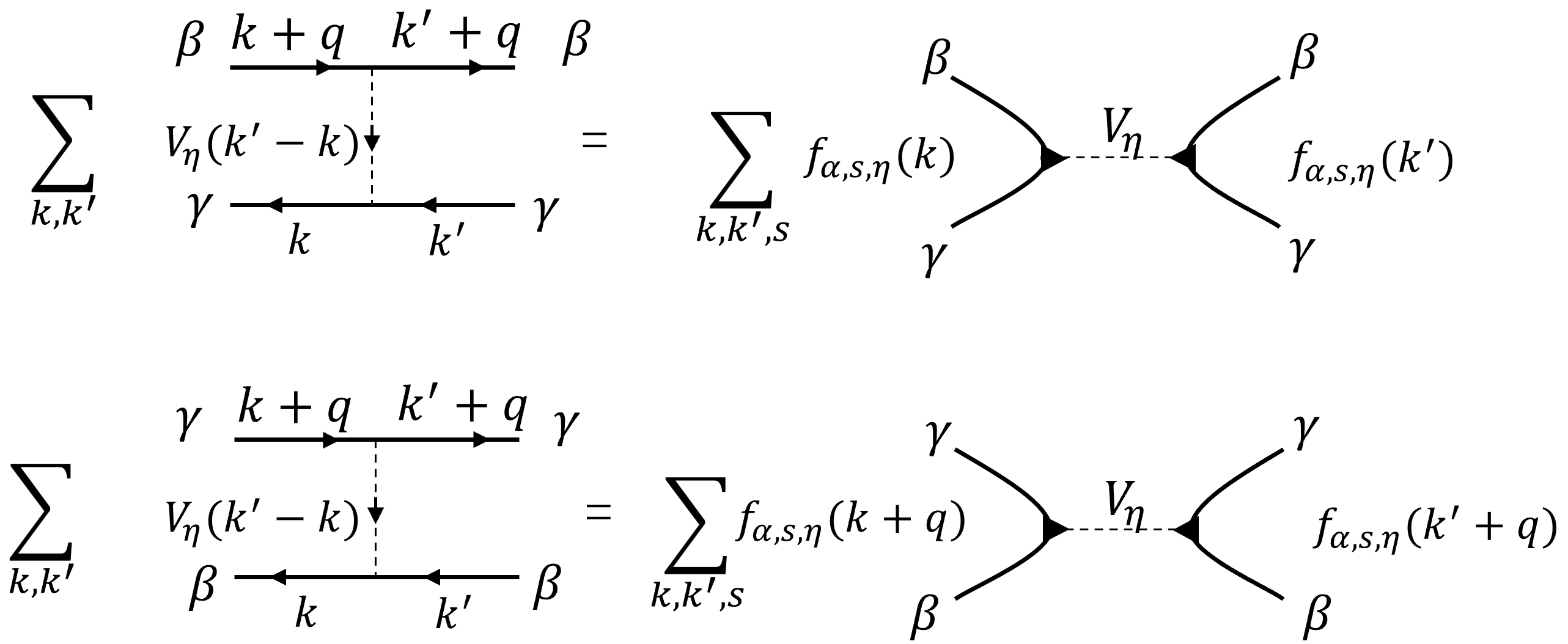}
    \caption{Decouple the ladder-type interaction with internal momentum into constant interaction with 2 bond operators. }
    \label{decouple}
\end{figure}

\begin{figure}
    \centering
    \includegraphics[scale=1.0]{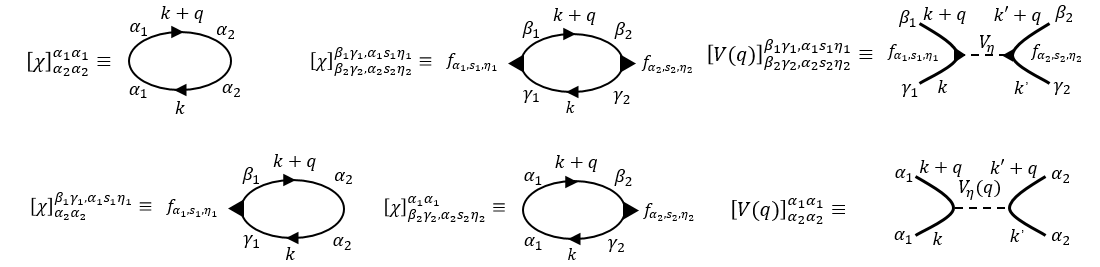}
    \caption{Definition of the susceptibility matrix and vertex indices. The momentum transfer in the form factor is decided by the order of orbital index. When $\epsilon_{\alpha\beta\gamma}=1$ we have $f_{\alpha,s,\eta}=f_{\alpha,s,\eta}(\bm{k})$. When $\epsilon_{\alpha\beta\gamma}=-1$, $f_{\alpha,s,\eta}$ represents $f_{\alpha,s,\eta}(\bm{k+q})$.}
    \label{definition}
\end{figure}
In order to analyze the charge instability beyond the mean field, we adopted the RPA formalism. At the RPA level, the charge susceptibility is usually obtained by considering the geometric summation of bubble-type type diagrams, where onsite and bond orders usually decoupled. To treat them on equal footing, we consider both bubble and ladder diagrams with interactions. However, as interactions are nonlocal and thus the internal interaction lines involve internal momentum, the analytic summation of such ladder diagrams are difficult. Here, the special properties of inter-site repulsion disscussed in the following allows to achieve it in a matrix form for our case~\cite{Romer2021}. 
%as shown in Fig.\ref{fig3.5}. 
%According to the spirit of RPA, all of diagrams must be expressed in the combination of $\chi_0$, $V$ and $\chi_{RPA}$ and the sum of inner momentum must be absorbed in susceptibilities to achieve a Dyson type equation. 
%in the 2nd column on the right hand side of Fig.\ref{fig3.5},

\begin{figure}
    \centering
    \includegraphics[scale=0.7]{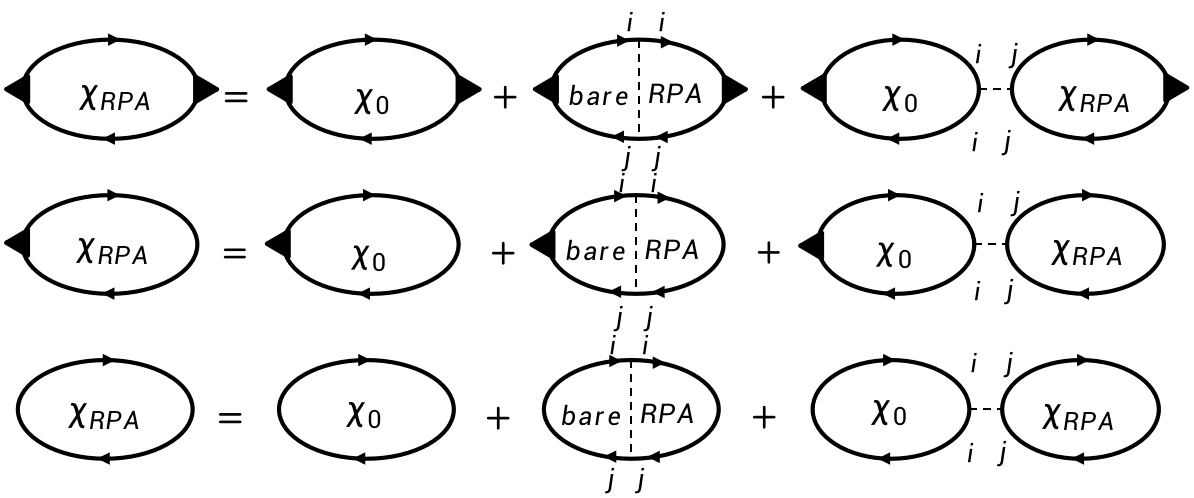}
    \caption{Full RPA diagram for susceptibilities in bond (upper), mixed (middle) and onsite (bottom) channels. 
    %vertecial line in 3rd column can be decoupled
    }
    \label{fig3.5}
\end{figure}

To deal with the interaction lines carrying internal momentum in ladder diagrams, we decouple their form factors into bond form factors using trigonometric identities,
\begin{align} 
    & \cos(\bm{(k-k')}\cdot \bm{l})= \cos(\bm{k \cdot l})\cos(\bm{k'\cdot l})+\sin(\bm{k \cdot l})\sin(\bm{k'\cdot l}), \label{Eqdec1}, \\
    & \cos(\bm{(k-k')}\cdot \bm{l})= \cos(\bm{(k+q) \cdot l})\cos(\bm{(k'+q)\cdot l})+\sin(\bm{(k+q) \cdot l})\sin(\bm{(k'+q)\cdot l}) \label{Eqdec2}.
\end{align} 
With this decoupling, two internal momenta are separated and two additional vertices appear, as schematically demonstrated in Fig.\ref{decouple}. We explicitly show it in the first-order ladder diagram and its formula is given by,
\begin{align}
    & \sum_{\bm{k,k'}}   f_{\alpha_1,s_1,\eta_1}(\bm{k})G^{\beta_1\beta_2}(\bm{k+q}) G^{\gamma_2\gamma_1}(\bm{k})
    V_{\eta_2}(\bm{k-k'})
    G^{\beta_2 \beta_3}(\bm{k'+q}) G^{\gamma_3 \gamma_2}(\bm{k'})f_{\alpha_3,s_3,\eta_3}(\bm{k'}) \nonumber\\ 
    &=   \sum_{\bm{k,k'},s_2} f_{\alpha_1,s_1,\eta_3)}(\bm{k})G^{\beta_1\beta_2}(\bm{k+q}) G^{\gamma_2 \gamma_1}(\bm{k}) f_{\alpha_2,s_2,\eta_2}(\bm{k}) V_{\eta_2} f_{\alpha_2,s_2,\eta_2}(\bm{k'}) G^{\beta_2 \beta_3}(\bm{k'+q}) G^{\gamma_3 \gamma_2}(\bm{k'})f_{\alpha_3,s_3,\eta_3}(\bm{k'})\nonumber\\
    & = \sum_{s_2}[\chi]^{\beta_1\gamma_1,\alpha_1 s_1 \eta_1}_{\beta_2,\alpha_2 \gamma_2,s_2 \eta_2}(\bm{q}) V_{\eta_2} [\chi]^{\beta_2 \gamma_2,\alpha_2 s_2 \eta_2}_{\beta_3 \gamma_3,\alpha_3 s_3 \eta_3}(\bm{q}).
\end{align}
Where $G^{\alpha\beta}$ is the Green's function involving sublattice $\alpha$ and $\beta$. Thus, a bond susceptibility in a specific channel can be coupled with those in other channels through the ladder diagrams.  
%to decouple the vertical interaction line into bond form factors. In this way, a vertical interaction line with momentum dependence can be turned into a constant interaction with two bond operators, as shown in Fig.\ref{decouple}. Using Eq\ref{Eqdec1} or Eq\ref{Eqdec2} is determined by the order of sublattice index $\beta$ and $\gamma$. So that we can deal with these ladder diagrams. For example:
While, the bubble diagram of bond susceptibilities will introduce a susceptibility in the mixed channel with only one vertex (as shown in Fig.\ref{definition}), which is the coupling between bond and onsite charge order. Then, the first-order bubble of this mixed susceptibility will involve the onsite susceptibility. The ladder and bubble diagrams couple the onsite and bond charge orders. The full RPA diagrams are displayed in Fig.\ref{fig3.5}, which can expressed,
%and bonmotivates us to introduce a general- ized susceptibility matrix 
%Thus the RPA susceptibility can be calculated, and full RPA susceptibility in Fig\ref{fig3.5} can be expressed as
\begin{align}
    & [\chi_{RPA}]^{\alpha_1 \beta_1,\gamma_1 s_1 \eta_1}_{\alpha_2 \beta_2,\gamma_2s_2 \eta_2}=[\chi_0]^{\alpha_1 \beta_1,\gamma_1 s_1 \eta_1}_{\alpha_2 \beta_2,\gamma_2 s_2 \eta_2}+\sum_{ij,ks\eta}[\chi_0]^{\alpha_1 \beta_1,\gamma_1 s_1 \eta_1}_{i j,ks \eta} [V_\eta(\bm{q})]^{i j,ks\eta}_{i j,ks\eta} [\chi_{RPA}]^{i j,k s \eta}_{\alpha_2 \beta_2}+[\chi_0]^{\alpha_1 \beta_1}_{i i} [V_\eta(\bm{q})]^{i i}_{j j} [\chi_{RPA}]^{j j}_{\alpha_2 \beta_2}, \\
    & [\chi_{RPA}]^{\alpha_1 \beta_1,\gamma_1 s_1 \eta_1}_{\alpha_2 \alpha_2}=[\chi_0]^{\alpha_1 \beta_1,\gamma_1 s_1 \eta_1}_{\alpha_2 \alpha_2}+\sum_{ij,ks\eta}[\chi_0]^{\alpha_1 \beta_1}_{i j,s \eta} [V_\eta(\bm{q})]^{i j,k s \eta}_{i j,k s \eta} [\chi_{RPA}]^{i j,k s \eta}_{\alpha_2 \alpha_2}+[\chi_0]^{\alpha_1 \beta_1,\gamma_1 s_1 \eta_1}_{i i} [V_\eta(\bm{q})]^{i i}_{j j} [\chi_{RPA}]^{j j}_{\alpha_2 \alpha_2}, \\
    & [\chi_{RPA}]^{\alpha_1 \alpha_1}_{\alpha_2 \alpha_2}=[\chi_0]^{\alpha_1 \alpha_1}_{\alpha_2 \alpha_2}+\sum_{ij,k s\eta}[\chi_0]^{\alpha_1 \alpha_1}_{i j,k s \eta} [V_\eta(\bm{q})]^{i j,k s \eta}_{i j,k s \eta} [\chi_{RPA}]^{i j,k s \eta}_{\alpha_2 \alpha_2}+[\chi_0]^{\alpha_1 \alpha_2}_{i i} [V_\eta(\bm{q})]^{i i}_{j j} [\chi_{RPA}]^{j j}_{\alpha_2 \alpha_2}. 
\end{align}
Here, $[\chi]^{\alpha_1 \beta_1,\gamma_1 s_1 \eta_1}_{\alpha_2 \beta_2,\gamma_2s_2 \eta_2}$, $[\chi]^{\alpha_1 \beta_1,\gamma_1 s_1 \eta_1}_{\alpha_2 \alpha_2}$ and $[\chi]^{\alpha_1 \alpha_1}_{\alpha_2 \alpha_2}$ represent susceptiblity in the bond, mixed and onsite channels, respectively. The upper (lower) indices are those in left (right) part of susceptibility bubble or interaction vertex as shown in Fig.\ref{definition}. $ \alpha,\beta,\gamma\in{1,2,3}$ are sublattice indices and the ($\gamma s \eta$) index denotes the form factor. 
This hierarchy structure motives us to treat oniste and bond charge order together in a RPA matrix form. Under the basis $\mathcal{O}_o$ defined in above section, the full $27\times27$ RPA susceptibility is given by
\begin{equation}
    \chiup_{RPA}(\bm{q})= (I+\chiup_0(\bm{q})\mathcal{U}_c(\bm{q}))^{-1} \chiup_0(\bm{q}),
\end{equation}
where the interaction matrix reads,
\begin{equation}
\begin{aligned}
\mathcal{U}_c(\boldsymbol{q}) & =\left(\begin{array}{ccc}
V_{\mathrm{1}}^c & 0 & 0 \\
0 & V_{\mathrm{2}}^c & 0 \\
0 & 0 & U^c(\boldsymbol{q})
\end{array}\right), \\
U^c(\boldsymbol{q}) & =\left(\begin{array}{ccc}
0 & V_{12} & V_{13} \\
V_{12} & 0 & V_{23} \\
V_{13} & V_{23} & 0
\end{array}\right), \\
V_\eta^c & =-\operatorname{diag}\left\{2 V_\eta, 2 V_\eta, \ldots\right\}, \\
V_{\beta \gamma} & =\sum_\eta 2 V_{\mathrm{\eta}} f_{\alpha,+, \mathrm{\eta}}(\boldsymbol{q}).
\end{aligned}
\end{equation}
Here $V^c_\eta$ is a $12\times12$ matrix for the $\eta$-th bond channel.

\section{Classic energy for the CDW with intra-unit charge modulation}
In this section, we discuss  the classical potential energy gain of the nSDM state, which mainly involves charge density modulation within three sublattices. Assuming the charge density in three sublattices is modified to $n_1=n+(\delta_2+\delta_3),\text{ }n_2=n-\delta_2,\text{ }n_3=n-\delta_3$, where $n$ is the normal charge density, we discussion the potential energy gain under different inter-site repulsion.

%Under onsite repulsion $U$, the energy difference between nSDM phase and uniform charge density is.
%\begin{equation}
%    \Delta E_U= U (n_1^2+n_2^2 +n_3^2-3n^2)=U[(\delta_2+\delta_3)^2+\delta_2^2+\delta_3^2]>0
%\end{equation}

Under the NN repulsion $V_1$, the energy difference relative to the normal state is
\begin{equation}
    \Delta E_{V_1}=2V_1 (n_1 n_2+n_2 n_3 +n_3 n_1-3n^2)=-2V_1(\delta_2\delta_3+\delta_2^2+\delta_3^2)<0.
\end{equation}
Under the NNN repulsion $V_2$, the energy difference relative to the normal state is
\begin{equation}
    \Delta E_{V_2}=2V_2 (n_1 n_2+n_2 n_3 +n_3 n_1-3n^2)=-2V_2(\delta_2\delta_3+\delta_2^2+\delta_3^2)<0.
\end{equation}
Under the NNNN repulsion $V_3$, which is a intra-sublattice repulsion, the energy gain is
\begin{equation}
    \Delta E_{V_3}=3 V_3 (n_1^2+n_2^2 +n_3^2-3n^2)=3V_3[(\delta_2+\delta_3)^2+\delta_2^2+\delta_3^2]>0.
\end{equation}
From the above expressions, we can conclude that $V_1 \text{ and } V_2$ will promote the nSDM phase while $ V_3$ will suppress it. The energy gain is increases linearly with the replsion $V_{1,2}$ and this explains that nSDM is favored when the repulsion is strong.

\section{Physical understanding of CDW order tendency for NN and NNN bonds}
In this section, we provide physical understanding about the favored orders on the NN and NNN bonds, using weak-coupling analysis. There are three Fermi surface nesting vectors $\bm{Q}_{1,2,3}$ connecting three pairs of opposite edges. The charge density wave orders with $\bm{Q}_{1,2,3}$ are in general degenerate and thus we can choose one charge order with $\bm{Q}_i$ to study its effect on the band structures.  Taking the charge order with $\bm{Q}_1$ as an example, the effective Hamiltonian in the band space with the inclusion of this CDW reads,
 \begin{eqnarray}
H_{\text{eff}}(\bm{k})&=& \left(\begin{array}{cc}  \epsilon_\nu(\bm{k})  & \Delta(\bm{k})  \\ 
 \Delta^\dag(\bm{k}) &  \epsilon_{\nu}(\bm{k}+\bm{Q}_1)   \\  \end{array}\right),
\end{eqnarray}
where the $ \epsilon_{\nu}(\bm{k})$ is the $\nu$-th band energy and $\Delta(\bm{k})$ is the CDW order parameter in the band space. In the weak-coupling limit ($ \Delta(\bm{k}) $ is small), the gap opening is prominent around the Fermi level, i.e. $\epsilon_\nu(\bm{k})\approx0$, and the relevant states are the Fermi segments connected by the $\bm{Q}_1$ ( shown in Fig.1(b) ). The instability of this CDW order is determined by the energy gain, which is directly proportional to the induced gap size on the Fermi surface. Thus, we can explore the gap size on the Fermi surfaces of various CDW orders to examine the CDW formation propensity. In the following, we study both onsite and bond CDW orders in the band space.

The onsite CDW operator is $n_\alpha(\bm{q})=\frac{1}{\sqrt{N}}\sum_{\bm{k}}c^\dag_{\alpha}(\bm{k}+\bm{q})c_\alpha(\bm{k})$. The corresponding CDW order parameter in the band space reads $\Delta^\nu_{n_\alpha}(\bm{k}) \propto \langle c^\dag_{\alpha}(\bm{k}+\bm{Q}_1)c_\alpha(\bm{k}) \rangle_\nu$, where $\langle\rangle_{\nu}$ is the average over the eigenvector of the $\nu$-th band. Due to the sublattice texture on the Fermi surface, $\Delta^\nu_n(\bm{k})$ is zero when $\bm{k}$ is located at the VHS and reaches the maximum at the midpoints between two VHSs. Thus, the VHSs with large DOS cannot be gapped by this onsite CDW order and this leads to a small energy gain.

The bond CDW operator on the $\beta\gamma$ bond is $\tilde{B}_{\alpha,\pm,\eta}(\bm{q})=\frac{1}{\sqrt{N}}\sum_{\bm{k}} \tilde{f}_{\alpha,\pm,\eta}(\bm{k}) c^\dagger_{\beta}(\bm{k}+\bm{q}) c_{\gamma}(\bm{k})$. The real and imaginary bond order operators in the symmetric and anti-symmetric channels are given by,
\begin{eqnarray}
    B_{\alpha,\pm,\eta}^{\prime(\prime\prime)}(\bm{q}) &=& \frac{1}{2(i)} \{\tilde{B}_{\alpha,\pm,\eta}(\bm{q})\pm [\tilde{B}_{\alpha,\pm,\eta}(\bm{q})]^\dag \},\\
    &=& \frac{1}{2(i)}\frac{1}{\sqrt{N}}\sum_{\bm{k}} [\tilde{f}_{\alpha,\pm,\eta}(\bm{k})  c^\dagger_{\beta}(\bm{k}+\bm{q}) c_{\gamma}(\bm{k}) \pm \tilde{f}^*_{\alpha,\pm,\eta}(\bm{k})   c^\dag_{\gamma}(\bm{k}) c_{\beta}(\bm{k}-\bm{q})]
\end{eqnarray}
Setting $\bm{q}=\bm{Q}_1$ and $(\alpha\beta\gamma)=(123)$, these operators can be further written as
\begin{eqnarray}
    B_{\alpha,\pm,\eta}^{\prime(\prime\prime)}(\bm{Q}_1) &=& \frac{1}{2(i)}\frac{1}{\sqrt{N}}\sum_{\bm{k}} [-\tilde{f}_{\alpha,\pm,\eta}(\bm{k}-\bm{Q}_1)c^\dagger_{\beta}(\bm{k}) c_{\gamma}(\bm{k}+\bm{Q}_1)  \mp \tilde{f}^*_{\alpha,\pm,\eta}(\bm{k})c^\dagger_{\gamma}(\bm{k}) c_{\beta}(\bm{k}+\bm{Q}_1)  ].
 \end{eqnarray}
 Due to $\bm{Q}_1 \cdot \bm{l}_{1,\text{NN}}=\pi$ and $\bm{Q}_1 \cdot \bm{l}_{1,\text{NNN}}=0$, the operators on the NN and NNN bonds can be written as,
\begin{eqnarray}
    B_{\alpha,\pm,\text{NN}}^{\prime(\prime\prime)}(\bm{Q}_1) &=& \frac{1}{2(i)}\frac{1}{\sqrt{N}}\sum_{\bm{k}} [\tilde{f}_{\alpha,\pm,\text{NN}}(\bm{k})c^\dagger_{\beta}(\bm{k}) c_{\gamma}(\bm{k}+\bm{Q}_1)  \mp \tilde{f}^*_{\alpha,\pm,\text{NN}}(\bm{k})c^\dagger_{\gamma}(\bm{k}) c_{\beta}(\bm{k}+\bm{Q}_1)  ],\\
    B_{\alpha,\pm,\text{NNN}}^{\prime(\prime\prime)}(\bm{Q}_1) &=& \frac{1}{2(i)}\frac{1}{\sqrt{N}}\sum_{\bm{k}} [-\tilde{f}_{\alpha,\pm,\text{NNN}}(\bm{k})c^\dagger_{\beta}(\bm{k}) c_{\gamma}(\bm{k}+\bm{Q}_1)  \mp \tilde{f}^*_{\alpha,\pm,\text{NNN}}(\bm{k})c^\dagger_{\gamma}(\bm{k}) c_{\beta}(\bm{k}+\bm{Q}_1)  ].
 \end{eqnarray}  

\begin{figure}
    \centering
    \includegraphics[scale=0.5]{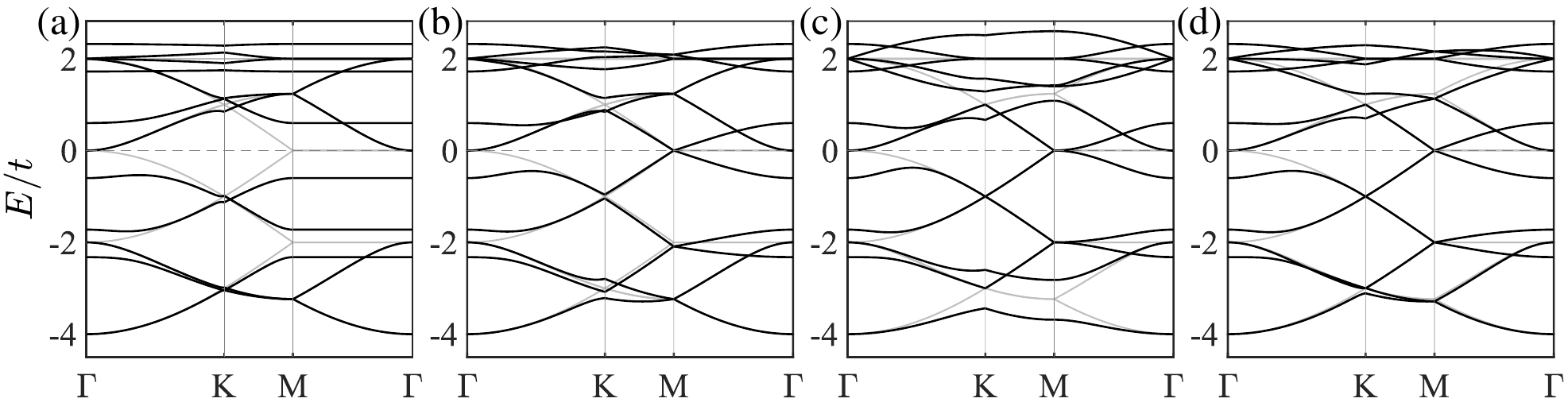}
    \caption{ Band structure with bond CDW on the NN and NNN bonds: 1Q real (a) and imaginary (b) bond order on the NN bond; 1Q real (a) and imaginary (b) bond order on the NNN bond. The gray and black curves denote the folded bands without and with CDW order. Here the bond order parameters are the same for four plots.}
    \label{CDWgap}
\end{figure}

 On the two Fermi surface segments, we have $\bm{k} \cdot \bm{l}_{\alpha,\text{NN}}=\pm\pi/2$ and $\bm{k}_{\beta,\gamma} \cdot \bm{l}_{\alpha,\text{NNN}}=\pm\pi/2$. The NN and NNN symmetric form factors vanish on the VHSs and thus the corresponding bond CDW order parameter $\Delta^\nu_B(\bm{k})$ on the Fermi surface is small, making them less favored. In the anti-symmetric channel, the form factor is imaginary and we have $\tilde{f}^*_{\alpha,-,\eta}(\bm{k})=-\tilde{f}_{\alpha,-,\eta}(\bm{k})$. The anti-symmetric CDW order parameters on the NN and NN bonds  can be rewritten as,
 \begin{eqnarray}
 \label{AS1}
    \Delta^\nu_{B^{\prime/\prime\prime}_{NN}}(\bm{k}) &\propto& \tilde{f}_{\alpha,-,\text{NN}}(\bm{k}) \langle c^\dagger_{\beta}(\bm{k}) c_{\gamma}(\bm{k}+\bm{Q}_1)  \pm c^\dagger_{\gamma}(\bm{k}) c_{\beta}(\bm{k}+\bm{Q}_1) \rangle_\nu ,\\
     \label{AS2}
    \Delta^\nu_{B^{\prime/\prime\prime}_{NNN}}(\bm{k})  &\propto& -\tilde{f}_{\alpha,-,\text{NNN}}(\bm{k}) \langle c^\dagger_{\beta}(\bm{k}) c_{\gamma}(\bm{k}+\bm{Q}_1)  \mp c^\dagger_{\gamma}(\bm{k}) c_{\beta}(\bm{k}+\bm{Q}_1) \rangle_\nu .
 \end{eqnarray} 
For the NN bond order parameter, $ \tilde{f}_{\alpha,-,\text{NN}}(\bm{k})=i$ for $\bm{k}$ on the Fermi surface segments. When $\bm{k}$ is around the VHSs, the CDW order parameters $\Delta^\nu_{B^{\prime/\prime\prime}_{NN}}(\bm{k}) $ are close in both channels due to the pure sublattice feature. When $\bm{k}$ moves away from the VHSs, the corresponding eigenstate is a mixture of two sublattices. This leads to enhanced $\Delta^\nu_{B^{\prime}_{NN}}(\bm{k}) $ but decreased $\Delta^\nu_{B^{\prime\prime}_{NN}}(\bm{k}) $ due to the addition and cancelation between two terms. Especially, $\Delta^\nu_{B^{\prime\prime}_{NN}}(\bm{k}) $ vanishes at the midpoint between two VHSs. Therefore, the real bond CDW is more favorable than the imaginary CDW in the anti-symmetric NN channel. For the NNN bond orders, $ \tilde{f}_{\alpha,-,\text{NN}}(\bm{k})$ varies with $\bm{k}$ and drops to zero at the midpoint between two VHSs.  The behaviors of  $\Delta^\nu_{B^{\prime\prime}_{NNN}}(\bm{k}) $ is the opposite with those of  $\Delta^\nu_{B^{\prime\prime}_{NN}}(\bm{k}) $ according to Eq.\ref{AS2}. Thus, the imaginary bond CDW is more favorable than the real CDW in the anti-symmetric NNN channel. These insights provide a physical understanding for our analysis of relative CDW fluctuations and CDW order tendency on these bonds in the main text. We further perform numerical calculations with different CDW orders, as shown in Fig.\ref{CDWgap}. From the case of bond order on the NN bond (a) (real) and (b) (imaginary), it is evident that the real bond order open large CDW gap on the Fermi surface. Similarly, the CDW gap for real and imaginary bond orders are comparable but around the M point, the CDW gap of the imaginary bond order is larger. All these results are consistent with our above analytical analysis. 

\section{ Charge patterns of CDW order and their impact on band structures }
\begin{figure}
    \centering
    \includegraphics[scale=0.5]{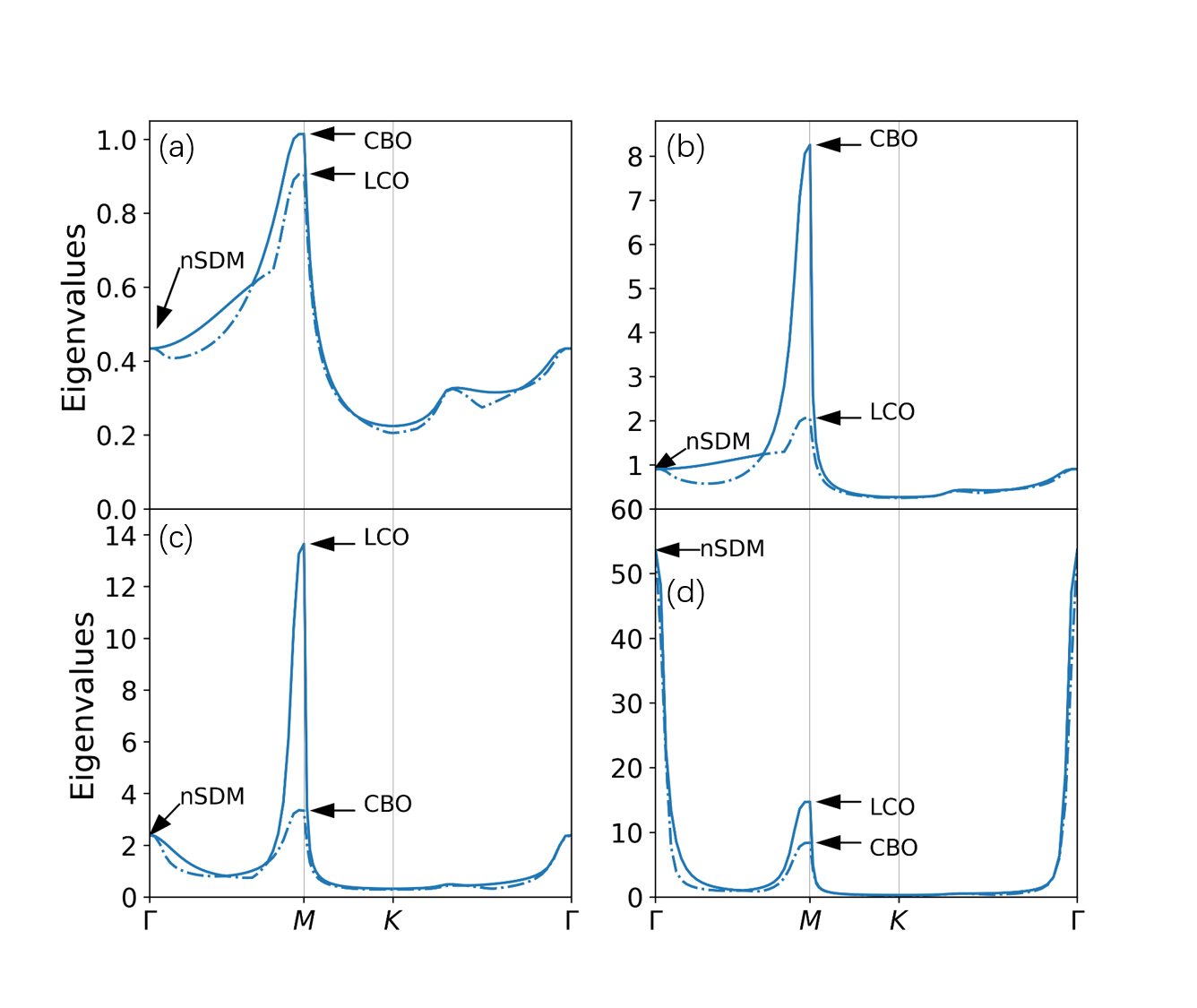}
    \caption{Eigenvalue of susceptibility matrix along high symmetry line. (a) Bare susceptibility (b) $V_{\text{1}}=0.6$, $V_{2}=0.0$. (c) $V_{1}=0.0$, $V_{2}=0.95$. (d) $V_{1}=0.5$, $V_{2}=0.75$. The temperatures in (a)-(c) and (d) are $\beta=200$ and $\beta=100$, respectively, same parameter with Fig2 in main text. Solid line and dashed line are 1st and 2nd largest eigenvalue.}
    \label{eigval}
\end{figure}

In this section, we study the CDW pattern of CDW instabilities and their impact on the electronic structures. With decreasing temperature, an eigenvalue of the RPA susceptibility $\chiup_{\text{RPA}}$ at $\Gamma$ or $\bf{M}$ diverges, signaling an instability at this vector. The CDW pattern is associated with the corresponding eigenvector. Fig.\ref{eigval} shows the largest eigenvalues of the susceptibility matrix along the high-symmetry path with different interactions at two temperatures. The soild and dashed lines denote 1st and 2nd largest eigenvalue, respectively, while the arrows show the corresponding CDW pattern at $\Gamma$ or $\bf{M}$ points. These results fully agree with our calculations presented in Fig.2 in the main text.

\begin{figure}
    \centering
    \includegraphics[scale=0.6]{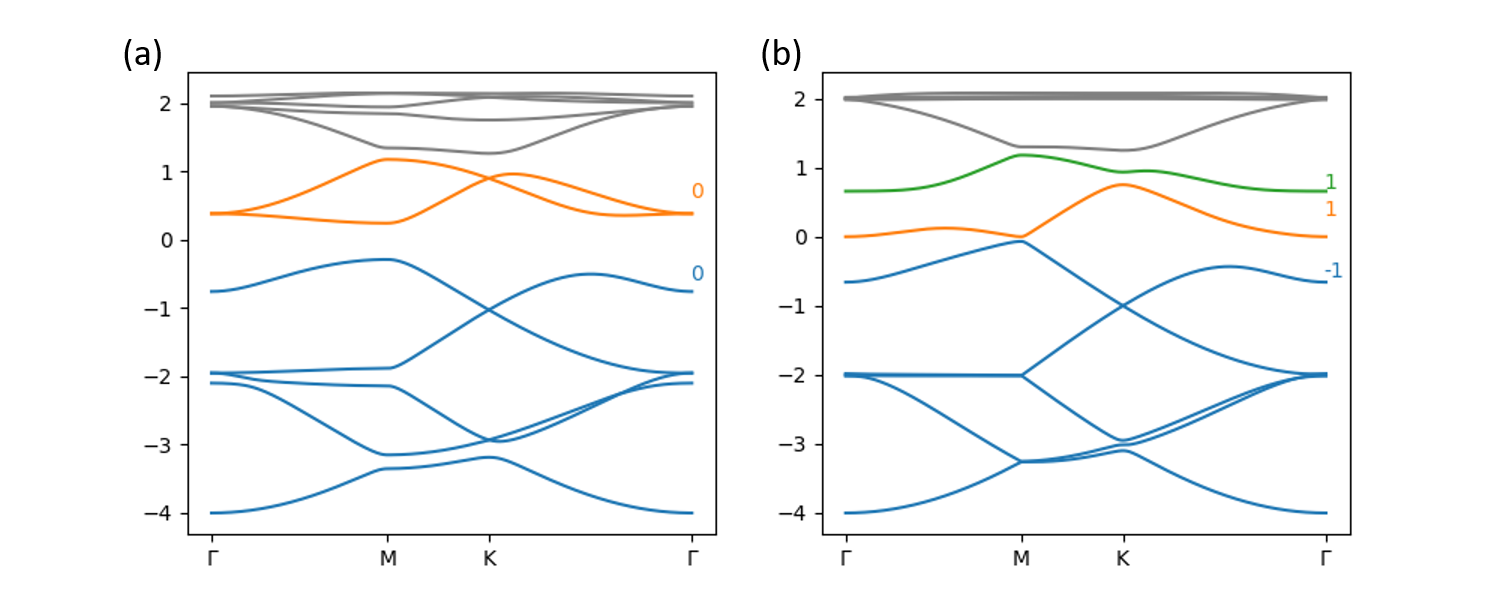}
    \caption{Energy band of CBO and LCO phase in folded BZ, numbers near the bands with same color are corresponding Chern number of the bands}
    \label{chern_band}
\end{figure}

We further study the effect of these orders on the band structures. Within a CDW state, its effect can be described by the symmetry-breaking mean-field Hamiltonian,
\begin{equation}
\begin{aligned}
H^{\text{MF}}_{\text{CDW}} & =\sum_{R,\alpha}[ (t+\Delta^{Re}_{\alpha,+,NN}(R)+i \Delta^{Im}_{\alpha,+,NN}(R) +\Delta^{Re}_{\alpha,-,NN}(R)+i \Delta^{Im}_{\alpha,-,NN}(R))c_{\beta,r}^\dagger c_{\gamma,r+a_\alpha/2}  \\
 & +(t+\Delta^{Re}_{\alpha,+,NN}(R)+i \Delta^{Im}_{\alpha,+,NN}(R)-\Delta^{Re}_{\alpha,-,NN}(R)-i \Delta^{Im}_{\alpha,-,NN}(R)) c_{\beta,r}^\dagger c_{\gamma,r - a_\alpha/2} \\
& +(\Delta^{Re}_{\alpha,+,NNN}(R)+i \Delta^{Im}_{\alpha,+,NNN}(R) +\Delta^{Re}_{\alpha,-,NNN}(R)+i \Delta^{Im}_{\alpha,-,NNN}(R))c_{\beta,r}^\dagger c_{\gamma,r+(a_\beta-a_\gamma)/2} \\
 & + (\Delta^{Re}_{\alpha,+,NNN}(R)+i \Delta^{Im}_{\alpha,+,NNN}(R)-\Delta^{Re}_{\alpha,-,NNN}(R)-i \Delta^{Im}_{\alpha,-,NNN}(R)) c_{\beta,r}^\dagger c_{\gamma,r - (a_\beta-a_\gamma)/2}-(\mu+\rho_\alpha(R))c^\dagger_{r,\alpha} c_{r,\alpha}]. 
\end{aligned}
\end{equation}
Here, $R$ is the coordinate of the unicell, $r$ is the coordinate of the corresponding sublattice, $t=-1$ is the hopping integral and $\Delta$ is the CDW order parameter. $\Delta_{\alpha,\pm,\eta}^{Re/Im}(R)=\Delta_{\alpha,\pm,\eta}^{Re/Im} \cos(\bm{Q\cdot R})$ is the order parameter of the symmetric or antisymmetric CBO/LCO, $\rho_\alpha(R)=\rho_\alpha \cos(\bm{Q\cdot R})$ is the onsite charge order and $\mu$ is the chemical potential at the p-type VH filling ($\mu=0$).

%To understand the possible instability induced by these fluctuations, we diagonalize the susceptibility matrix and find the eigenvector corresponds to the largest eigenvalue. 
We show some representative CDW patterns from the eigenvector of RPA susceptibility. In the $V_1$-dominant regime, the real bond fluctuation at M point is strong, as shown in Fig.\ref{eigval}(b), and the CDW pattern mainly involves NN and NNN bonds. The corresponding order parameters are $\Delta_{\alpha,-,NN}^{Re}:\Delta_{\alpha,-,NNN}^{Re}\approx -0.0604:0.0366$. In the $V_2$-dominant regime, the imaginary bond fluctuation at M point is strong, as shown in Fig.\ref{eigval}(c), and the CDW pattern mainly involves NN and NNN bonds. The corresponding order parameters are $\Delta_{\alpha,-,NN}^{Im}:\Delta_{\alpha,-,NNN}^{Im}\approx-0.0469:0.0528$. In both regimes, the 3Q will be more energetically favored according to our analysis.  With strong repulsion, onsite CDW fluctuation at $\Gamma$ point is dominant and the corresponding CDW pattern involving symmetric bond fluctuations is two-fold. The two-fold order parameters are: $\rho_1:\rho_2:\rho_3=2:-1:-1$, $\Delta_{2,+,NN}=\Delta_{3,+,NNN}$ and $\rho_1:\rho_2:\rho_3=0:1:-1$, $\Delta_{2,+,NN}=-\Delta_{3,+,NNN}$. According to our analysis in the main text, this state forms a nematic order, breaking the six-fold rotational symmetry.

%At a low temperature, the largest eigenvalue of the susceptibility matrix peeks at $\bm{Q}=\textbf{M}$ point the result eigenvector indicate these flucluations will drive the system to CBO phase with coexist nonzero NN and NNN order $\Delta_{\alpha,-,NN}^{Re}:\Delta_{\alpha,-,NNN}^{Re}\approx -0.0604:0.0366$ when $V_1$ is large and LCO phase with coexist NN and NNN order $\Delta_{\alpha,-,NN}^{Im}:\Delta_{\alpha,-,NNN}^{Im}\approx-0.0469:0.0528$ when $V_2$ is large. Under a higher temperature, if $V_1$ and $V_2$ are comparable and the eigenvalue will peak at $\bm{Q}=\bm{\Gamma}$ point, indicting a two-degenerate onsite nSDM phase with nonzero order parameter $\rho_1:\rho_2:\rho_3=2:-1:-1$ and $\Delta_{2,+,NN}=\Delta_{3,+,NNN}$ which breaks C6 rotation symmetry. Fig.\ref{eigval} shows the largest eigenvalues of the susceptibility matrix along the high symmetry line, the soild and dashed lines are 1st and 2nd largest eigenvalue, and arrows show the phase correspond to the eigenvalue and $\Gamma$ or $M$ points. this result is in agree with the argument about Fig2 in the main text. 

In the main text, we show the unfolded band structure in 3Q CBO and 3Q LCO and nSDM states in Fig.4 of the main text. The adopted order parameters used are (a) $\Delta^{Re}_{\alpha,-,NN}=-0.12, \Delta^{Re}_{\alpha,-,NNN}=0.07$ for CBO phase, (b) $\Delta^{Im}_{\alpha,-,NN}=-0.09, \Delta^{Im}_{\alpha,-,NNN}=0.10$ for LCO phase and (c) $\rho_1=0.2 \text{, }\rho_2=\rho_3=-0.1$, $\Delta^{Re}_{2,+,NN}=\Delta^{Re}_{3,+,NN}=0.05$ for nSDM phase.
The corresponding folded band structures of 3Q CBO and LCO phase are displayed in Fig.\ref{chern_band}. The CBO opens a isotropic full gap around the Fermi energy. For the LCO phase, the gap is nearly closed at the M point folded Brillouin zone. Since the LCO phase breaks the time reversal symmetry, which introduces non-trivial Chern numbers in band structures, shown in Fig.\ref{chern_band}. The total Chern number for the occupied bands is -1 in the LCO phase. The nonzero total Chern number in LCO phase can lead to an anomalous hall effect and the LCO can generate orbital magnetism, which may make it possible to distinguish CBO and LCO in experiment.
%\end{equation}
%\begin{figure}
%    \centering
%    \includegraphics{fm_diagram_div.pdf}
%    \caption{Caption}
%    \label{decouple}
%\end{figure}
\section{Orbital magnetic moment of the loop current order}
\begin{figure}
    \centering
    \includegraphics[width=0.5\textwidth]{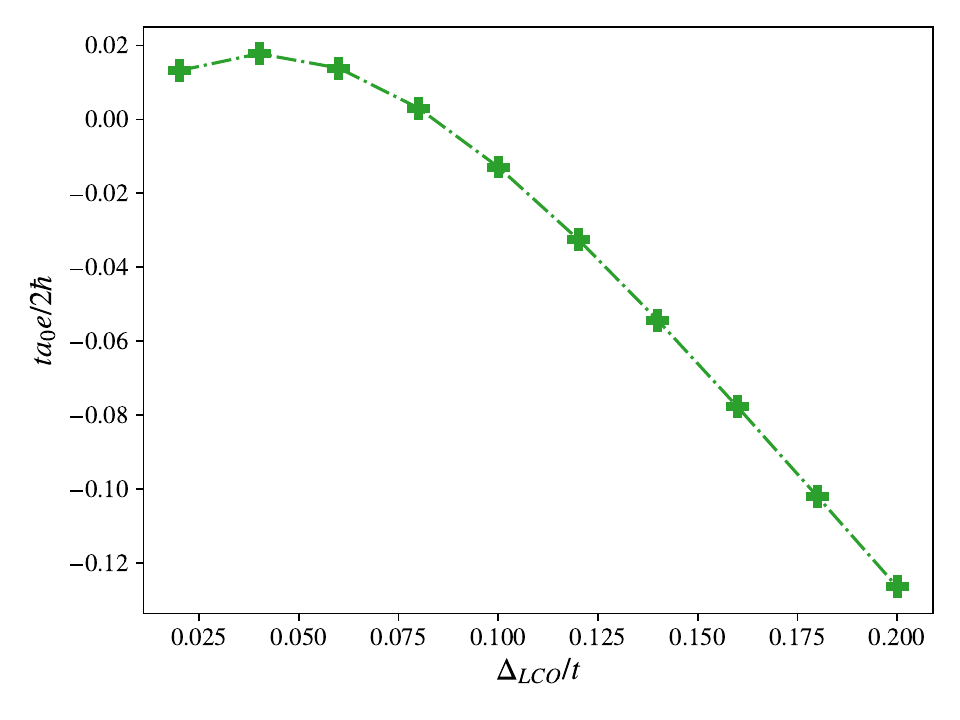}
    \caption{Orbital maganzation with LCO $1.5\Delta_{LCO}=1.5\Delta_{LCO}^{NN}=-\Delta_{LCO}^{NNN}$ per $2\times 2$ supercell.}
    \label{FigM}
\end{figure}
There is no local moment in the loop current phase since there is no net flux. However, as this loop current order spontaneously breaks the time-reversal symmetry. It still lead to a non-zero orbital magnetization, which is given by\cite{RevModPhys.82.1959}
\begin{equation}
    \bm{M}=\sum_n \int_{BZ} \frac{d\bm{k}}{(2\pi)^2} [\bm{m}_n(\bm{k}) -\frac{e}{\hbar}\epsilon_{n\bm{k}}\Omega_n(\bm{k})]f(\epsilon_{n\bm{k}}).
\end{equation}
$\bm{m}_n(\bm{k})$ is the orbital moment of $n$-th band and $\Omega_n(\bm{k})$ is the berry curvature. They are given by 
\begin{equation}
    \bm{m}_{n\bm{k}}=-i\frac{e}{2\hbar}(\langle\frac{\partial u_{nk}}{\partial k_x}|(H_{\bm{k}} -\epsilon_{nk}) |\frac{\partial u_{nk}}{\partial k_y} \rangle -H.c.)\hat{z}
\end{equation}
and 
\begin{equation}
    \Omega_n(\bm{k})=i(\langle\frac{\partial u_{nk}}{\partial k_x}|\frac{\partial u_{nk}}{\partial k_y} \rangle - H.c.).
\end{equation}

The numerical result of orbital magnetization is shown in Fig.\ref{FigM}. In $\text{CsV}_3\text{Sb}_5$ we use lattice constant $a_0 \simeq 5.4 \AA$ and $t=0.5 \text{eV}$, then the unit $\frac{ta_0^2 e}{2\hbar}\simeq 1.9\mu_B$. Then the total magnetization is about $10^{-2} \mu_B$.

\section{Effect of next-nearest hopping}
\begin{figure}
    \centering
    \includegraphics[width=0.9\textwidth]{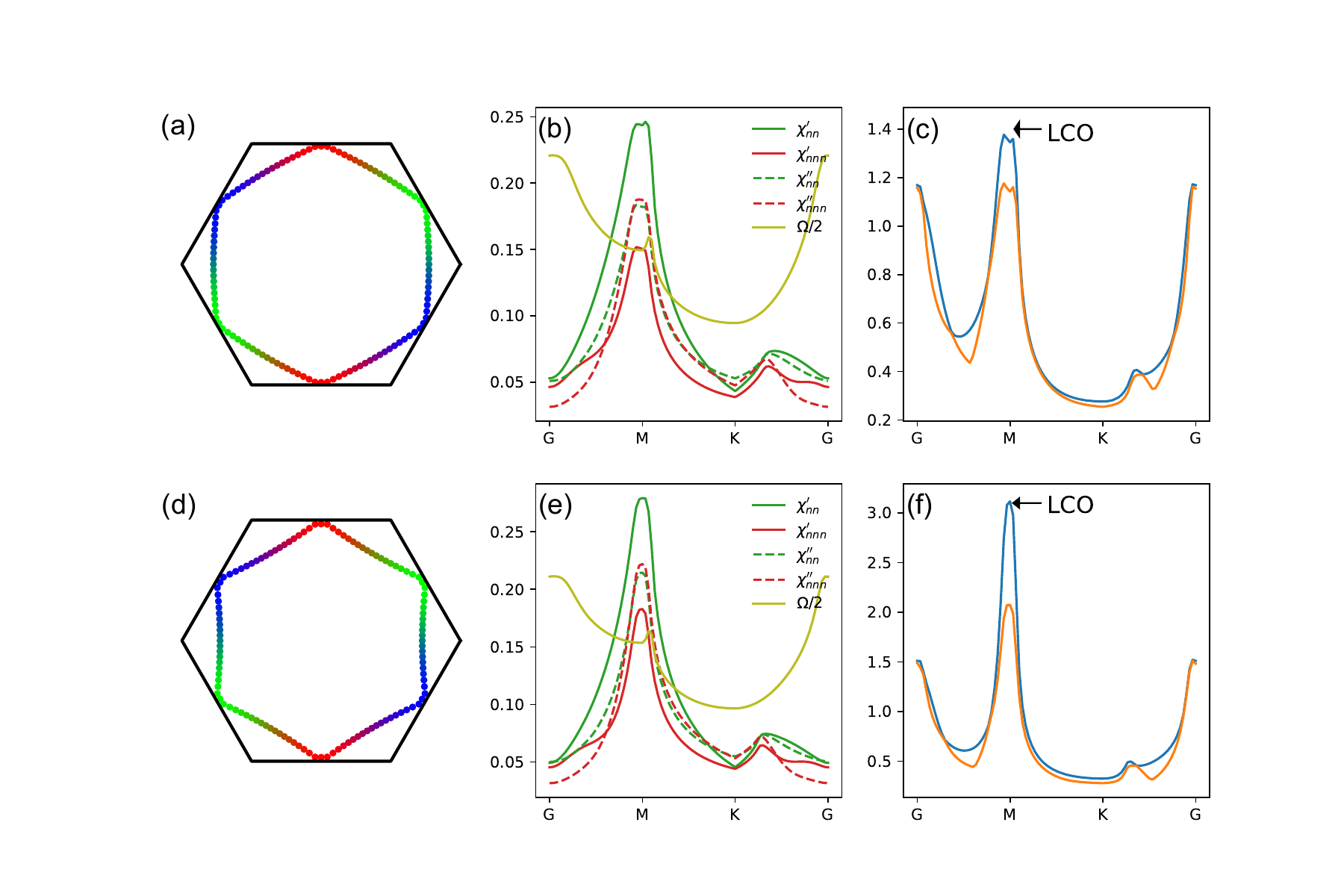}
    \caption{Fermi surface and susceptibilities with finite $t'$ (a)-(c) $t'=0.05$ (d)-(f) $t'=-0.05$. (a) and (c) are the fermi surfaces. (b) and (e) are bare susceptibilities along the high-symmetry line. (c) and (f) are the first two largest eigenvalues under interaction (c): $V_1=0.0\text{ } V_2=0.7$, (d): $V_1=0.0\text{ }V_2=0.85$. Susceptibilities are calculated under $\beta=200$}
    \label{Figt2}
\end{figure}

\begin{figure}
    \centering
    \includegraphics[width=0.5\textwidth]{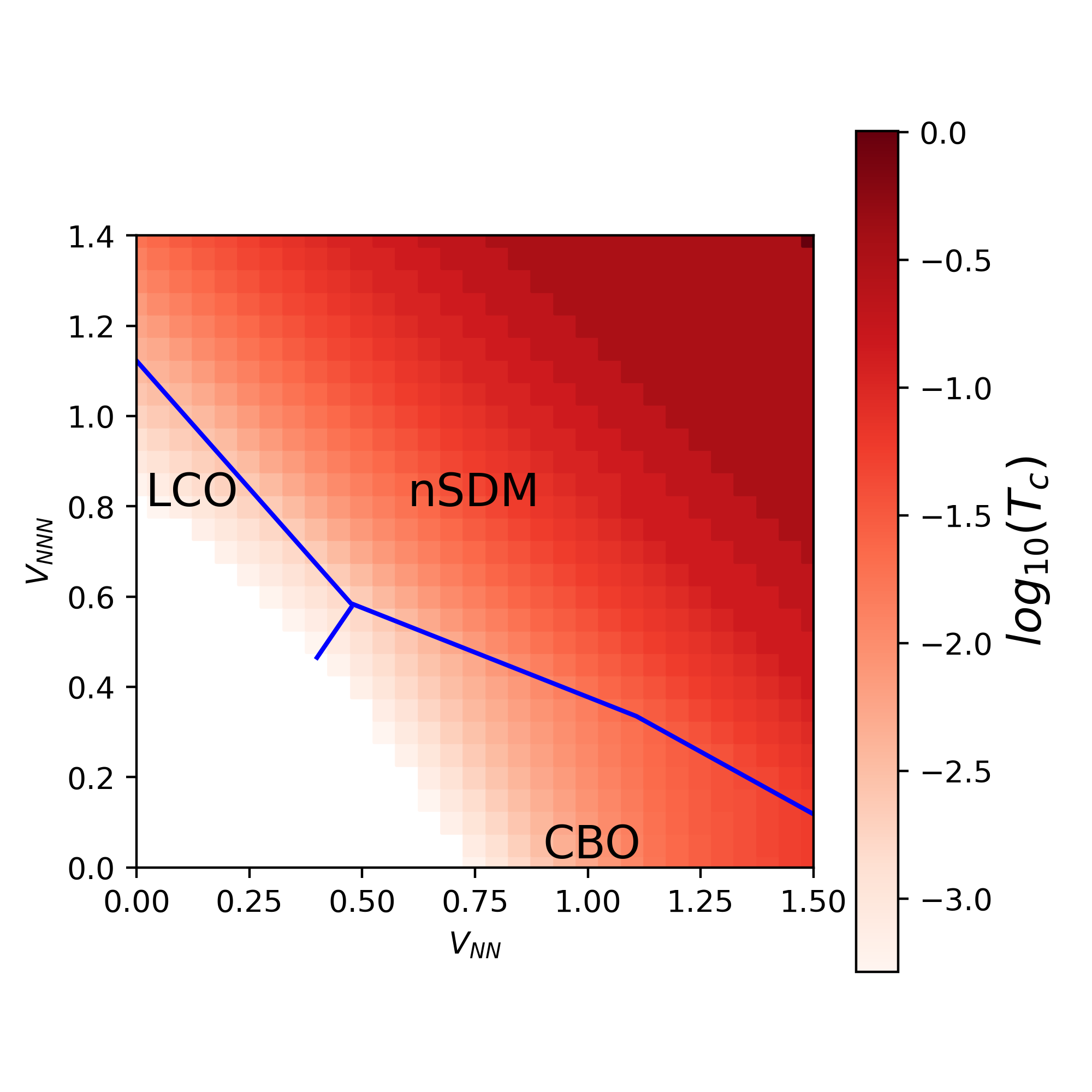}
    \caption{RPA phase diagram of charge orders when $t'=-0.1$.}
    \label{phase_diag}
\end{figure}
In the real material, next-nearest hopping $t'$ will be non-zero. In our model, a finite $t'$ will change the shape of the Fermi surface, and suppress the perfect nesting at $M$ point as shown in Fig.\ref{Figt2}. However, the LCO can still be the leading instability under suitable interaction. Our calculation shows that when $t'$ is negative, the Fermi surface turns to a shrinked hexagon. This shape will make the nesting vector between points aside to two VHSs closer to $\mathbf{M}$, which compensates the violation of perfect nesting and makes the $\mathbf{M}$ point susceptibility peak more stable. In the real 135 kagome material the Fermi surface is closer to the negative $t'$ case. It leads to a robust loop current phase. As shown in Fig.\ref{phase_diag}, this LCO phase can survive in $t'=-0.1$.

\section{Effective pairing interaction from onsite and bond charge fluctuations }
\begin{figure}
    \centering
    \includegraphics[width=1.0\textwidth]{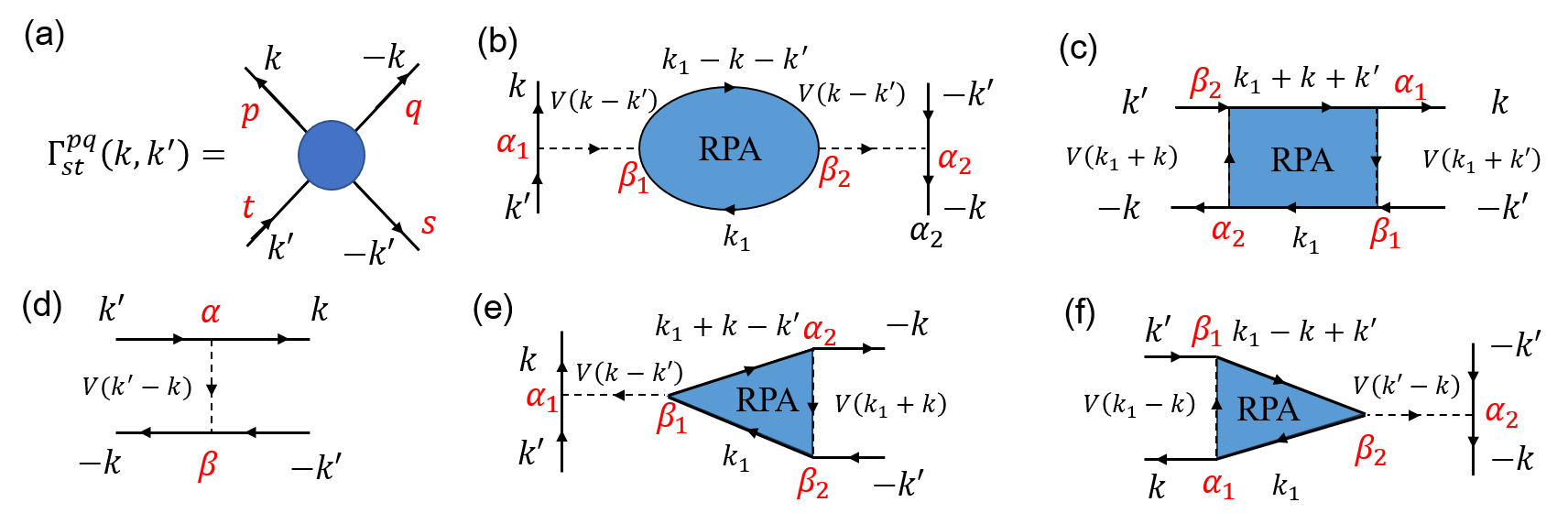}
    \caption{Superconductivity pairing given under RPA approximation. (a) Definition of the pairing kernel $\Gamma^{pq}_{st}$. (d) effective pairing at tree level. (b)(c)(e)(f) 2nd order Effective interaction vertex in SC pairing (a) bubble vertex (b) ladder vertex (c-d) vertex corrections. Red characters are sublattice indices.}
    \label{FigC}
\end{figure}
%In the spinless model, there is no limitation from spin index when write down a effective interaction vertex. For a given interaction vertex $H_I=\sum_{\alpha,\eta}V_{\alpha,\eta} c_\beta^\dagger c_\beta c_\gamma^\dagger c_\gamma$ we can assume the $\beta$ leg have a spin index and $\gamma$ leg have another spin index, so we can not connect them directly. Then we can get the true spinless vertex after anti-symmtrize the vertex with the assumed spin. In this section $\tilde{\Gamma}$ and $\Gamma$ represent the pairing vertex without or with this assumed spin index.

When the Fermi level moves away from VHSs, the Fermi surface nesting weakens, leading to the suppression of both onsite and bond charge orders. However, these charge fluctuations can promote particle-particle instabilities, i.e. superconductivity. In this section, we study the  superconducting pairing from these fluctuations of both onsite and bond charge orders based the Feynman diagrams. In the sublattice space, the general effective Cooper pair scattering interaction is
\begin{equation}
V^{\text{eff}}=\sum_{pqst} \tilde{\Gamma}^{pq}_{st}(\bm{k},\bm{k}') c^\dagger_p(\bm{k}) c^\dagger_q(\bm{-k}) c_s(\bm{-k'}) c_t(\bm{k'}),
\end{equation}
where $p,q,s,t$ are the sublattice indices and the $\tilde{\Gamma}^{pq}_{st}(\bm{k},\bm{k}')$ is anti-symmetric in the spinless case.
In the following, we deal with the effective interactions using the conventional diagrams technique and then do the antisymmetrization to obtain effective interaction for the spinless case. The bare interaction can be written as in a compact form 
\begin{equation}
    H_I=\frac{1}{N}\sum_{\gamma,\eta}\sum_{\bm{k,k',q}}V_{\gamma,\eta}(\bm{q}) c_\alpha^\dagger (\bm{k}) c_\alpha(\bm{k+q}) c_\beta^\dagger(\bm{k'+q}) c_\beta(\bm{k'}),
\end{equation}
with $\alpha,\beta,\gamma$ being the sublattice indices and $\eta=1,2$ represents NN or NNN repulsion. We have  $V_{\gamma,\eta}(\bm{q})=V_{\gamma,\eta}\cos(\bm{q}\cdot{l_{\gamma,\eta}})$ and $V_{1,\eta}=V_{2,\eta}=V_{3,\eta}=V_\eta$ for normal isotropic Colomb repulsion.

All the diagrams that contribute to the effective interactions are  displayed in Fig.\ref{FigC}(b-f).% where both bubble and ladder diagrams get involved. Here the pairing interaction is mediated by fluctuations in the onsite, bond and mixed channels. 
 The effective pairing interaction from the first-order diagrams reads,
\begin{align*}
    \Gamma^{\alpha \beta(1)}_{\beta \alpha}(\bm{k},\bm{k'})=\Gamma^{\beta \alpha(1)}_{\alpha \beta}(\bm{k},\bm{k'})= V_{\gamma,1}\cos(\bm{q\cdot l_{\gamma,1}})+V_{\gamma,2}\cos(\bm{q\cdot l_{\gamma,2}}) \equiv V_\gamma(\bm{q})=V(\bm{q}),
\end{align*}
with $\bm{q}=\bm{k-k'}$. 
At second-order both bubble and ladder diagrams get involved. Here the pairing interaction is mediated by fluctuations in the onsite, bond and mixed channels. For the bubble diagram shown in Fig.\ref{FigC} (b), the effective interaction comes from onsite charge fluctuations, which is given by,
\begin{align*}
    \Gamma^{\alpha_1 \alpha_2}_{\alpha_2 \alpha_1}(\bm{k,k'}) & = V_{\gamma_1}(\bm{q}) [\chi(\bm{q})]^{\beta_1 \beta_1}_{\beta_2 \beta_2} V_{\gamma_2}(\bm{q}).
\end{align*}
For the vertex correction (VC) and ladder diagrams, we need to choose suitable decoupling for the  internal-momentum dependent interaction, based on the indices of the vertices. For the VC diagram in Fig.\ref{FigC} (e), the interaction reads, 
\begin{eqnarray}
\Gamma^{\alpha_1 \alpha_2}_{\beta_2 \alpha_1}(\bm{k,k'}) =
\begin{cases}
 \sum_{s_2,\eta_2} s_2 V_{\gamma_1}(\bm{q}) [\chi(\bm{q})]^{\beta_1 \beta_1}_{\alpha_2 \beta_2,\gamma_2 s_2 \eta_2} V_{\gamma_2,\eta_2}f_{\gamma_2,s_2,\eta_2}(\bm{k }) & \epsilon_{\alpha_2\beta_2\gamma_2}=1 \vspace{2ex}\\
\sum_{s_2,\eta_2} V_{\gamma_1}(\bm{q}) [\chi(\bm{q})]^{\beta_1 \beta_1}_{\alpha_2 \beta_2,\gamma_2 s_2 \eta_2} V_{\gamma_2,\eta_2}f_{\gamma_2,s_2,\eta_2}(\bm{k'}) &  \epsilon_{\alpha_2,\beta_2,\gamma_2}=-1 \\
\end{cases}
\nonumber\\
\end{eqnarray}
The relevant indices are provided in VC in Fig\ref{FigC} (e).
Similarly, for the VC diagram in Fig\ref{FigC} (f), the interaction reads, 
\begin{eqnarray}
\Gamma^{\alpha_1 \alpha_2}_{\alpha_2 \beta_1}(\bm{k,k'}) =
\begin{cases}
\sum_{s_1,\eta_1} s_1 V_{\gamma_1,\eta_1} f_{\gamma_1,s_1,\eta_1}(\bm{k}) [\chi(\bm{q})]^{\beta_1 \alpha_1,\gamma_1 s_1 \eta_1}_{\beta_2 \beta_2} V_{\gamma_2}(\bm{q}) & \epsilon_{\beta_1,\alpha_1,\gamma_1}=1, \vspace{2ex}\\
\sum_{s_1,\eta_1} V_{\gamma_1,\eta_1} f_{\gamma_1,s_1,\eta_1}(\bm{k'}) [\chi(\bm{q})]^{\beta_1 \alpha_1,\gamma_1 \eta_1}_{\beta_2 \beta_2} V_{\gamma_2}(\bm{q}) &  \epsilon_{\beta_1,\alpha_1,\gamma_1}=-1 \\
\end{cases}
\nonumber\\
\end{eqnarray}
According our calculations, we find the these VC diagrams are numerically small but nonzero.

For the ladder diagram shown in Fig.\ref{FigC} (c), the interaction reads
\begin{eqnarray}
 \Gamma^{\alpha_1 \alpha_2}_{\beta_1 \beta_2}(\bm{k},\bm{k'}) =
\begin{cases}
\sum_{s_1,s_2,\eta_1,\eta_2} V_{\gamma_1,\eta_1} V_{\gamma_2,\eta_2} (s_1 s_2 f_{\gamma_2,s_2,\eta_2}(\bm{k})  [\chi(\bm{q'})]^{\beta_2 \alpha_2,\gamma_2 s_2 \eta_2}_{\alpha_1 \beta_1,\gamma_1 s_1 \eta_1} f_{\gamma_1,s_1,\eta_1}(\bm{k'})) & \epsilon_{\beta_2,\alpha_2,\gamma_2}=1, \epsilon_{\alpha_1,\beta_1,\gamma_1}=1 \vspace{2ex}\\
\sum_{s_1,s_2,\eta_1,\eta_2} V_{\gamma_1,\eta_1} V_{\gamma_2,\eta_2} (s_1 f_{\gamma_2,s_2,\eta_2}(\bm{k})  [\chi(\bm{q'})]^{\beta_2 \alpha_2,\gamma_2 s_2 \eta_2}_{\alpha_1 \beta_1,\gamma_1 s_1 \eta_1} f_{\gamma_1,s_1,\eta_1}(\bm{k})) &  \epsilon_{\beta_2,\alpha_2,\gamma_2}=1$, $\epsilon_{\alpha_1,\beta_1,\gamma_1}=-1 \vspace{2ex} \\
\sum_{s_1,s_2,\eta_1,\eta_2} V_{\gamma_1,\eta_1} V_{\gamma_2,\eta_2} (s_2 f_{\gamma_2,s_2,\eta_2}(\bm{k'})  [\chi(\bm{q'})]^{\beta_2 \alpha_2,\gamma_2 s_2 \eta_2}_{\alpha_1 \beta_1,\gamma_1 s_1 \eta_1} f_{\gamma_1,s_1,\eta_1}(\bm{k})) &   \epsilon_{\beta_2,\alpha_2,\gamma_2}=-1$, $\epsilon_{\alpha_1,\beta_1,\gamma_1}=1 \vspace{2ex} \\
\sum_{s_1,s_2,\eta_1,\eta_2} V_{\gamma_1,\eta_1} V_{\gamma_2,\eta_2} ( f_{\gamma_2,s_2,\eta_2}(\bm{k'})  [\chi(\bm{q'})]^{\beta_2 \alpha_2,\gamma_2 s_2 \eta_2}_{\alpha_1 \beta_1,\gamma_1 s_1 \eta_1} f_{\gamma_1,s_1,\eta_1}(\bm{k})) &
\epsilon_{\beta_2,\alpha_2,\gamma_2}=-1$, $\epsilon_{\alpha_1,\beta_1,\gamma_1}=-1 %\vspace{2ex}\\
%0 & \text{others}
\end{cases}
\nonumber\\
\end{eqnarray}
with $\bm{q'}=\bm{k+k'}$ in above equations. Among above cases $\epsilon_{\alpha_a,\beta_a,\gamma_a}=0$ is not allowed. By projecting the effective interaction onto the Fermi surface, the pairing vertex in the band space is given by \cite{Graser_2009,Kemper_2010,Wu2015,Dürrnagel2022},
\begin{equation}
    \Gamma_{\mu\nu}(\bm{k,k'})=\sum_{pqst} \Gamma^{pq}_{st}(k,k') a^{p*}_\mu(\bm{k}) a^{q*}_\mu(-\bm{k}) a^{s}_\nu(-\bm{k}') a^{t}_\nu(\bm{k}'),
\end{equation}
where $a^{q}_\mu(\bm{k})$ is the eigen state on the $\mu$ band at the Fermi point $\bm{k}$. The pairing vertex in our spinless case can be obtained by performing the anti-symmetrization on this effective interaction,
\begin{equation}
    \tilde{\Gamma}(\bm{k,k'})=\Gamma(\bm{k,k'})-\Gamma(\bm{k,-k'})
\end{equation} 
Near the transition temperature T$_c$, the gap function can be obtained by solving the following linearized gap equation,
\begin{equation}
-\int_{\mathrm{FS}} \frac{d \bm{k}^{\prime}}{V_G \left|v_F({\bm{k}^{\prime}})\right|} \tilde{\Gamma}\left(\bm{k}, \bm{k}^{\prime}\right) \Delta\left(\bm{k}^{\prime}\right)=\lambda \Delta(\bm{k}) ,
\end{equation}
where where $v_F({\bm{k}^{\prime}})$ is the Fermi velocity at the momentum $\bm{k}'$ on the Fermi surface and $V_G=\frac{8\pi^2}{\sqrt{3}}$ is the area the of Brillouin zone. Here $\lambda$ is the pairing strength and $\Delta(\bm{k})$ is the corresponding gap function on the Fermi surface. The dominant pairing is represented by the gap function associated with the largest positive pairing eigenvalue.

\section{Role of Bond and Onsite Fluctuation in determining pairing symmetry }
\begin{figure}
    \centering
    \includegraphics[scale=0.7]{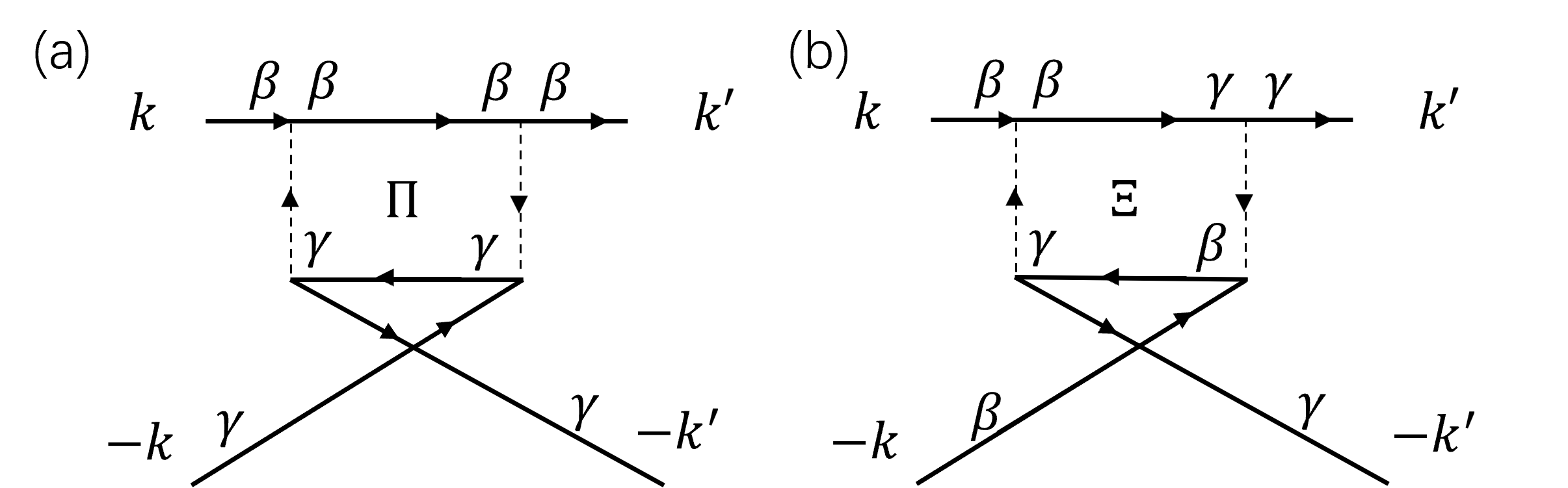}
    \caption{Pairing vertex mediated by two different bond fluctuations: $\Pi$-type (a) and $\Xi$-type fluctuations. }
    \label{sc_bond}
\end{figure}

\begin{figure}
    \centering
    \includegraphics[scale=0.42]{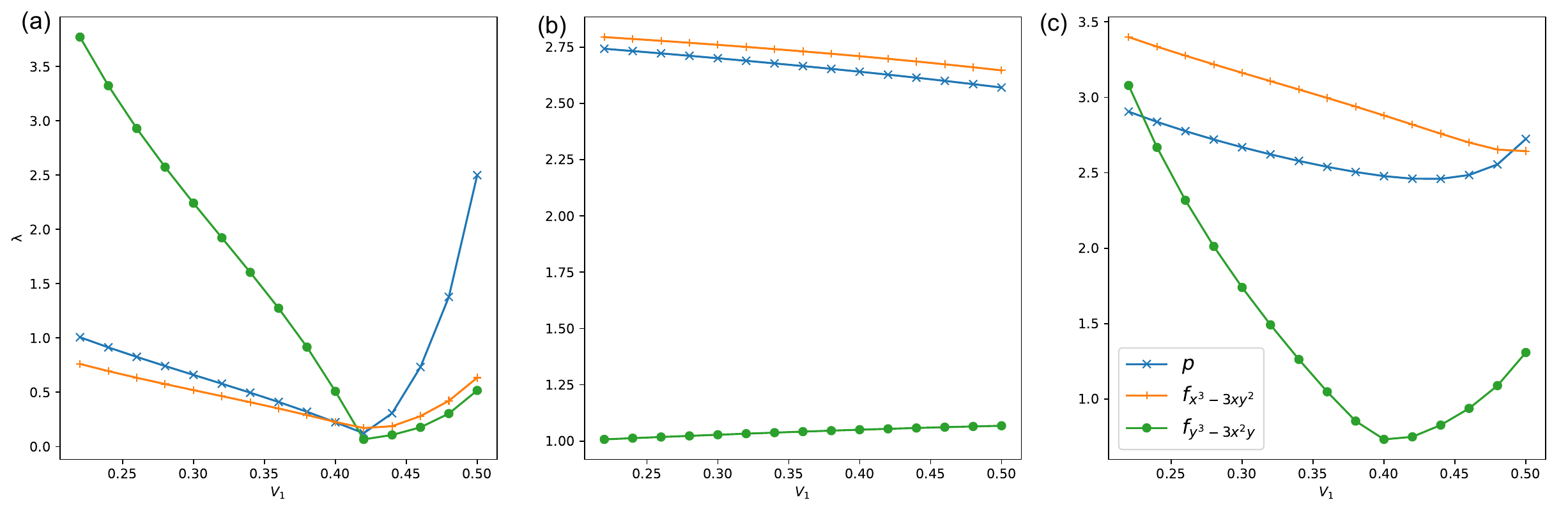}
    \caption{Leading pairing symmetry under $\mu=0.01$ and $\beta=150$ along the line $V_1+V_2=1.15$, when considering contribution of (a) ladder diagrams (b) bubble diagrams (c) full effective interaction}
    \label{pairing}
\end{figure}

%When considering about the pairing symmetry, the VC term in effective interaction is negligible as coupling between bond order and onsite order is small, the additional antisymmetric form factor cancels out when summing over the BZ. 
%The ladder type interaction shown in Fig.\ref{sc_bond} is attributed to bond susceptibilities which peaks at the $\textbf{M}$ points. So it contributes significantly to scattering between $\bn{M}$ points. In the undoped case, if considering scattering between $\bm{k=M_2}$ and $\bm{k'=M_3}$ as an example, in Fig.\ref{sc_bond}(a) case $\bm{k}$ and $\bm{k'}$ need same sublattice component and it is not allowed by p-type vHSs.

Dominant pairing gap is dictated by the pairing interactions. Now we focus on the Cooper pair scattering between states near VHSs, which is mainly mediated by bond fluctuations as shown in Fig.\ref{sc_bond}. Due to the unique sublattice texture on the Fermi surface, pairing scattering between different VHSs are primarily attributed to the $\Xi$-type fluctuation according to Fig.\ref{sc_bond} (b). Taking the Cooper pair scattering between $\bm{k}=\textbf{M}_2$ and $\bm{k'}=\textbf{M}_3$ as an example,
the pairing vertex is contributed by the $\Xi$-type fluctuation and it is given by,
%\begin{equation}
%    \Gamma(\bm{M_2,M_3}) \approx \sum_{\eta} V_{\eta} V_{\eta} (- f_{1,-,\eta}(\bm{M_3})  [\chi(\bm{q'})]^{2 3,1 - \eta}_{32,1 - \eta} f_{1,-,\eta}(\textbf{M_3}))\propto -((V_1)^2[\chi(\bm{q'})]^{2 3,1 - 1}_{32,1 - 1}+(V_2)^2[\chi(\bm{q'})]^{2 3,1 - 2}_{32,1 - 2}),
%\end{equation}
\begin{equation}
    \Gamma(\textbf{M}_2,\textbf{M}_3) \approx \sum_{\eta} V_{\eta} V_{\eta} (- f_{1,-,\eta}(\textbf{M}_3)  [\chi(\bm{q'})]^{2 3,1 - \eta}_{32,1 - \eta} f_{1,-,\eta}(\textbf{M}_3))\propto -((V_1)^2[\chi(\bm{q'})]^{2 3,1 - 1}_{32,1 - 1}+(V_2)^2[\chi(\bm{q'})]^{2 3,1 - 2}_{32,1 - 2}).
\end{equation}
This can be further written as $\Gamma(\textbf{M}_2,\textbf{M}_3)\propto -(V_{NN}^2\Xi_{22}+V_{NNN}^2\Xi_{88})$ using the notations in the main text. Therefore, the sign of this effective interaction is determined by the sign of the $\Xi$ susceptibility, which is the opposite for NN and NNN bonds. In the spinless case, this interaction should vanish at time reversal invariant points. In the main text, we consider the case with slight doping and the Fermi level moves away from VHSs. We consider two representative Fermi points near $\textbf{M}_{2,3}$, $\bm{k}=(1-\delta)\textbf{M}_2$ and $\bm{k'}=(1-\delta)\textbf{M}_3$ with $0<\delta\ll 1$. So the momentum transfer in $\Xi$ is shifted to $\bm{q}'=\bm{k}+\bm{k}'=-(1-\delta)\textbf{M}_1$, which is located on $\textbf{M}-\bm{\Gamma}$ line. In the antisymmetric interaction $\Gamma(\bm{k},-\bm{k}')$, the momentum transfer in $\Xi$ is shifted to $\bm{q}=\bm{k}-\bm{k}'=\bm{M_1}+\delta(\bm{M_3}-\bm{M_2})$, which is located on $\textbf{M}-\bm{K}$ line. From the Fig.2 in main text, we observe that the susceptibilities decays much faster on the $\textbf{M}-\bm{K}$ direction than the $\textbf{M}-\bm{\Gamma}$ direction, implying $|\Gamma(\bm{k},\bm{k}')|> |\Gamma(\bm{k},-\bm{k}')|$. Therefore, the pairing spinless vertex is dominantly determinated  by $\Gamma(\bm{k},\bm{k}')$. With this, the NN $\Xi$-type bond fluctuation induces a repulsive scattering interaction but the NNN $\Xi$-type bond fluctuation induces an attractive between k points near $\textbf{M}_2$  and $\textbf{M}_3$. When $V_2$ is dominant, the gap function on around $\textbf{M}_2$ and $\textbf{M}_3$ should have the same sign and this leads to the $f_{y^3-3x^2y}$-wave pairing shown in the main text. Since the antisymmetrization of interaction cancels most of the vertex $ \Gamma(\textbf{M}_2,\textbf{M}_3)$, the scattering is prominent when the susceptibility is nearly divergent. This explains why the $f_{y^3-3x^2y}$-wave pairing only emerges in a small regime of the pairing phase diagram (Fig.6(b) in the main text).
%The pairing instability $\lambda$ goes up quickly in Fig\ref{pairing}(a) when the interaction goes two opposite sides.

Away from VHSs, espetially near the midpoint between two VHSs (P$_2$ in the main text), the contribution of bubble-type diagrams becomes important. This bubble-type diagram contains onsite charge susceptibility peaking at $\bm{q}=0$. Near the $\pm$P$_2$ point, the contribution from bubble diagram can overcome that of ladder diagram when $V_2$ is dominant, as shown in Fig.5 (d),(f) in the main text, and this results a repulsive interaction between two opposite edges of the Fermi surfaces. This leads to  $f_{x^3-3xy^2}$-wave pairing. When $V_1$ is dominant, the dominant real bond fluctuations generate an attractive interaction between two opposite edges, leading to a $p$-wave pairing, as shown in the main text. In Fig.\ref{pairing}, we plot the pairing strength from ladder-type (a), bubble-type and total contribution as function of $V_1$ in the $V_1+V_2=1.15$ line regime. It is evident that the bond fluctuations favor the $f_{y^3-3x^2y}$-wave and $p$-wave pairing for the $V_2$-dominant and $V_1$-dominant regimes, respectively. In contrast, the onsite charge fluctuations favors the $f_{x^3-3xy^2}$-wave pairing. Increasing $V_1$ will drive a phase transition from the  $f_{x^3-3xy^2}$-wave pairing to the $p$-wave pairing, as shown in Fig.\ref{pairing} (c).

\section{Suppression of triplet superconductivity under non-magnetic impurity}
Since for triplet pairing symmetry, $\Delta(k)=-\Delta(-k)$, then it must be nodal to have a sign change along the Fermi surface. Small amount of non-magnetic disorder will change the superconducting gap size and critical temperature. For instance, if we consider a pairing with finite angular momentum $n$ in the electron gas \cite{SpringerLectureNotes}.

\begin{equation}
H =\sum_k \epsilon_k c_k^\dagger c_k + \Delta_n \cos(n\phi)  c_k^\dagger c_{-k}^\dagger + \text{H.c.}
\end{equation}

and small amount of onsite impurity, with onsite potential:
\begin{equation}
V(\mathbf{r}) = \sum_i u_0 \delta(\mathbf{r}-\mathbf{r}_i)
\end{equation}
where $\mathbf{r}_i$ is the location of the impurity atom. The Hamiltonian in Nambu space reads $H=\epsilon_k \tau_3+\Delta_ncos(n\phi)\tau_1$. Then we consider this problem in the $T$-matrix regime:
\begin{align}
G &= G_0 + G_0 T G_0 \\
T_{kk'} &= \frac{V_{k''k'}}{1 - V_{kk''} \sum_k G_{0,k}}
\end{align}
where $G_0 = (i\omega_n - H)^{-1}$ is the $2\times 2$ Green's function matrix for superconductor. $G$ is the full Green's function dressed by impurities.

For a finite $n$,
\begin{equation}
g_0 = \sum_k G_{0,k} = \frac{i\omega_n \tau_0}{\sqrt{\omega^2 + \Delta_0^2 \cos^2(n\phi)}}.
\end{equation}
Where $\tau_i$ is the Pauli matrix span the particle-hole space. As $\int_0^{2\pi} \cos(n\phi) d\phi = 0$, the $\Delta_n \tau_1$ term in $g_0$ vanishes. This is the main difference between SC with angular momentum and s-wave SC. The $\epsilon_k \tau_3$ term also vanishes for a constant DOS near the Fermi surface. When impurity scattering is weak, self-energy can be written as $\hat{\Sigma}(i\omega_n) = n_{\text{imp}} \langle \hat{T}(i\omega_n) \rangle_k$. Thus, the self-energy term can only modify the $\omega_n$ term in the Green's function. 

After considering disorder, the self-consistent equation changes to:
\begin{equation} \label{eq:gap}
\Delta_n = 2g_n \pi T \sum_{i\omega_n} \int \frac{d\phi'}{2\pi} 
\frac{\Delta_n \cos^2(n\phi)}{\sqrt{ \omega_n^2 (1 + \eta_{\omega_n})^2 + \Delta_n^2 \cos^2(n\phi) } } 
\end{equation}
where
\begin{equation}
\eta_{\omega_n} = \frac{1}{2\Gamma} \int \frac{d\phi'}{2\pi} 
\frac{1 + \eta_{\omega_n}}{\sqrt{ \omega_n^2 (1 + \eta_{\omega_n})^2 + (\Delta_n \cos(2\phi'))^2 }},
\end{equation}
with $\Gamma$ being the life time of quasi-particle $\frac{1}{2\Gamma}=-\text{Im}\Sigma(0)$.
Since the $\Delta_n$ term enters the modification of $\omega_n$, the structure of gap equation changes, which is different from s-wave superconductivity. Thus equation (\ref{eq:gap}) suggests that nonmagnetic impurity scattering is pairing breaking and can suppress the transition temperature as long as the Cooper pair carries non-zero angular momentum. This leads to the famous Abrikosov-Gorkov result:
\begin{equation}
\ln \frac{T_c}{T_{c0}} = \psi\left(\frac{1}{2}\right) - \psi\left( \frac{1}{2} + \frac{1}{4\pi\Gamma T_c} \right)
\end{equation}
$\psi$ is the digamma function and $T_{c0}$ is the transition temperature of a pure system. Then $T_c$ decreases when increasing impurity scattering. 

Since all triplet pairing symmetries are nodal, this result applies to all triplet cases in our calculation. So all obtained states in our RPA calculation will have a narrower gap and decreased transition temperature when considering non-magnetic disorder effects.

\end{document}